\documentclass{JFM-FLM_Au}

\usepackage{cancel}
\usepackage{upgreek}

\lefttitle{F. Steinbrenner, B. Turan, H. Teng and H. Xiao}
\righttitle{Turbulence generation and data assimilation with a latent diffusion model}

\title{Turbulence generation and data assimilation in wall-bounded flows with a latent diffusion model}

\author{Fabian Steinbrenner\aff{1,2,$\ast$}, Baris Turan\aff{1,2,$\ast$}, Hao Teng\aff{1,2}  \and Heng Xiao\aff{1,2}}

\affiliation{\aff{1}Stuttgart Center for Simulation
	Science, University of Stuttgart, 70569 Stuttgart,
	Germany\\
	\aff{2}Institute of Aerospace Thermodynamics,
	University of Stuttgart, 70569 Stuttgart, Germany \\
	\aff{$\ast$}Equal contribution
}

\corresau{Hao Teng, \email{ddsim.hao.teng@gmail.com} and Heng Xiao \email{heng.xiao@simtech.uni-stuttgart.de}}

\begin{document}
	\maketitle
	
\begin{abstract}
	Wall-bounded turbulent flows are chaotic and multiscale, rendering fast prediction at high Reynolds numbers computationally prohibitive in applications such as wind farms. Classical data assimilation is based on repeated solutions of the governing equations and thus inherits this cost. Generative models learn the probability distribution of flow states, enabling scalable probabilistic reconstruction. Our generative framework couples a variational autoencoder with a diffusion transformer to generate four-dimensional spatiotemporal samples. Bayesian conditioning enables data assimilation without retraining and allows statistical constraints to be imposed through sampling. The framework is applied to a subdomain of turbulent plane Couette flow, where the corresponding DNS in this generation region requires $O(10^6)$ spatial degrees of freedom. Using $O(10)$  latent spatial degrees of freedom, the model achieves a compression ratio of $O(10^5)$, which is one to two orders of magnitude above prior reports. It reproduces single-point statistics up to fourth order and the energy spectra, as well as the intermittency and phase-sensitive structure captured by velocity-increment and $Q$--$R$ statistics. Two assimilation scenarios demonstrate that, when observations are statistically consistent with the prior, conditional diffusion models with the proposed sampling strategy preserve complex turbulent statistics in the posterior. However, enforcing these constraints while preserving physical fidelity and sample diversity introduces an inherent trade-off. Excessive conditioning can distort the learned prior, paralleling limitations of classical ensemble-based data assimilation, where this can likewise degrade the prior covariance. These results highlight both the promise of diffusion models as probabilistic surrogates and the challenges of conditioning them.
\end{abstract}

\section{Introduction}
\label{sec:introduction}
Prediction and uncertainty quantification of unsteady turbulent flows in complex systems remains a fundamental challenge in fluid mechanics. This challenge is most severe for high-Reynolds-number, wall-bounded flows, where turbulence is multiscale, strongly non-stationary, and only sparsely observed. Accurate, time-resolved prediction of the evolving flow dynamics is therefore essential for monitoring, control, and decision-making.
This requirement goes beyond traditional engineering objectives that focus on time-averaged quantities. Emerging control and forecasting strategies instead demand fast and reliable predictions of transient flow behaviour under uncertain and rapidly changing conditions. Such demands arise in natural and industrial settings, including urban climate, atmospheric boundary layers, extreme weather events, and wind power plants \text{\citep{Bauer2015TheQR,veers2019grand}}, often coupled.

Wind farms provide a canonical example of this broader class of problems. They couple natural atmospheric turbulence with engineered systems and operate under highly unsteady inflow conditions \citep{veers2019grand}. Real-time turbine control strategies therefore rely on accurate predictions of the evolving flow field to optimize power generation and ensure stable operation \citep{Meyers2022WindFF}. Yet the core inputs, such as terrain, boundary conditions and atmospheric boundary layer stability, are often imprecisely known or vary over time. Moreover, turbulence at high-Reynolds numbers is inherently uncertain: it exhibits chaotic dynamics across a wide range of interacting scales. Modelling uncertainties such as subgrid-scale errors in large-eddy simulations  and discretization errors of the numerical solvers further contribute to the overall uncertainty of the system.  

Wind farms produce extensive observational data during operation through \text{meteorological} masts, unmanned aerial vehicles (UAV) \citep{Molter_2020} and light detection and ranging (LiDAR) systems \citep{guo2022space}. This data can be used to mitigate the aforementioned predictive uncertainty through data assimilation. Data assimilation combines incomplete observations of a system with the predictions of the numerical model to interpolate its state. Traditionally, data assimilation frameworks such as ensemble Kalman filter (EnKF) and four-dimensional variational methods have been coupled to a high-fidelity numerical solver, making their application to wind farm flows challenging due to the associated computational cost.

Recent advances in machine learning have spurred growing interest in data-driven approaches to address the limitations of traditional methods. For instance, \citet{brajard2021combining} combined a neural network-based model for estimating a correction to a reduced-order model and data assimilation based on EnKF. More recently, \citet{ozalp2026} similarly incorporated EnKF based data assimilation into a reduced-order model, utilising  EnKF to update the latent state of the reduced-order model in an augmented state-space formulation. Their framework performs stable and accurate predictions of chaotic systems like the Kuramoto–Sivashinsky equation and 2D Kolmogorov flow. 
\citet{valero2025improved} applied random forest regression to state estimation in turbulent channel flow, using sparse observations from a high-fidelity body-fitted simulation to correct the predictions of a low-fidelity immersed boundary method solver. 

\begin{figure}[t!]
	\centering
	\includegraphics[width=0.9\textwidth]{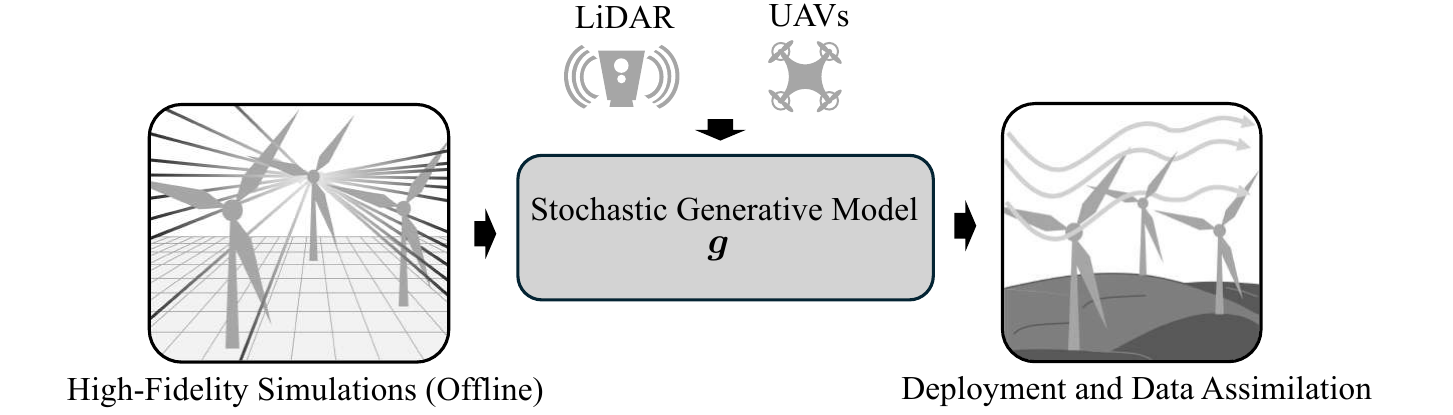}
	\caption{Our goal is to enable real-world data assimilation of wind farm flow fields under operating conditions, using high-fidelity simulations to train a stochastic generative model, and field observations from unmanned aerial vehicles (UAVs) and light detection and ranging (LiDAR).}
	\label{fig:vision-schematic}
\end{figure}

Among various machine learning frameworks, generative models offer great potential for addressing the challenges of real-time data assimilation. 
Stochastic generative models are data-driven approaches that learn an underlying data probability distribution and are capable of producing new samples that statistically resemble those observed in the training data. Building on these capabilities, we envision a data assimilation framework based on a stochastic generative surrogate model for the real-time reconstruction of the full wind-farm flow field from observational data. Here, ``real-time'' refers to the aspirational deployment target of
reconstructing the flow field rapidly enough to inform turbine control (e.g.\ rotor blade
pitch adjustment), and describes the long-term vision rather than the present framework. A schematic of this vision is outlined in figure \ref{fig:vision-schematic}.
In a first offline step, \textit{high-fidelity simulations} are carried out under several representative operating conditions such as different wind speeds, terrains and atmospheric boundary stabilities. Then, a \textit{stochastic generative model} is trained on this simulation data.  
The trained model can efficiently generate samples from the data distribution, replacing the costly numerical solvers used in traditional data assimilation frameworks. 
During \textit{deployment}, the surrogate model is exposed to operating conditions that differ, at least slightly, from those encountered during training. In this setting, the trained model provides a prior over plausible flow fields, which is combined with real-time observations through posterior sampling (conditional generation). The observations are intended to steer the stochastic generative model towards the actual operating conditions, which would allow it to adapt its predictions beyond the training distribution. 
\begin{figure}[t!]
	\centering
	\includegraphics[width=\textwidth]{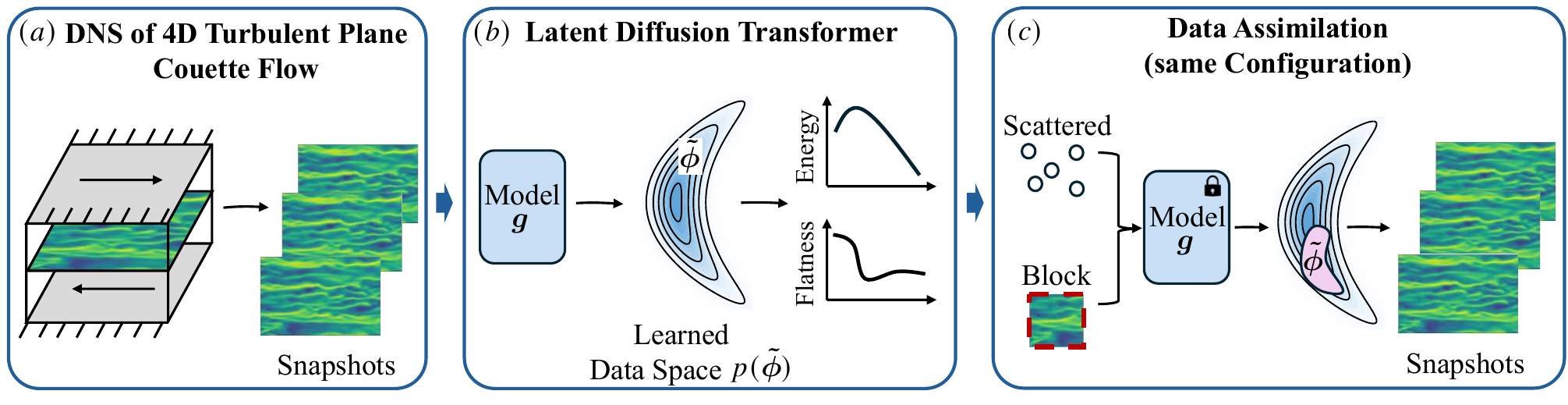}
	\caption{Schematic of the proposed three-step data assimilation framework using stochastic generative models. In this work, we establish the foundation for this vision. 
		$(a)$ We first construct a data distribution from \textit{DNS of four-dimensional turbulent plane Couette flow}. $(b)$ A \textit{latent diffusion transformer} model is then trained to generate statistically consistent flow-field samples from a compact sample space, faithfully reproducing the associated turbulent statistics. $(c)$ Finally, the model performs \textit{data assimilation for the same flow configuration} using diffusion posterior sampling with two observation types: scattered observations and a localized rectangular data block. 
	}
	\label{fig:scope-schematic}
\end{figure}

Diffusion models have recently emerged as the state-of-the-art in generative modelling, surpassing the previous generation of generative adversarial networks with their stable training and high sample quality \citep{dhariwal2021diffusion}. In diffusion models, Gaussian noise is first incrementally added to data samples e.g., flow field snapshots in the forward process, ultimately degrading them into pure noise. The added noise at a given step of the forward process is estimated by a neural network during training. The trained network is used to progressively remove the added noise in the reverse process, ultimately generating a clean sample. 
A key feature of diffusion models is their ability to generate conditional samples within a Bayesian-like framework, which enables the generation of samples consistent with specified conditions \citep{zhan2024conditional}. These conditions could be multi-source observations obtained from UAVs or LiDARs in the case of wind farms, allowing diffusion models to be used in data assimilation tasks. 

Despite the potential benefits, diffusion models have not yet been explored for data assimilation in three-dimensional, high-Reynolds-number, wall-bounded turbulent flows, which are relevant for wind farm applications. Indeed, their use in turbulence generation has been largely confined to simplified settings, primarily in two dimensions, including two-dimensional Kolmogorov flow snapshots \citep{SHU2023111972,shysheya2024conditional} and limited three-dimensional spatiotemporal generation \citep{gao2024bayesian}. 
Posterior sampling  from sensor data has been demonstrated mostly in one-dimensional and two-dimensional configurations \citep{li2024learning}. 
In all of these studies, diffusion models are applied directly in physical space without dimensionality reduction. However, in all the other cited works, the lack of dimensionality reduction limits their applicability to simplified scenarios due to the extremely high number of degrees of freedom required to resolve realistic high-Reynolds-number turbulence in four-dimensional spatiotemporal applications.

This limitation is addressed by latent diffusion models projecting the data to a low-dimensional latent space using a separate neural network \citep{rombach2022high}. For such a neural network, various architectures have been explored in the literature. One of these architectures is the conventional autoencoder, which can provide rich latent representations and superior reconstruction accuracy. The variational autoencoder (VAE) \citep{kingma2022autoencodingvariationalbayes}, on the other hand, regularises the latent space by imposing a Gaussian prior, leading to a smooth and compact latent space \citep{solera2024beta}. The~$\beta$-VAE offers further improvement on a smooth latent space by encouraging disentangled, i.e., uncorrelated and independent, latent representations \citep{higgins2017beta}. \citet{du2024conditional} introduced a latent diffusion framework based on conditional neural fields capable of generating four-dimensional spatiotemporal trajectories on unstructured grids. Similarly to \cite{li2024learning}, they demonstrate posterior sampling using the method of diffusion posterior sampling of~\citet{chung2023diffusion}. Recently, \citet{amoros2026guiding} employed a hybrid convolutional neural network (CNN)–transformer architecture, termed P3D, to reconstruct snapshots of two- and three-dimensional turbulence from sparse observations.

Guided by the overarching goal of using on-the-fly observations and diffusion models for data assimilation, in this work, we aim to lay the groundwork for such a deployment. Specifically, we focus on a simpler problem and propose a conditional latent diffusion model that can efficiently generate four-dimensional (i.e., three spatial dimensions and one time dimension) spatiotemporal trajectories of turbulent plane Couette flow at $Re_h=1300$ from a single direct numerical simulation (DNS). We use a diffusion model coupled with a $\beta$-VAE as the stochastic generative model. The scope and contributions of this paper are shown in figure \ref{fig:scope-schematic}. We first create our high-fidelity dataset from the \textit{DNS of a four-dimensional turbulent plane Couette flow} simulation. We then train the latent diffusion model on this dataset. 
The trained model serves as a prior over four-dimensional spatiotemporal turbulent flow fields and is shown to reproduce key turbulent statistics. 
The same dataset is then used for the \textit{data assimilation} of time series observations.

We adopt the \textit{diffusion transformer} (DiT) as the neural backbone of the architecture for our diffusion model. DiT is based on the transformer architecture of~\cite{vaswani2017attention}, which has been successfully applied to predict temporal dynamics in fluid flows~\citep{solera2024beta, gao2024generative}.  DiT has achieved state-of-the-art results in computer vision tasks \citep{Peebles_2023_ICCV}. However, it has been scarcely explored in the field of fluid mechanics, where previous works mainly used traditional U-net-based architectures~\citep{du2024conditional, gao2024bayesian}, with the notable exceptions of~\citet{zhou2025text2pde,amoros2026guiding}. 
The self-attention mechanism of the transformer can capture long-range dependencies more effectively than convolutional U-nets, which is crucial for incompressible flows, where pressure acts as a global Lagrange multiplier enforcing the divergence-free condition globally. 
We couple the DiT with a $\beta$-VAE in a two-stage framework, where the $\beta$-VAE learns a compact latent representation and the DiT models its spatiotemporal evolution. After training, the model enables posterior sampling when observational data are available. The original DiT architecture is extended to generate 4D spatiotemporal turbulent flows.

We train models with fewer than 96 degrees of freedom and show that all models down to 16 degrees of freedom can capture key turbulent phenomena and statistics with near DNS-level accuracy. The reduction in dimensionality from $O(10^6)$ spatial degrees of freedom to $O(10)$ in the latent space represents a compression ratio of $O(10^5)$ and is of an order of magnitude higher than most current 
data-driven reduced-order or latent diffusion models for turbulent flows. \citep{eivazi2022towards, linot2023dynamics, du2024conditional, solera2024beta, vinograd2025reduced}.
We also demonstrate consistency with the turbulent statistics when conditioned on pointwise time-series observations.
In practical scenarios, one can only expect low-order statistics such as mean or Reynolds stress derived from incomplete experimental observations \citep{mons2021ensemble}. 
These observations define marginal distributions, while the full flow field corresponds to a joint distribution in a high-dimensional state space with unobserved degrees of freedom marginalized implicitly. 

Current diffusion-based frameworks enforce statistical quantities by retraining and corrector steps during sampling \citep{jacobsen2025cocogen} or defining the quantity of interest as an observation operator and using automatic differentiation \citep{gao2024bayesian,liu2025confildinlet}. In our method, we instead condition the diffusion model on pointwise time-series observations drawn from the flow, so that consistency with the turbulent statistics emerges through Bayesian conditioning on samples of the underlying distribution rather than through differentiation of statistical operators. We demonstrate this approach on two data assimilation tasks resembling real-world observations. 
Because both the prior and the assimilated observations 
are derived from the same DNS, the present study constitutes an 
in-distribution proof of concept; demonstrating generalisation to 
observations whose statistics depart from the prior is left for future 
work. We demonstrate that, 
even in this in-distribution regime, 
simultaneously maintaining the turbulent statistics 
encoded in the prior, consistency with the observations, and the 
stochastic nature of the diffusion model represents a set of conflicting 
objectives and is by no means trivial. 
Specifically, we show that spatially dense observations, which are 
therefore strongly correlated, can distort the learned diffusion prior 
and lead to departures from correct turbulent statistics. This behaviour 
parallels known challenges in traditional data assimilation methods such 
as the EnKF, where the analysis update can violate physical constraints, 
including conservation laws. The framework introduced 
here provides a foundation on which generalisation to flow conditions 
absent from the training data could be built; demonstrating such 
generalisation requires assimilating observations whose statistics depart 
from the prior and is beyond the scope of the present study.

Our approach differs from the majority of previous latent-space generative models, which typically validate results only qualitatively or with limited turbulent statistics. In the present work, we assess performance quantitatively and comprehensively, benchmarked against the standard established for conventional DNS \citep{kim1987turbulence} and extended
beyond it. We evaluate single-point statistics up to fourth order together with the premultiplied
energy spectra, and further examine two phase-sensitive measures that lie outside that standard:
the scale-dependent PDF of the longitudinal velocity increment, which probes intermittency, and the joint PDF of the velocity-gradient invariants $Q$ and $R$, which probes the alignment between strain and rotation at the small scales. These metrics are used to determine the minimal latent dimension, a challenging task for high-Reynolds-number wall-bounded turbulence that is often omitted in previous frameworks combining learned encoders and forecasters.
In evaluating the conditionally generated turbulence, we focus on turbulent statistics in addition to the root mean squared error (RMSE) relative to instantaneous trajectories. While conditioning may reduce RMSE compared with prior sampling, physical consistency is ultimately assessed through the reproduction of statistical quantities, which are the primary descriptors of turbulent dynamics.
Furthermore, to our knowledge, the present work represents the first use of $\beta$-VAE for reduced-order modelling in latent diffusion models for turbulent flows. Previous work has demonstrated the effectiveness of $\beta$-VAE to extract a minimal set of latent variables in simpler flow scenarios \citep{eivazi2022towards, solera2024beta}. 
Such an ability of the $\beta$-VAE likely underlies the high compression ratio described above for the fully turbulent wall-bounded flow considered here.

The paper is organised as follows. The methodology and the training dataset are described in \S\ref{sec:methods} and \S\ref{sec:Comp_setup} respectively. Results on dimension reduction, turbulence generation, and data assimilation are presented in \S\ref{sec:results}. Conceptual aspects of generative models for turbulent flows are discussed in \S\ref{sec:discussions}. Conclusions and implications are given in \S\ref{sec:conclusions}.

\section{Methods}
\label{sec:methods}
\begin{figure}[t!]
	\centering
	\includegraphics[width=\textwidth]{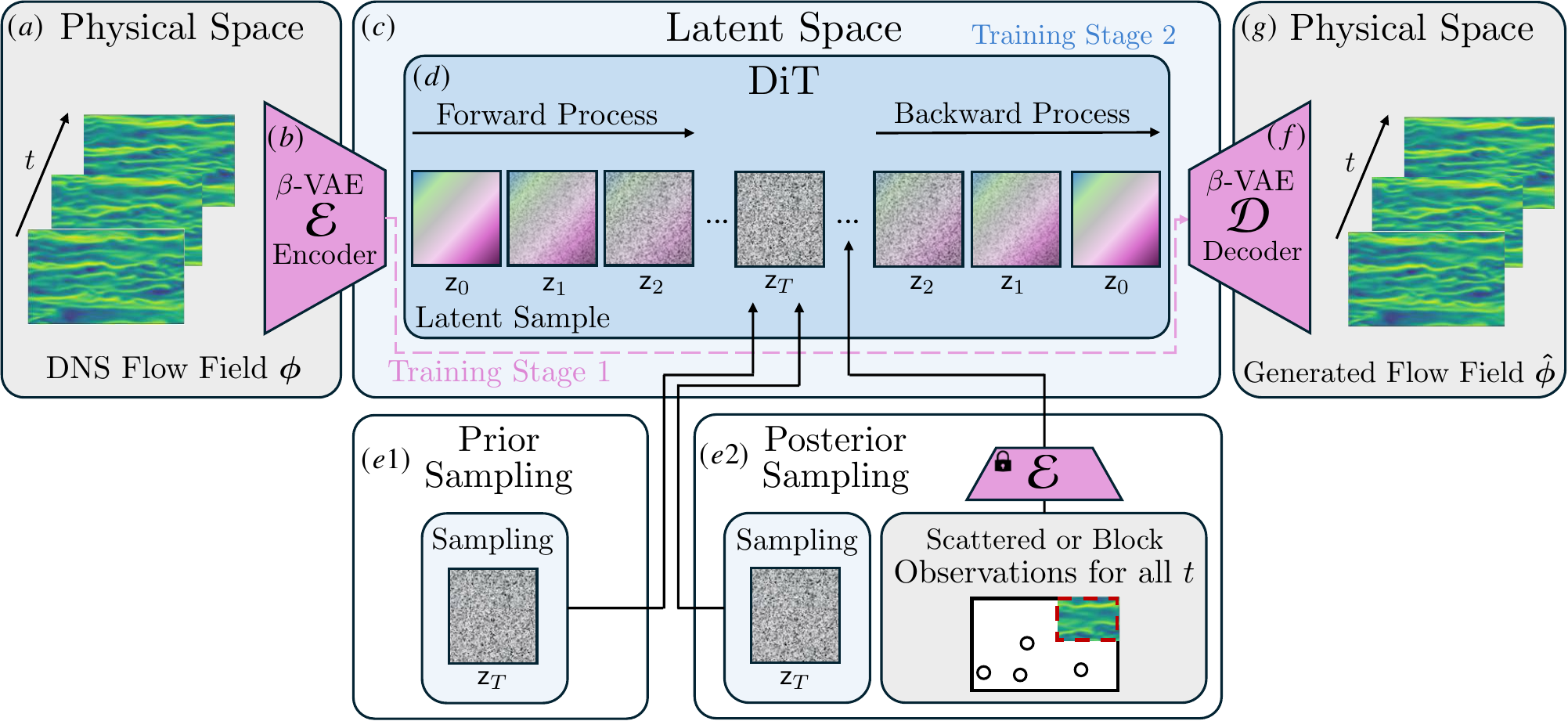}
	\caption{Schematic of the latent conditional diffusion framework. ($a,g$) The input and output of the framework is a four-dimensional spatiotemporal flow field in physical space. ($b$) The encoder of a $\beta$-VAE maps a temporal sequence of DNS snapshots $\boldsymbol{\phi}$ of velocities $u, v, w$ and pressure $p$ into a low-dimensional latent space $\mathsf{z}$ $(c)$. ($d$) A diffusion transformer (DiT) learns the latent dynamics. The learned DiT generates new a sequence of latents by prior sampling ($e1$) or posterior sampling ($e2$, provided observations in physical space), where $\mathsf{z}_0$ denotes the clean latent sample and $\mathsf{z}_T$ the fully diffused one. ($f$) The decoder then reconstructs four-dimensional spatiotemporal flow fields $\boldsymbol{\hat{\phi}}$ from the generated latents. Training of the $\beta$-VAE and the DiT is performed in two separate stages.}
	\label{fig:PCF_framework}
\end{figure}

Our generative learning framework aims to construct a reduced-order model of turbulent flows capable of generating temporal sequences of three-dimensional turbulent flow snapshots. We achieve this through a two-stage latent diffusion approach capable of learning the underlying probability distribution of the four-dimensional turbulent flow. An overview of the complete pipeline is shown in figure~\ref{fig:PCF_framework}. Given three-dimensional flow-field snapshots in physical space (figure~\ref{fig:PCF_framework}$a$), a $\beta$-VAE encoder (figure~\ref{fig:PCF_framework}$b$) compresses the input into a compact latent representation (figure~\ref{fig:PCF_framework}$c$), and a diffusion transformer (DiT) models the temporal evolution in this latent space (figure~\ref{fig:PCF_framework}$d$). 
The learned DiT generates new sequence of latents by prior sampling (figure~\ref{fig:PCF_framework}$e1$) or posterior sampling (figure~\ref{fig:PCF_framework}$e2$, provided observations in physical space).
The decoder of the $\beta$-VAE (figure~\ref{fig:PCF_framework}$f$) then maps these sequences back to the physical domain (figure~\ref{fig:PCF_framework}$g$).

We train the two components sequentially. First, the $\beta$-VAE learns a low-dimensional encoding of the turbulent snapshots (see figures~\ref{fig:PCF_framework}$b, f$). Next, the DiT learns the distribution of the latent trajectories (see figure~\ref{fig:PCF_framework}$c,d$).
After training, the framework enables data assimilation through posterior sampling when observation data are available, as described in \S\ref{subsec:Methods_cDM}.

\subsection{Dimensionality Reduction with $\beta$-VAE}
\label{subsec:Methods_ROM}
Let $\boldsymbol{\Phi}(\boldsymbol{x},t)=[u(\boldsymbol{x},t), v(\boldsymbol{x},t), w(\boldsymbol{x},t), p(\boldsymbol{x},t)]\in\mathbb{R}^4$ be an incompressible turbulent flow field with velocities $u$, $v$, $w$, and pressure $p$, and $\boldsymbol{x}=[x,y,z]$ be the spatial coordinate. We discretise the spatial domain using a structured Cartesian grid of $d=d_1 \times d_2 \times d_3$ points, with $d_1$, $d_2$, and $d_3$ points in the $x$-, $y$-, and $z$-directions, respectively.
We define a snapshot of the flow at time $t$ as the array
\begin{equation}
	\boldsymbol{\phi}(t)=[\boldsymbol{\Phi}(x_1,y_1,z_1,t),\, \cdots\,,\boldsymbol{\Phi}(x_{d1},y_{d2},z_{d3},t)]\in\mathbb{R}^{d_c\times d} ,
\end{equation}
where $d_c=4$ is the number of input channels, corresponding to each of the four field variables. 

\begin{figure}[t!]
	\centering
	\includegraphics[width=\linewidth]{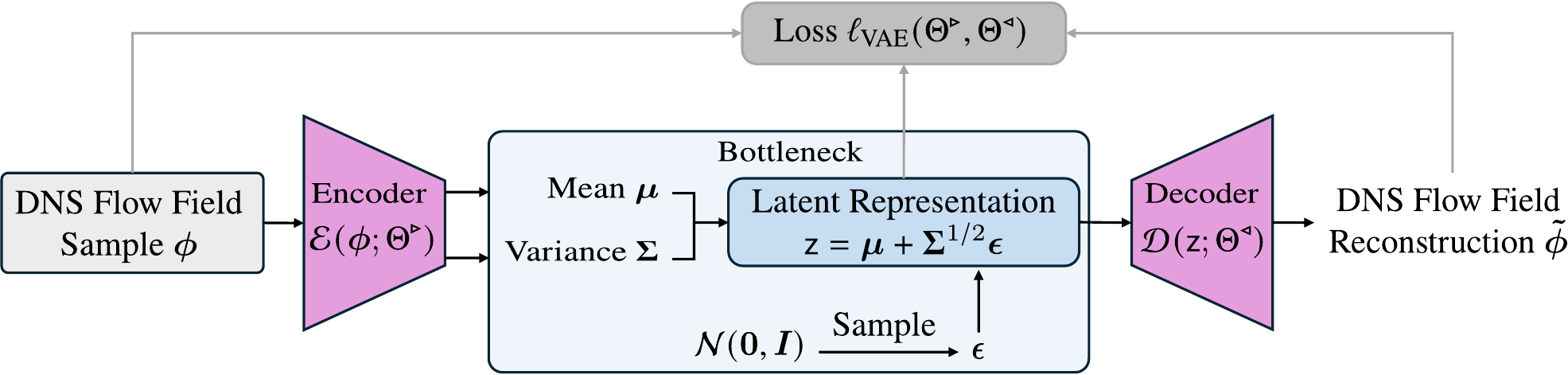}
	\caption{Schematic of a VAE architecture. The encoder takes a DNS flow field sample and predicts the mean $\boldsymbol{\mu}$ and variance $\boldsymbol{\Sigma}$ of the variational distribution. In the bottleneck, a sample is drawn from a normal distribution and the latent representation is constructed as a sample from the variational distribution. The decoder reconstructs the input data. The training is conducted by minimising the loss function in equation~\ref{eq:vae_loss}. The detailed $\beta$-VAE architecture is given in appendix~\ref{subsec:app_VAE}.
	}
	\label{fig:VAE_schematic}
\end{figure}

We reduce the dimensionality of the turbulent flow fields using a $\beta$-variational autoencoder ($\beta$-VAE)~\citep{higgins2017beta}.
The $\beta$-VAE is essentially a variational autoencoder (VAE) with an additional regularisation in the loss function, whose strength is controlled by a hyperparameter $\beta \in \mathbb{R}$ (see equation~\ref{eq:vae_loss} later). A schematic of the VAE architecture used in the first training stage is shown in figure~\ref{fig:VAE_schematic}.
A VAE consists of an encoder and a decoder, each composed of multiple nonlinear blocks, which together learn a probabilistic latent-variable representation of the data in the bottleneck.
Specifically, given an input snapshot~$\boldsymbol{\phi} \in \mathbb{R}^{d_c \times d}$ at time $t$ over all spatial locations, the encoder $\mathcal{E}$ maps the snapshot to a low-dimensional latent representation $\mathsf{z} \in \mathbb{R}^{\tilde{d}_c \times \tilde{d}}$,
where $\mathsf{z}=\mathcal{E}(\boldsymbol{\phi};\Theta^\triangleright)$ defines a function $\mathcal{E}: \mathbb{R}^{d_c \times d}\rightarrow \mathbb{R}^{\tilde{d_c} \times \tilde{d}}$ with the reduced physical dimension $\tilde{d} = \tilde{d}_{1} \tilde{d}_{2} \tilde{d}_{3}$ and the number of latent channels $\tilde{d}_c$. The decoder~$\mathcal{D}$ reconstructs the input field $\boldsymbol{\phi}$ from the latent bottleneck according to the mapping $\mathcal{D}:\mathbb{R}^{\tilde{d_c} \times \tilde{d}}\rightarrow \mathbb{R}^{d_c \times d},\tilde{\boldsymbol{\phi}} = \mathcal{D}(\mathsf{z}; \Theta^\triangleleft)$.
The dimensionality of the latent space $\tilde{d}_c \times \tilde{d}$ is typically much smaller than that of the input space, $d_c \times d$, by a factor of $10^5$ as will be shown later.
Since the spatial dimensionality of the data is significantly higher than the temporal dimensionality, we apply the encoder and decoder exclusively along the spatial dimensions, treating each time instance $\boldsymbol{\phi}(t)$ as an independent snapshot. 
In general, the temporal dimension can also be incorporated into autoencoder-based dimensionality reduction to achieve a more efficient representation. However, decoupling spatial compression from temporal modelling is a common and well-motivated strategy in reduced-order modelling of turbulent flows~\citep{eivazi2022towards,du2024conditional,solera2024beta}, as it allows each sub-problem to adopt a specialised architecture suited to its structure. The $\beta$-VAE employs three-dimensional convolutional networks (CNNs), which exploit local spatial inductive bias~\citep{lecun2002gradient,battaglia2018relational} to construct structured latent representations of coherent flow structures~\citep{Maulik2021,LEE2020108973,murata2020nonlinear}. The transformer-based DiT is adopted as the neural backbone of the diffusion model for temporal modelling in latent space, as transformers represent the state-of-the-art for sequence modelling and temporal predictions, with their self-attention mechanism enabling the capture of long-range temporal dependencies~\citep{solera2024beta,wang2024towards}. Spatiotemporal autoencoder-based latent representations feeding a diffusion model remain largely unexplored in turbulence reduced-order modelling and represent an interesting direction for future work. 

The encoder and decoder mappings are represented by neural networks with trainable parameters~$\Theta^{\triangleright}$ and $\Theta^{\triangleleft}$, respectively. A detailed description of the encoder and decoder model architectures and the choice of hyperparameters is provided in appendix~\ref{subsec:app_VAE}. The VAEs model the data distribution $\mathsf{p}(\boldsymbol{\phi})$ indirectly through the joint distribution~$\mathsf{p}(\boldsymbol{\phi},\mathsf{z}) = \mathsf{p}(\boldsymbol{\phi}\mid \mathsf{z})\,\mathsf{p}(\mathsf{z})$ and obtain $\mathsf{p}(\boldsymbol{\phi})$ through marginalization. This formulation enables nonlinear dimensionality reduction because simple low-dimensional expressions for the conditional distribution $\mathsf{p}(\boldsymbol{\phi}\mid\mathsf{z})$ and the prior $\mathsf{p}(\mathsf{z})$ can describe complex high-dimensional data. The encoder and decoder are trained jointly by minimising the following loss function
\begin{equation}    
	\ell_{\text{VAE}}(\Theta^\triangleright, \Theta^\triangleleft)
	= \lVert \boldsymbol{\phi} - \boldsymbol{\tilde{\phi}} \rVert^2
	- \beta \frac{1}{2}\left(
	\text{tr}(\boldsymbol{\Sigma})
	+ \boldsymbol{\mu}^\top \boldsymbol{\mu}
	- d_\mathsf{z}
	- \log(\det(\boldsymbol{\Sigma}))
	\right),
	\label{eq:vae_loss}
\end{equation}
where tr and det indicate trace and determinant, respectively; $\boldsymbol{\mu}$ and $\boldsymbol{\Sigma}$ denote the mean and covariance of the variational normal distribution
\(
\var(\mathsf{z}\mid \boldsymbol{\phi};\Theta^{\triangleright}) = \mathcal{N}(\boldsymbol{\mu},\boldsymbol{\Sigma}),
\)
and the prior is $\mathsf{p}(\mathsf{z})=\mathcal{N}(\boldsymbol{0},\mathsfbi{I})$. Here, $d_\mathsf{z}\equiv\tilde{d}_c\tilde{d}$ denotes the total number of degrees of freedom in the latent space. The loss is commonly known as negative evidence lower bound (ELBO). The first term in the loss enforces reconstruction accuracy by minimising the $L^2$-norm $\lVert \cdot\lVert$ between the reconstructed and input samples. The second term is the Kullback-Leibler divergence $D_\text{KL}\left( \var(\mathsf{z} | \boldsymbol{\phi}; \Theta^\triangleright) \, \lVert \, \mathsf{p}(\mathsf{z}) \right)$ and measures how similar the variational distribution $\var(\mathsf{z} | \boldsymbol{\phi}; \Theta^\triangleright)$ is to the prior $\mathsf{p}(\mathsf{z})$. During training, the encoder computes the mean $\boldsymbol{\mu}$ and the covariance $\boldsymbol{\Sigma}$ from the input snapshot, as shown in figure \ref{fig:VAE_schematic}. Then, we draw a sample $\boldsymbol{\epsilon} \sim \mathcal{N}(\boldsymbol{0},\mathsfbi{I})$ and construct the latent variable as $\mathsf{z}=\boldsymbol{\mu} + \boldsymbol{\Sigma}^{1/2} \boldsymbol{\epsilon}$ to draw from the intended variational distribution. The latent sample $\mathsf{z}$ is then fed to the decoder, which reconstructs the input snapshot $\boldsymbol{\tilde{\phi}}$. The reconstructed snapshot is used to compute the loss, and an optimization algorithm updates the parameters of the encoder and decoder through backpropagation and autodifferentiation of the gradients.

The number of latent channels $\tilde{d}_c$ (ranging from 1 to 4; see table~\ref{tab:vae_details} in appendix~\ref{subsec:app_VAE}) and the reduced physical dimension $\tilde{d}$ are design choices that determine the capacity of the bottleneck and consequently the efficiency of the latent representation. The latent channels are not restricted to any of the $d_c = 4$ channels for the physical variables (velocities $u, v, w$, or pressure $p$); instead, they jointly encode the whole flow information. Restricting the latent dimensionality imposes a bottleneck on the information that can pass through the network, leading to a trade-off between reconstruction fidelity and dimensionality reduction. In combination with an appropriate choice of the hyperparameter $\beta$, this trade-off can yield efficient latent representations that balance information preservation with channel capacity limitations.
Ideally, each latent channel would correspond to an interpretable flow factor, a property associated with a high degree of disentanglement in the latent space \citep{higgins2017beta}. Achieving such interpretability for flows, however, remains challenging and has so far only been demonstrated for low-Reynolds-number regimes \citep{solera2024beta} or two-dimensional cases \citep{eivazi2022towards}.

\subsection{Prior Sampling with Diffusion Transformer (DiT)} 
\label{subsec:Methods_DiT}
We now describe the second stage of the framework: modelling the temporal evolution of the latent representations. A Denoising Diffusion Probabilistic Model (DDPM) \citep{ho2020denoising} is trained to generate temporal sequences $\mathsf{z}_0$ of $d_t$ latent variables, which are subsequently decoded into physical space by the $\beta$-VAE decoder to yield the generated flow field $\hat{\boldsymbol{\phi}} = \mathcal{D}(\mathsf{z}_0; \boldsymbol{\Theta}^{\triangleleft})$. Note that $\hat{\boldsymbol{\phi}}$ is distinct from the VAE reconstruction $\tilde{\boldsymbol{\phi}}$ introduced in \S\ref{subsec:Methods_ROM}, which is obtained by directly encoding and decoding a DNS snapshot; here, $\hat{\boldsymbol{\phi}}$ represents a genuinely new sample drawn from the learned distribution of turbulent flow fields.

To train the DDPM, the sequences of latent variables are first progressively perturbed with Gaussian noise in the forward process
This process is defined by the transition kernel 
\begin{equation}
	\var(\mathsf{z}_\tau|\mathsf{z}_{\tau-1})=\mathcal{N}\!\left(\mathsf{z}_{\tau};\sqrt{1-\gamma_\tau}\mathsf{z}_{\tau-1}, \gamma_\tau\mathsfbi{I}\right).
\end{equation}
Here, $\mathsf{z}_\tau$ denotes the intermediate noised sample at diffusion time $\tau=1,\dots,T$ and $\gamma_\tau$ is the variance of the added noise at each step, determined by the variance schedule. In this work, we use $T=1000$ and a linear variance schedule with $\gamma_1=1\times10^{-4}$ and $\gamma_{1000}=0.02$, following \citet{ho2020denoising}. Using the reparametrization trick, $\mathsf{z}_\tau$ can be expressed as
\begin{equation}
	\mathsf{z}_\tau=\sqrt{1-\gamma_\tau}\mathsf{z}_{\tau-1}+\sqrt{\gamma_\tau}\boldsymbol{\epsilon},
\end{equation}
with $\boldsymbol{\epsilon}\sim\mathcal{N}\!(\boldsymbol{0},\mathsfbi{I})$.

The noised latent $\mathsf{z}_\tau$ at an arbitrary time $\tau$ can then be sampled in closed form as
\begin{equation}
	\mathsf{z}_\tau=\sqrt{\bar{\alpha}_\tau}\mathsf{z}_0+\sqrt{1-\bar{\alpha}_\tau} \boldsymbol{\epsilon}_\tau,
\end{equation}
where $\alpha_\tau=1-\gamma_\tau$, $\bar{\alpha}_\tau=\prod_{i=0}^\tau\alpha_\tau$ and noise $\boldsymbol{\epsilon}_\tau\sim\mathcal{N}\!(\boldsymbol{0},\mathsfbi{I})$.
A neural network is trained to reverse this process by predicting the noise $\epsilon_\tau$ from the input $\mathsf{z}_\tau$ at each diffusion step, yielding the reverse transition  
\begin{equation}
	\mathsf{p}(\mathsf{z}_{\tau-1}|\mathsf{z}_\tau;\theta) = 
	\mathcal{N}\!\left(\mathsf{z}_{\tau-1};
	\tfrac{1}{\sqrt{\alpha_\tau}}
	\left(\mathsf{z}_\tau - \tfrac{1-\alpha_\tau}{\sqrt{1-\bar{\alpha}_\tau}} \hat{\boldsymbol{\epsilon}}(\mathsf{z}_\tau, \tau; \theta)\right),
	\, \gamma_\tau \mathsfbi{I}
	\right),
\end{equation}
where $\theta$ denotes the trainable parameters of the network for noise prediction. 
The model is trained by minimising the mean squared error 
\begin{equation}
	\ell_{\text{DDPM}}=\mathbb{E}_{\mathsf{z}_0,\boldsymbol{\epsilon}_\tau}\left[ \lVert \boldsymbol{\epsilon}_\tau - \hat{\boldsymbol{\epsilon}}(\mathsf{z}_\tau,\tau; \theta)\rVert^2\right]  
\end{equation}
between true noise $\boldsymbol{\epsilon}_\tau$ and predicted noise $\hat{\boldsymbol{\epsilon}}$ by the denoising network parameterised by~$\theta$. With a trained diffusion model, the reverse transition kernel $\mathsf{p}(\mathsf{z}_{\tau-1}|\mathsf{z}_\tau;\theta)$ can be sampled iteratively to synthesise new sequences of latents from i.i.d. Gaussian noise. Each Gaussian noise sample leads to a distinct sequence, with no constraints (conditions) imposed on the generated samples. Hence, this process is referred to as unconditional generation in the generative modelling literature \citep{ho2020denoising}. In this work, we use the term prior sampling to be consistent with data assimilation terminology. 

The reverse process can also be expressed by a stochastic differential equation using the score \( \boldsymbol{s}_\tau(\mathsf{z}_\tau,\tau)=\nabla_{\mathsf{z}_\tau}\log{\var(\mathsf{z}_\tau)} \) of the data distribution $\var(\mathsf{z}_\tau)$. The score function is the vector pointing towards the direction in this state space where the rate of change \(\var(\mathsf{z}_{\tau})\) is highest. This formulation forms the basis for score-based diffusion models \citep{song2019generative, song2021scorebased}, where the neural network approximates the score function $\hat{\boldsymbol{s}}(\mathsf{z}_\tau,\tau;\theta)\approx \boldsymbol{s}_\tau(\mathsf{z}_\tau,\tau)$ instead of the noise. 
However, the marginal score $\nabla_{\mathsf{z}_\tau}\log{\var(\mathsf{z}_\tau)}$ is usually intractable. 
Instead, denoising score matching exploits the identity~\( \nabla_{\mathsf{z}_\tau}\log \var(\mathsf{z}_\tau)
= \mathbb{E}_{\mathsf{z}_0 \mid \mathsf{z}_\tau}
\!\left[
\nabla_{\mathsf{z}_\tau}
\log \var(\mathsf{z}_\tau \mid \mathsf{z}_0)
\right],\)
and trains the network using the tractable conditional score.
Since $\var(\mathsf{z}_\tau|\mathsf{z}_0)$ is Gaussian, its score has a closed-form expression given by
\begin{equation}
	\nabla_{\mathsf{z}_\tau}\log{\var(\mathsf{z}_\tau|\mathsf{z}_0)}=-\frac{\boldsymbol{\epsilon}_\tau}{\sqrt{1-\bar{\alpha}_\tau}}.
\end{equation}  
Since the conditional score is linearly proportional to the noise, we parameterise the score network via noise prediction as
\begin{equation}
	\hat{\boldsymbol{s}}(\mathsf{z}_\tau,\tau;\theta)=-\frac{\hat{\boldsymbol{\epsilon}}(\mathsf{z}_\tau, \tau; \theta)}{\sqrt{1-\bar{\alpha}_\tau}}~,
\end{equation}
which clearly indicates that estimating the noise added to the latent sample at diffusion time $\tau$ is equivalent to estimating the score function $\boldsymbol{s}$. In both DDPMs and score-based frameworks, the neural network serves as a mapping from an i.i.d. Gaussian distribution to a complex joint distribution in a high-dimensional space of the flow state. In this sense, the reverse process can be viewed as a typical regression problem, and DDPMs and score-based models are merely two different parameterizations of this problem. 

We use the diffusion transformer (DiT) as the neural backbone of our diffusion model. DiT is based on the transformer architecture of \cite{vaswani2017attention}. In a transformer, the input is first embedded into a sequence of tokens. The sequence is then fed to the attention layer, which forms the key component of the transformer. For flow problems, attention allows a model to focus on the most relevant spatial and temporal regions of a flow, enabling it to capture long-range interactions and coherent structures effectively. In our setting, the temporal generation horizon of DiT is approximately half of the integral time scale of the flow at $y^+=13$. Additional details about the DiT and hyperparameters used in this work are provided in appendix~\ref{subsec:app_DiT}. 

\subsection{Data Assimilation by Posterior Sampling}
\label{subsec:Methods_cDM}
Posterior sampling enables diffusion models to generate samples adhering to prescribed conditions $\Psi$, allowing them to be used for data assimilation. In the context of data assimilation, conditions describe observations of the system such as scattered sensor observations. These observations can be formally described as $\Psi=\mathcal{F}(\mathsf{z})+\boldsymbol{\varepsilon}$, where $\mathcal{F}$ is the observation operator and $\boldsymbol{\varepsilon}\sim\mathcal{N}(\boldsymbol{0},\sigma^2\mathsfbi{I})$ is the observation noise.
The observation operator $\mathcal{F}: \mathbb{R}^{\tilde{d}_c \times \tilde{d}} \rightarrow \mathbb{R}^{d_c \times N_\text{obs}}$ maps latent samples $\mathsf{z}$ in latent space to observations~${\Psi}$ in the physical space, where $N_\text{obs}$ denotes the number of observations. In our case the observation operator $\mathcal{F} \equiv \mathcal{H} \circ  \mathcal{D}$ is a composition of the decoder $\mathcal{D}$ of the $\beta$-VAE and the operator $\mathcal{H}$ (e.g., extraction of velocities at a location or a Fourier transform of velocity field to energy spectrum). 

Posterior sampling requires an estimate of the conditional score
\begin{equation}    
	\nabla_{\mathsf{z}_\tau}\log{\mathsf{p}(\mathsf{z}_\tau|\Psi)}=\nabla_{\mathsf{z}_\tau}\log{\mathsf{p}(\mathsf{z}_\tau)} +\nabla_{\mathsf{z}_\tau} {\log \mathsf{p}(\Psi|\mathsf{z}_\tau)}.
	\label{eq:cdm-bayes}
\end{equation}
This identity follows directly from Bayes' rule. The first term corresponds to the prior (unconditional) score, which is learned by the trained unconditional diffusion model. The second term is the gradient of the log-likelihood and acts as a correction to the prior score. In the reverse-time formulation of the diffusion process, this likelihood term modifies the drift of the reverse-time stochastic differential equation (SDE), effectively performing posterior sampling in latent space by biasing the dynamics towards regions consistent with the imposed conditions.
However the likelihood term is intractable due to its dependence on diffusion time $\tau$. We approximate this term as 
\begin{equation}
	\nabla_{\mathsf{z}_\tau}\log \mathsf{p}(\Psi|\mathsf{z}_\tau)\simeq -\rho\nabla_{\mathsf{z}_\tau}\lVert\Psi-\mathcal{F}(\hat{\mathsf{z}}_0)\rVert,
	\label{eq:dps_alt}    
\end{equation}
following the diffusion posterior sampling approach of \citet{chung2023diffusion}. Here, $\rho$ is the conditioning strength parameter, ranging from 0.3 to 1, controlling the effect of conditioning on the generation process, and $\hat{\mathsf{z}}_0=\mathbb{E}_{\mathsf{z}_0\sim p\left(\mathsf{z}_0 \vert \mathsf{z}_\tau\right)}[\mathsf{z}_0]$ is the denoising estimate of the clean latent variable $\mathsf{z}_0$. The gradient $\nabla_{\mathsf{z}_\tau}\lVert\Psi-\mathcal{F}(\hat{\mathsf{z}}_0)\rVert$ is computed using automatic differentiation. A detailed derivation of equation~\ref{eq:dps_alt} is provided in Appendix~\ref{subsec:derivations}.

In principle, diffusion posterior sampling can handle any observational data as long as it can be expressed as a differentiable function, whether it be scattered velocity observations or energy spectra. However, using it to directly impose statistical quantities of interest in turbulence can be challenging. While simpler functions such as mean velocities and Reynolds stresses (from variances) are straightforward, matching more complicated observational data such as energy spectra requires backpropagation through functions like Fourier transforms.

\begin{figure}[t!]
	\centering
	\includegraphics[width=0.9\textwidth]{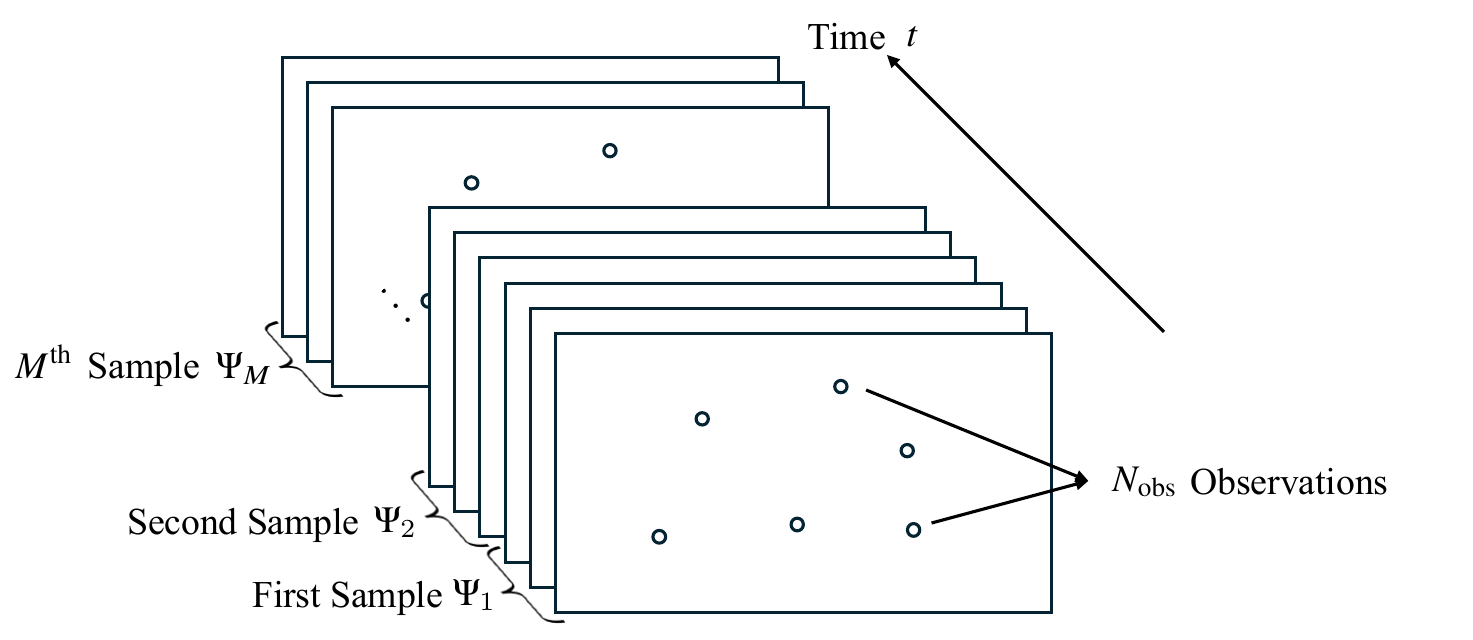}
	\caption{Schematic of the sampling strategy used to impose statistical quantities. The time series is divided into $M$ segments $\Psi_1, \Psi_2, \cdots,  \Psi_M$. Each generated sample sees one of the $M$ segments.}
	\label{fig:sampling}
\end{figure}

We propose an alternative approach that avoids backpropagation through statistical operators: we condition the model on time-series segments of pointwise observations, so that consistency with the underlying flow statistics emerges from Bayesian conditioning on samples of the marginal distribution rather than from explicit statistical targets.
We generate samples using synchronous time series observations of the flow field~$\boldsymbol{\Phi}(\boldsymbol{x}_i,t)\in \mathbb{R}^{d_c}, \, i=1, \dots, d$ at $N_\text{obs}$ locations. 
We divide the time series into $M$ segments of length $d_t$ as shown in figure \ref{fig:sampling}. We impose one of these $M$ segments $\Psi_i \in \mathbb{R}^{d_c \times N_\text{obs}\times d_t}, \, i=1, \dots, M$ (consisting of $N_\text{obs}\times d_t$ observations in total) on each generated sample as conditions through equation~\ref{eq:dps_alt}. In our work, the full field is in a space $\mathbb{R}^{d_c \times d \times d_t}$ with $d = 128\times 64 \times 128$ for the generation domain and a segment time length $d_t = 10$. The samples $\Psi_i$ reside in an observation space $\mathbb{R}^{d_c \times N_\text{obs} \times d_t}$ with, e.g., $N_\text{obs} = 10000$ for a task with this number of sensor observations. At a given location, each of the $d_t$ observations in a segment constitutes a realisation of the distribution of the flow field at that location. 
The combination of the $N_\text{obs} \times d_t$ observations then becomes a realisation $\Psi_i$ of the marginal distribution of the entire four-dimensional flow field. 
Diffusion posterior sampling then combines this marginal distribution with the prior obtained from the unconditional model to guide the diffusion process towards a joint distribution consistent with both the learned prior and the imposed observations. 
Because the flow is statistically stationary, the ensemble $\{\Psi_i\}_{i=1}^M$ inherits the statistics of the flow at the sensor locations; these enter the posterior through pointwise Bayesian conditioning, and the resulting samples can reproduce high-order statistics such as two-point correlations and energy spectra without differentiation through statistical operators. In the experiments reported here, the observations are drawn from the same DNS that produced the training data, so the posterior statistics largely reflect the prior. Whether this construction can also steer the posterior towards statistics that depart from the prior is an open question that we do not address in the present work.
Each posterior sample sees a different realisation in the observation space (marginal distribution). 
This ensemble-based sampling procedure admits a natural interpretation within probabilistic data assimilation and is conceptually analogous to the ensemble Kalman filter (EnKF) \citep{evensen2009data, zhang2022ensemble}. 
In the EnKF, an ensemble of model states represents possible realizations of the system, and its mean and covariance approximate the posterior distribution after assimilating observations. Similarly, the present framework generates an ensemble of samples from a learned diffusion-based prior, and each realisation represents a draw from the posterior distribution conditioned on the observations. The ensemble thus embodies uncertainty and statistical structure, with correlations implicitly encoded in the learned diffusion prior rather than propagated explicitly through covariance updates. A key distinction is that EnKF ensembles evolve in physical time according to governing equations, whereas here the evolution occurs in algorithmic diffusion time, beginning from noise and converging towards data-consistent realizations. The diffusion prior therefore plays a role analogous to the forecast model in ensemble data assimilation, while the conditioning step acts as a probabilistic update. It should be noted that matching marginal statistics alone does not guarantee faithful reproduction of the full joint distribution. In Appendix~\ref{app:Additional_results}, we provide additional validation of the joint distribution structure through two-point velocity correlations and temporal autocorrelations for the conditionally generated flow fields. Systematic validation of higher-order joint statistics remains an important direction for future work.

From a broader perspective, the unconditional generator represents a prior distribution learned from high-fidelity simulations under nominal operating conditions. 
In principle, when observations exhibit statistical characteristics that differ from those of the training data, posterior sampling should bias the diffusion process towards a posterior consistent with the measured flow. Whether this idealised picture is realised in practice (in particular the strength of the steering effect and the extent to which it can support controlled extrapolation beyond the training manifold) depends on observation density, spatial configuration, and the structure of the learned prior, and is not established by the experiments reported here. The wind-farm vision outlined in \S\ref{sec:introduction} is therefore a guiding aspiration rather than a demonstrated capability of the framework as presented.

\section{Dataset and Training Description}
\label{sec:Comp_setup}
We evaluate our framework on turbulent incompressible plane Couette flow. We train the $\beta$-VAE in the first stage of the training on instantaneous flow-field snapshots ${\boldsymbol{\phi}}$ of three-dimensional subdomains that are 1/32 the size of the computational domain of the DNS. The subsequent DiT is trained on sequences of 10 snapshots. The snapshots in these sequences are the encoded latent instantaneous flow fields from the fixed $\beta$-VAE, which is frozen at this stage. After the entire model is trained, new samples can be generated from the model through inference, with each inference generating a sequence of 10 snapshots of a single subdomain.

The incompressible plane Couette flow has a Reynolds number of $Re_h=1300$, based on the channel half-height $h$. The dataset is obtained from a spectral DNS of~\citet{teng2018turbulent} in a domain $40\pi \times 2 \times 6\pi$ on a $1024 \times 129 \times 256$ grid and with a timestep $\Updelta t = 0.01h/U_\text{w}$, where $U_\text{w}$ is the wall velocity at $y=0$ and $2$. 
Periodic boundary conditions were applied in the streamwise and spanwise directions using the homogeneity of plane Couette flow in these directions.
After reaching statistical stationarity, the spatiotemporal flow fields consisting of the velocity fields $\boldsymbol{u}=(u,v,w)$ and the pressure field $p$ are sampled over $20{,}000 \Updelta t$ at intervals of $10 \Updelta t$, yielding 2000 snapshots of the field~$\boldsymbol{\phi}$.
We generate only 1/32 of the DNS domain by exploiting the top--bottom symmetry as well as the streamwise ($x$) and spanwise ($z$) homogeneity of the turbulent statistics in plane Couette flows. 
\begin{figure}[t!]
	\centering
	\includegraphics[width=\textwidth]{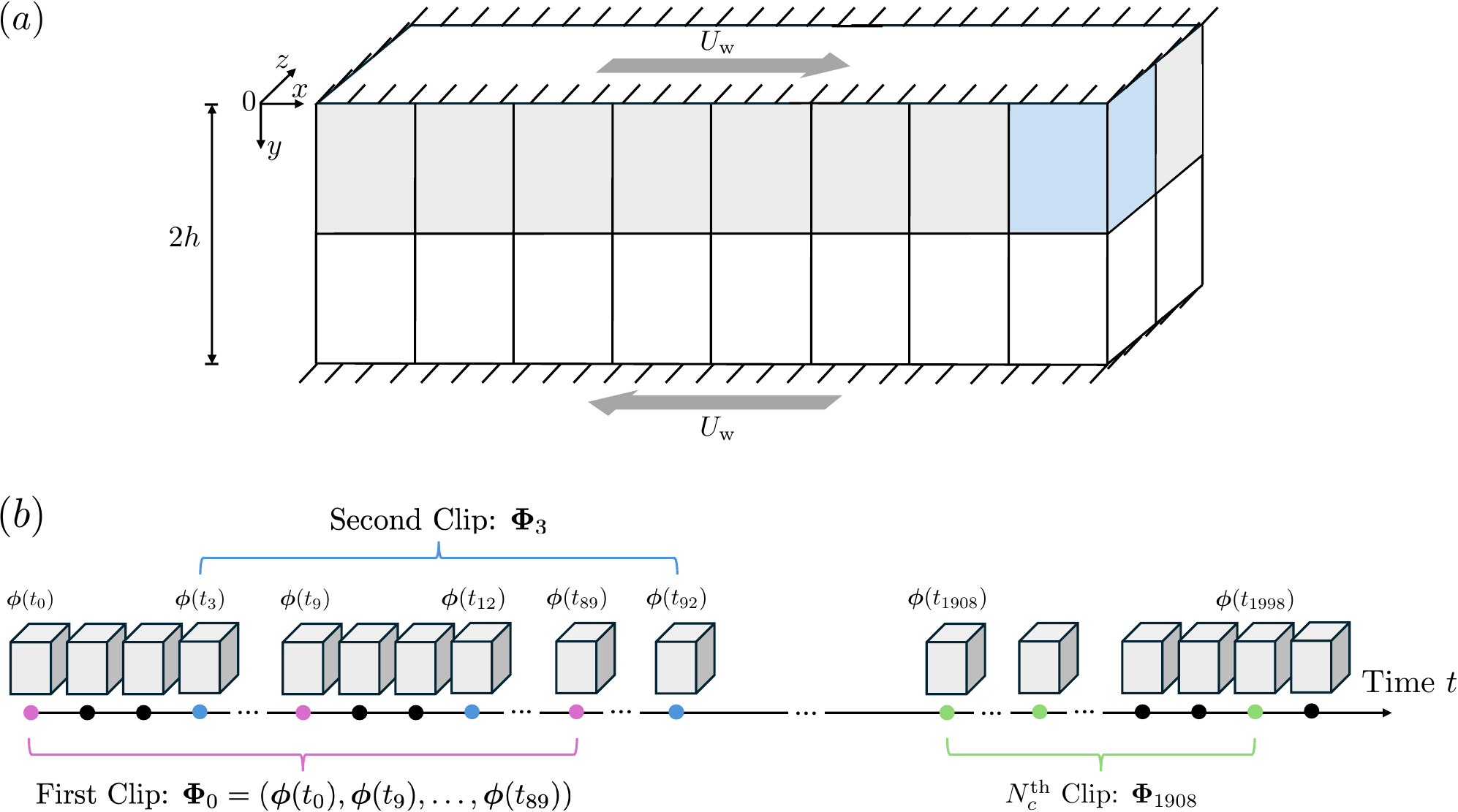}
	\caption{DNS of turbulent plane Couette flow is used to construct the spatiotemporal dataset used for training. $(a)$ First, The computational domain of the DNS is divided at the channel centreline. We use the top half with 16 subdomains (grey) and generate one subdomain (one snapshot) with 1/32 size of the computational domain of the DNS (blue). Depicted is the plane Couette flow domain with channel half-height $h$; both walls move in streamwise direction $x$ with wall velocity~$U_\text{w}$. 
		$(b)$ Second, for the training of the Diffusion Transformer (DiT) each trajectory  is segmented into clips. Illustrated is the segmentation of the 2000-snapshot trajectory of the DNS for one subdomain ${\boldsymbol{\phi}}(t_i)$ with $t_i = i \times 10\Updelta t$. Each snapshot is a three-dimensional image (grey box). Every~${\boldsymbol{\Phi}}_i$ is a clip including 10 snapshots with timestep $\Updelta t_{\mathrm{gen}}$ starting at $t_i$. A stride of $30\Updelta t$ produces $N_\text{c}=637$ clips for one subdomain and $10{,}192$ clips in total for all~16 subdomains in the top half of the simulation domain.}\label{fig:dataset_domain_construction}
\end{figure}
To train such a generator, we divide the top half of the computational domain of the DNS into 16 subdomains with $1/32$ of the size of the original DNS domain (figure~\ref{fig:dataset_domain_construction}$a$). This reduced subdomain size reduces GPU memory usage and increases the number of training samples. 
As we show in \S\ref{sec:results} our models successfully learn the dynamics of the plane Couette flow by using the top half for training. We note that symmetry-based data augmentation with the bottom half is possible. However, we have not carried out such an augmentation, as we have not observed any indication of overfitting during training.
Each subdomain has the size of~$L_x \times L_y \times L_z =  5\pi \times 1 \times 3\pi$ with~$d=128 \times 64 \times 128$ grid points. The subdomains are decorrelated across the streamwise and spanwise direction (refer to \S\ref{subsec:ROM_TurbulenceGen}), allowing each to serve as an independent sample and enabling the extraction of many training trajectories from a single DNS. 
The use of a reduced physical domain reflects a conceptual distinction between probabilistic generative modelling and conventional DNS requirements. As this difference is central to interpreting the training and generation strategy adopted here, we provide a detailed discussion in \S\ref{sec:discussion_domain}.

The trained model generates sequences of 10 snapshots (clips) with a timestep $\Updelta t_{\text{gen}}=100\Updelta t$, 
\[
\boldsymbol{\Phi}_i 
= \big[
\boldsymbol{\phi}(t_i),\;
\boldsymbol{\phi}(t_i + \Updelta t_{\mathrm{gen}}),\;
\dots,\;
\boldsymbol{\phi}(t_i + 9\,\Updelta t_{\mathrm{gen}})
\big], \]
with $t_i = i \times 10\Updelta t$. The resulting generation window is approximately half of the integral timescale of the flow. 
To create training samples, we divide the 2{,}000-snapshot trajectory obtained from the DNS into sequences of $d_t=10$ snapshots, as shown in figure~\ref{fig:dataset_domain_construction}$(b)$.
This procedure yields a total of $10{,}192$ clips, which constitute our dataset
\(
\{\, \boldsymbol{\Phi}_i \,\}_{i=0}^{10{,}191}
\) used to train the models described
in \S\ref{subsec:ROM_TurbulenceGen}. For conditional generation in \S\ref{subsec:DA}, the dataset is split in half along the streamwise direction. The model is then trained with the first half and tested with samples chosen from the second half. To ensure that training and test samples are fully decorrelated, test samples are chosen sufficiently far away in the streamwise direction from the training dataset. 

\section{Results}
\label{sec:results}
\subsection{Dimensionality Reduction and Turbulence Generation}
\label{subsec:ROM_TurbulenceGen}
\begin{figure}[t!]
	\centering
	\includegraphics[width=\textwidth]{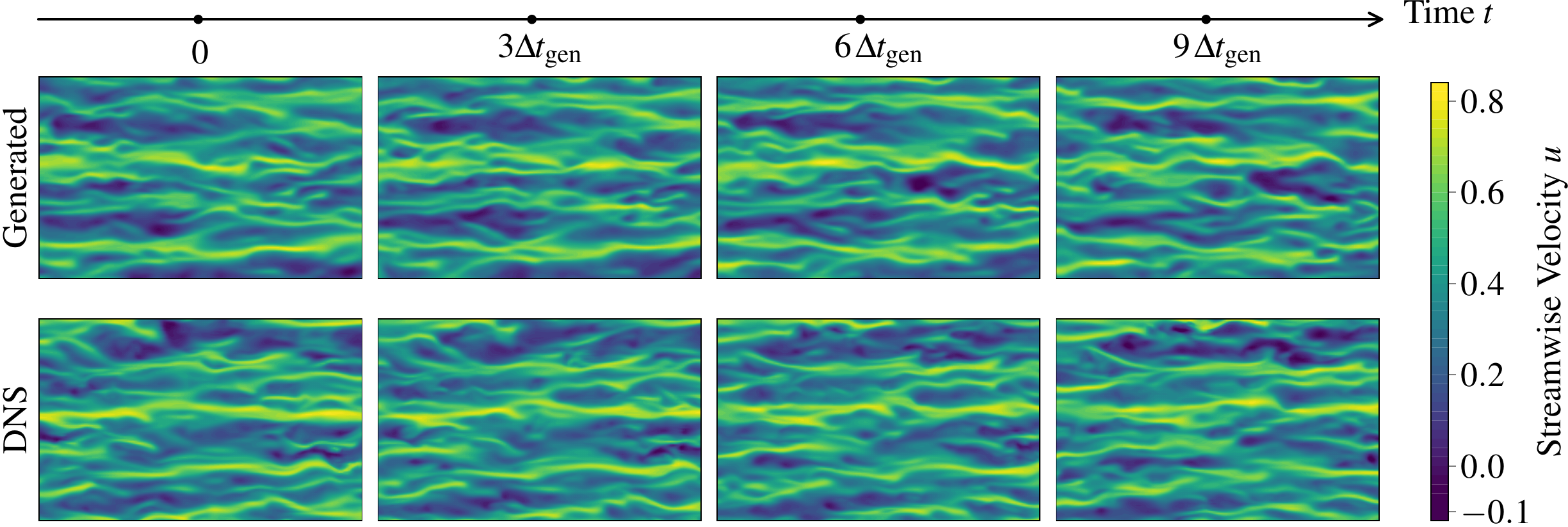}
	\caption{Prior sampling (unconditional generation) of plane Couette flow. Streamwise velocity $u$ at $y^+=15$ for a sequence generated with 16 latent degrees of freedom (top) compared to DNS data (bottom) at four selected time steps.}
	\label{fig:contour_plots_16_degrees of freedom_unconditional}
\end{figure}

\begin{figure}[t!]
	\centering
	\includegraphics[width=\textwidth]{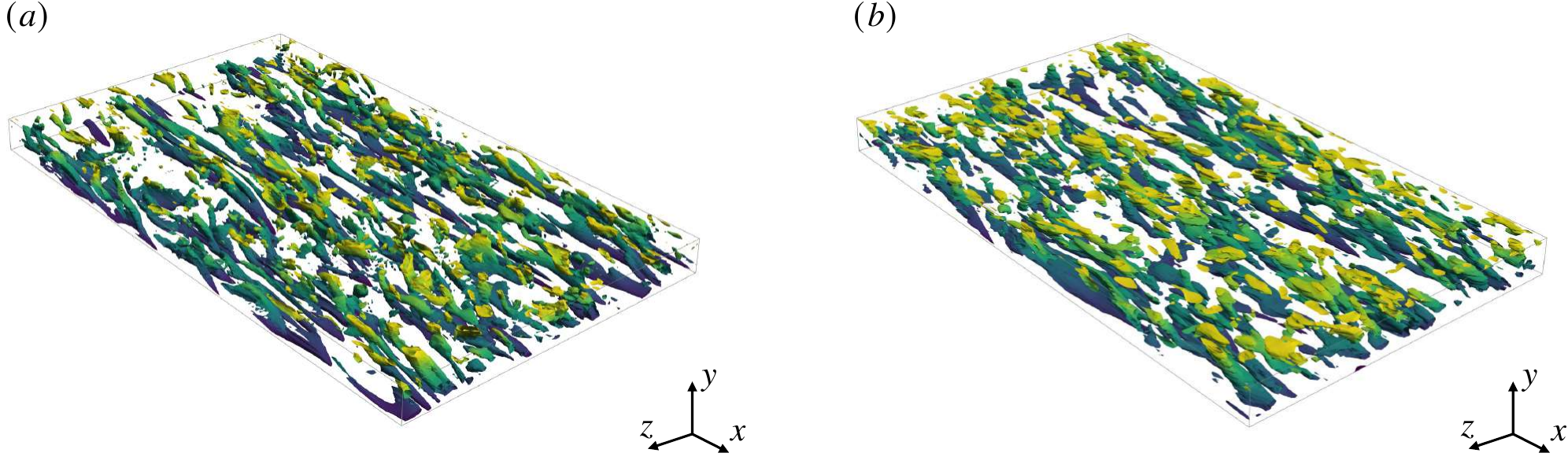}
	\caption{Vortical structures visualised by instantaneous iso-surfaces of the $Q$-criterion at $Q = 0.1$ for the $(a)$ DNS and $(b)$ the model with $d_{\mathsf{z}} = 16$. The colour coding indicates the wall distance.}
	\label{fig:q-criteria}
\end{figure}

We first evaluate the performance of our model in the (unconditional) prior sampling of plane Couette flow, focusing especially on dimensionality reduction.
We train multiple $\beta$-VAEs with different numbers of latent degrees of freedom $d_\mathsf{z}$. For each configuration, a corresponding DiT is trained in the latent space. We evaluate six configurations with latent degrees of freedom of 4, 8, 16, 32, 64, and 96. The detailed shapes of the trained latent representations are provided in table~\ref{tab:vae_details} of appendix~\ref{subsec:app_VAE}. Higher-dimensional latent spaces ($d_{\mathsf{z}} > 96$) were found to be difficult to train, likely due to the increasing complexity of the represented dynamics, and are not pursued further.
Identifying an efficient latent representation is essential for dimensionality reduction \citep{eivazi2022towards}. However, determining the minimal latent dimension for complex, multiscale flows remains challenging, and no universal criterion exists. Prior work has used various indicators \citep{linot2022data, vinograd2025reduced,zeng2024autoencoders}. Here, we select the latent dimension at which errors in the statistics level off, following the strategy of \citet{linot2023dynamics}.
The reconstructed velocity fields resemble the DNS in terms of both vortical structures and mean quantities.
The prior samples reproduce the long streamwise streaks characteristic of plane Couette flow. These streaks undergo waviness and subsequent breakdown, consistent with the DNS behaviour. Figure \ref{fig:contour_plots_16_degrees of freedom_unconditional} compares these characteristics for the DNS and the prior samples. We focus on the model with $d_{\mathsf{z}}=16$ in this figure, as later results indicate that turbulence statistics begin to degrade below this latent size.
The DNS contains numerous quasi-streamwise vortices typical of wall-bounded turbulence. The generated fields exhibit fewer small-scale vortical structures. 
The reduced abundance of fine-scale vortical structures reflects the limited degrees of freedom imposed by the latent representation. Consequently, the generated flow primarily captures the dominant large-scale motions, while smaller-scale turbulent motions are weakened or absent due to the constrained latent representation. This results in flow fields that are qualitatively less rich in small-scale turbulent activity, as illustrated by the instantaneous $Q$-criterion iso-surface in figure \ref{fig:q-criteria}.

While even low latent dimensions can reproduce general flow features and the mean velocity profile, only higher-dimensional latent spaces can faithfully represent the second- and higher-order statistics.
We assess model performance and justify the selected latent dimension by evaluating standard turbulence statistics of the generated flow fields up to fourth order. We consider statistically stationary plane Couette flow and decompose the velocity field as 
$\boldsymbol{u} = \boldsymbol{\overline{u}} + \boldsymbol{u'}$, where 
$\overline{(\cdot)}$ denotes the mean velocity and 
$(\cdot)'$ the fluctuations. The velocity components 
$(u, v, w)$ correspond to the streamwise, wall-normal, and spanwise directions. All statistics are computed from an ensemble of 300 generated sequences and are averaged in time and over the homogeneous streamwise and spanwise directions.

\begin{figure}[t!]
	\centering
	\includegraphics[width=\textwidth]{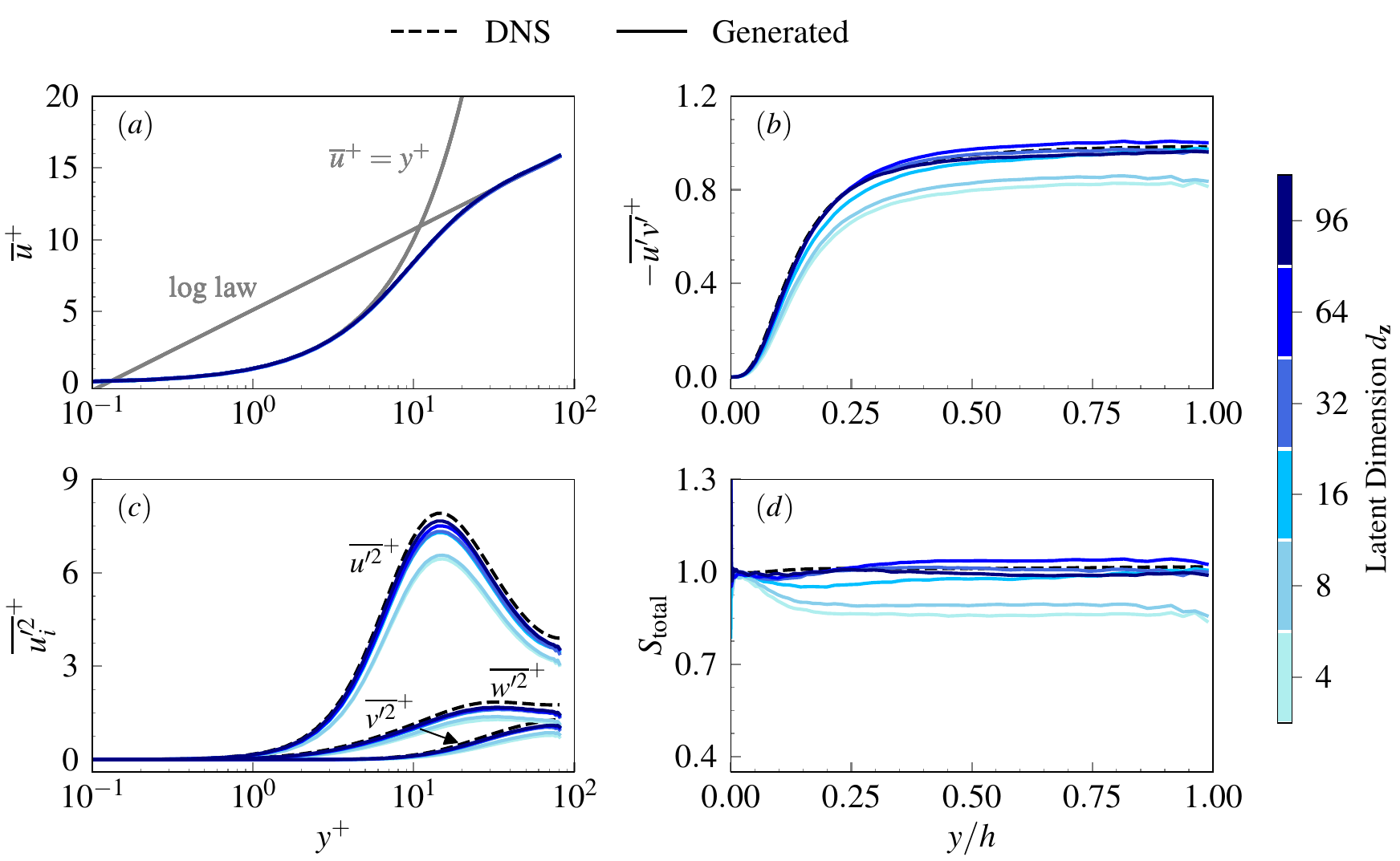}
	\caption{First- and second-order turbulent statistics of the prior samples. $(a)$ Mean streamwise velocity profiles compared with the law of the wall (grey), $(b)$ Reynolds shear stress, $(c)$ Reynolds normal stresses, $(d)$ total shear stress as the sum of mean and Reynolds shear stress. We show the statistics for flows generated with varying latent dimensions $d_{\mathsf{z}}$ from 4 to 96 and compare against the DNS data. Lighter to darker shades of blue indicate increasing latent dimensions.}
	\label{fig:mean_velo_stress}
\end{figure}

We begin our assessment with the first order mean streamwise velocity profile.
All models and the DNS can follow the logarithmic law of the wall with $\overline{u}^+=(1/\kappa)\ln{y^+}+C$ with $\kappa=0.41$ and $C=5.1$ \citep{avsarkisov2014turbulent}, where $\overline{u}^+ = (\overline{u} + U_\text{w})^+$. The profile of the mean streamwise velocity in wall coordinates is shown in figure \ref{fig:mean_velo_stress}$(a)$. Models with lower latent dimensions correspond to lighter shades of blue in this figure and throughout the section. Here we note that all normalisation of the model results into wall coordinates~$(\cdot)^+$ is done with friction velocity $u_\tau=\sqrt{\nu (\p \overline{u} / \p y)|_{y=0}}$ of the DNS. We observed that evaluating the streamwise velocity gradient at the wall ($y = 0$) produced significantly higher gradient values. This behaviour is attributed to the extremely fine mesh resolution near the wall, which increases numerical sensitivity in the gradient calculation, even though the predicted velocity field remains accurate.

First-order statistics alone do not reveal accuracy differences among models, including those with latent dimension 
$d_{\mathsf{z}} = 4$. Second-order statistics, however, show a clear degradation for 
$d_{\mathsf{z}} < 16$. Figure~\ref{fig:mean_velo_stress}$(b)$ presents the components of Reynolds normal stresses, the squared velocity fluctuations. Models with $d_{\mathsf{z}} \geq 16$ reproduce all three velocity fluctuation components accurately. Models with smaller latent dimensions significantly underestimate fluctuation intensities. This is evidenced by a sharp drop, most evident in the streamwise component $\overline{u'^2}^+$ where the peak in the buffer layer is significantly underestimated, although the location of the peak is captured by all models. This peak also hints at the location of the most energetic eddies in the turbulent flow, characterised by the maximum turbulent kinetic energy $\overline{K}^+ = 0.5 \left( \overline{u'^2}^+ + \overline{v'^2}^+ + \overline{w'^2}^+ \right)$ at $y^+=15$ in the buffer layer. As latent dimension increases, the generated statistics converge towards the DNS. These results indicate that latent spaces with sufficient dimensionality preserve the dominant flow dynamics. Diffusion models trained in such spaces generate prior samples that decode into physically consistent flow fields.
A central principle governing turbulent flows is overall force balance. Figure~\ref{fig:mean_velo_stress}$(d)$ shows the total stress, defined as the sum of the mean shear stress $\mathrm{d}\overline{u}^+/\mathrm{d}y^+$ and the Reynolds stress $-\overline{u^{\prime}v^{\prime}}^+$. Models with $d_{\mathsf{z}} \geq 16$ reproduce a total stress close to the DNS unity across the entire channel, indicating that fully stationary states are achieved \citep{teng2018turbulent}. In contrast, models with $d_{\mathsf{z}} < 16$ fail to maintain this balance, primarily due to an underestimation of the Reynolds shear stress (figure~\ref{fig:mean_velo_stress}$b$), consistent with the reduced velocity fluctuations (figure~\ref{fig:mean_velo_stress}$c$). Overall, it is evident that the models with $d_\mathsf{z}<16$ show a clear drop-off in the accuracy of the second-order statistics, indicating that $d_\mathsf{z}=16$ is the minimum latent dimension for our setup. 

\begin{figure}[t!]
	\centering
	\includegraphics[width=\textwidth]{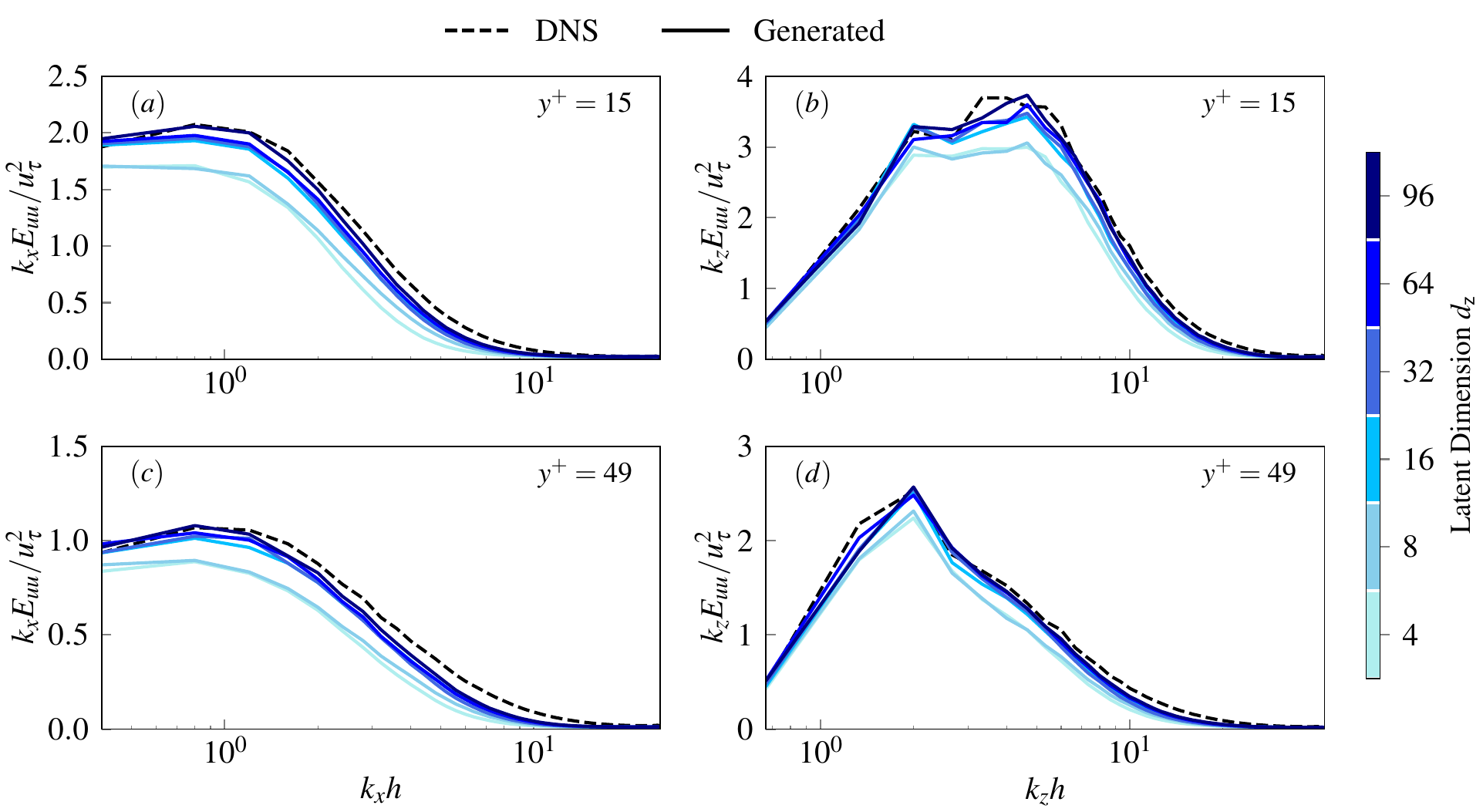}
	\caption{Premultiplied energy spectra of the prior samples. We evaluate the energy spectra at two wall-normal locations: $(a,b)$ $y^+=15$, and $(c,d)$ $y^+=49$. The panels show the pre-multiplied streamwise energy spectra as a function of the streamwise $(a,c)$ and the spanwise $(b,d)$ wavenumber. We report the statistics for flows generated with different latent dimensions $d_{\mathsf{z}}$ and compare with the DNS data.}
	\label{fig:premult_energy_spectra_spanwise}
\end{figure}

We quantify the distribution of energy across scales and the scales of the coherent structures with the premultiplied energy spectra.
All models reproduce the peak and plateau locations of the energy spectra, confirming that
the characteristic length scales of the coherent structures are preserved in the
generated samples. The generated spectra do, however, show that models with $d_\mathsf{z}<16$ systematically underestimate the peak value. 
In figure~\ref{fig:premult_energy_spectra_spanwise} we present the premultiplied streamwise and spanwise energy spectra $k_xE_{uu}$ and $k_zE_{uu}$ at locations \(y^+=15\) and \(y^+=49\). Here, \(k_{x}=2\pi/\lambda_{z}\) and \(k_{z}=2\pi/\lambda_{z}\) are the streamwise and spanwise wavenumber, and \(\lambda_{x}\) and \(\lambda_{z}\) are the respective wavelengths.
The energy spectra $E_{uu}$ is obtained from the amplitude of the discrete Fourier transform of the velocity fluctuation in the respective direction. They give the decomposition of the turbulent fluctuations in the wavenumber space. As the energy spectrum and the two-point correlation of the velocity fluctuations form a Fourier transform
pair, the premultiplied spectra of figure~\ref{fig:premult_energy_spectra_spanwise} and the correlations in appendix~\ref{app:two-point} contain the same information: a
minimum of $R_{uu}$ at separation $\Updelta \boldsymbol{x}_i$ corresponds to a spectral peak at
wavelength $\lambda_{\boldsymbol{x}_i} \approx 2\Updelta \boldsymbol{x}_i$. 
Near-wall coherent structures play a central role in sustaining turbulence in plane Couette flow. In the buffer layer, low-speed streaks form parallel to the wall and
undergo waviness and breakdown, which in turn generates larger-scale vortices in the
outer region of the flow
\citep{hamilton1995regeneration, pirozzoli2011large, pirozzoli2014turbulence,
	kline1990turbulent}. Previous numerical studies have shown that these streaks exhibit
universal characteristics across wall-bounded turbulent flows
\citep{pirozzoli2011large}. The long streamwise streaks are clearly visible in
figure~\ref{fig:contour_plots_16_degrees of freedom_unconditional}, and their spectral signature is
the broad plateau of $k_x E_{uu}$ at low streamwise wavenumbers in figure~\ref{fig:premult_energy_spectra_spanwise}($a,c$).
This broad plateau extends over the lowest resolved modes, retaining more than
$80\%$ of their peak value, and therefore
encompasses the streamwise extent of the buffer-layer streaks which drive the regeneration cycle. The long
large-scale coherent structures characteristic of plane Couette flow, with streamwise extents of
order $70h$ \citep{teng2018turbulent}, exceed the streamwise period of the domain and
are consequently not resolved. This truncation is a property of the DNS dataset rather
than of the generative model: both the reference and the generated fields are subject
to the same periodic domain, and the absence of a zero crossing in
$R_{uu}(\Updelta x)$ in appendix~\ref{app:two-point} is its physical-space counterpart.
The spanwise spectra on the other hand
(figure~\ref{fig:premult_energy_spectra_spanwise}\textit{b},\textit{d}) are sharply peaked well inside
the resolved range and have fallen to roughly one fifth of their peak value at the
largest resolved wavelength. So the dominant spanwise scales are fully contained in
the generated domain. At $y^+ = 15$ two peaks of in the spectrum are observed,
with a first peak at $k_z h \approx 2$ and a second peak at $k_z h \approx 4.5$
corresponding to $\lambda_z \approx 1.4h$, or $\lambda_z^+ \approx 120$: the classical
spanwise spacing of the near-wall streaks and quasi-streamwise vortices. In the logarithmic layer, the peak is caused by the so-called ``superstructures'' \citep{hutchins2007evidence} or ``large-scale motions'' \citep{pirozzoli2014turbulence}. For plane Couette flow, \cite{pirozzoli2014turbulence} report a spanwise wavelength of~\(\lambda_z\approx5h\) for these large-scale structures. The spanwise spectrum of the streamwise velocity exhibits a peak around \(k_zh\approx2\) or \(\lambda_z\approx\pi h\) for all models at \(y^+=49\) (figure \ref{fig:premult_energy_spectra_spanwise}$b$), similar to this value. As indicated by the qualitative agreement of the visualizations of the streamwise velocity field and Q-criterion iso-surfaces, the dominant coherent structures and their characteristic length scales are preserved in the generated samples. The pre-multiplied-energy spectra further supports this conclusion and provide the primary quantitative evidence.

\begin{figure}[t!]
	\centering
	\includegraphics[width=\textwidth]{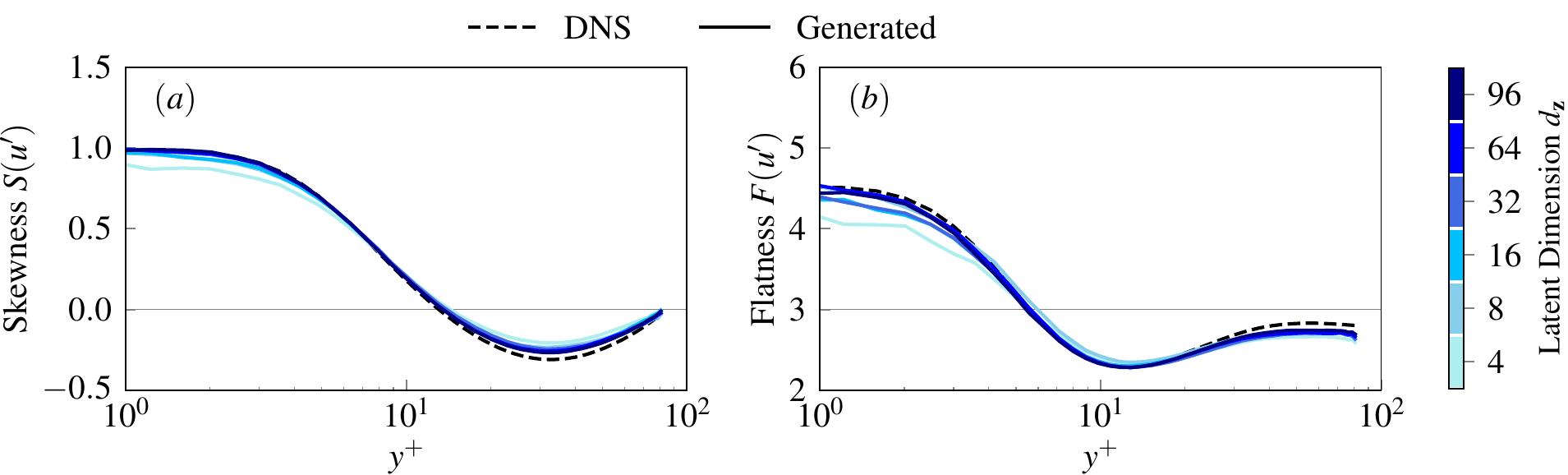}
	\caption{Third- and fourth-order turbulent statistics of prior samples. $(a)$ Skewness $S(u^{\prime})$ and $(b)$ flatness $F(u^{\prime})$ of the streamwise velocity fluctuation. We show the statistics for flows generated with varying latent dimensions $d_{\mathsf{z}}$ and compare against the DNS data.}
	\label{fig:flattness_skewness}
\end{figure}

All models with a latent dimension of eight or larger accurately reproduce the skewness and flatness of the streamwise velocity. This result indicates that the models capture key aspects of intermittency and asymmetry in the near-wall region.
Figure \ref{fig:flattness_skewness} shows that both skewness and flatness deviate strongly from Gaussian values near the wall, where a Gaussian distribution would yield $S(u')=0$ and $F(u')=3$. We compute the skewness as
$S(u') \equiv \overline{u^{\prime 3}}/\overline{(u^{\prime 2})}^{3/2}$
and the flatness as
$F(u') \equiv \overline{u^{\prime 4}}/\overline{(u^{\prime 2})}^{2}$.
As the wall is approached, skewness and flatness converge to approximately $1.0$ and $4.5$, respectively. These values indicate strong asymmetry and pronounced intermittency of the streamwise velocity, consistent with observations for wall-bounded turbulence \citep{kim1987turbulence}. Toward the channel centre, both statistics approach Gaussian behaviour.
These findings are noteworthy and somewhat surprising; even though the low-order statistics, such as Reynolds stresses, are not captured well by the models with a latent dimension smaller than~16, all models can accurately reproduce these higher-order statistics. This behaviour arises because skewness and flatness are normalised higher-order moments. In contrast, TKE is an absolute variance (second moment) and depends on the amplitude and representation of small-scale, decorrelated fluctuations; these are the first to be lost when the latent bottleneck is too small. As was seen in figure~\ref{fig:mean_velo_stress}$(c)$, the models produce fluctuations with the same shape but with a systematically smaller amplitude, skewness and flatness can remain nearly unchanged while the absolute variance (and thus TKE) decreases.

Turbulence validation of generative models has largely
rested on single-point statistics: the Reynolds stresses (second order) and,
less commonly, the skewness and flatness (third and fourth order) reported
above, extended by \citet{li2024synthetic} to generalized flatness up to eighth
order. Such single-point moments constrain the amplitude and shape of the
velocity distribution at a point, but they carry no spatial information and
cannot certify that a model has reproduced the spatial organisation of the flow.
That organisation is the essence of intermittency, which is associated with
coherent, spatially concentrated structures rather than with the distribution of
energy across scales \citep{she1990intermittent}. Two-point second-order
statistics, namely the premultiplied energy spectrum or equivalently its Fourier
dual the two-point correlation (Appendix~\ref{app:two-point}), do characterise
how energy is distributed across scales, but the spectrum is determined by the
Fourier amplitudes alone and discards the Fourier phases. The consequence is
concrete: a synthetic field obtained by retaining the DNS energy spectrum while
randomising the phases reproduces the spectrum, the two-point correlation, and
every single-point moment exactly, yet contains no coherent structures and looks
nothing like turbulence. The structural organisation that separates turbulence
from such phase-scrambled noise resides in the phase relationships, and resolving
it requires statistics sensitive to them.

\begin{figure}[t!]
	\centering
	\includegraphics[width=\textwidth]{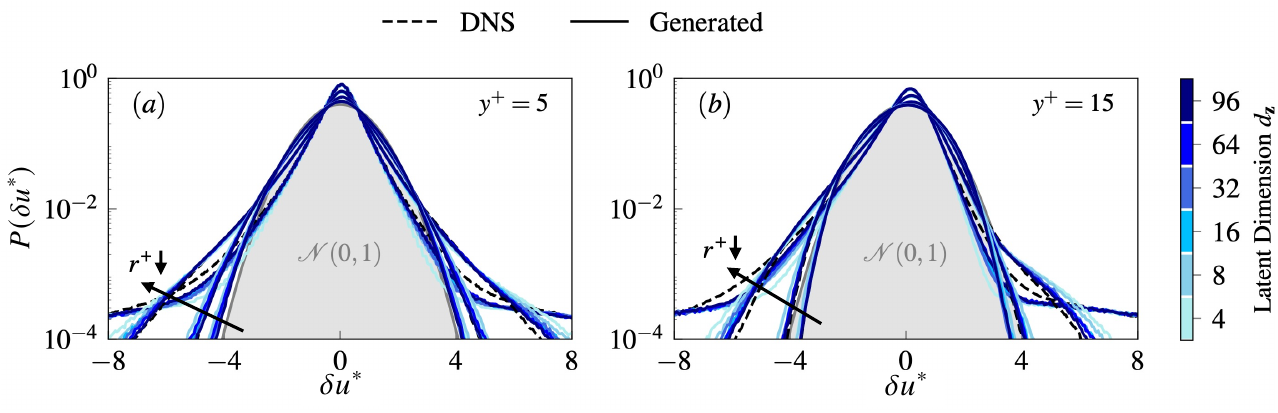}
	\caption{Probability density function of the standardized longitudinal velocity increment $\delta u^*$ for the prior (unconditional) samples, at wall-normal stations (a) $y^+ = 5$ and (b) $y^+ = 15$. Each curve is the $\delta u^*$
		distribution for one streamwise separation $r^+ = 350, 120, 40$, and $10$, standardized to zero mean and unit variance;
		the arrow indicates decreasing $r^+$. We show the PDFs for flows generated with varying latent dimensions $d_{\mathsf{z}}$ and compare against the DNS data.}
	\label{fig:uncond_velocity_increment}
\end{figure}

\begin{figure}[t!]
	\centering
	\includegraphics[width=\textwidth]{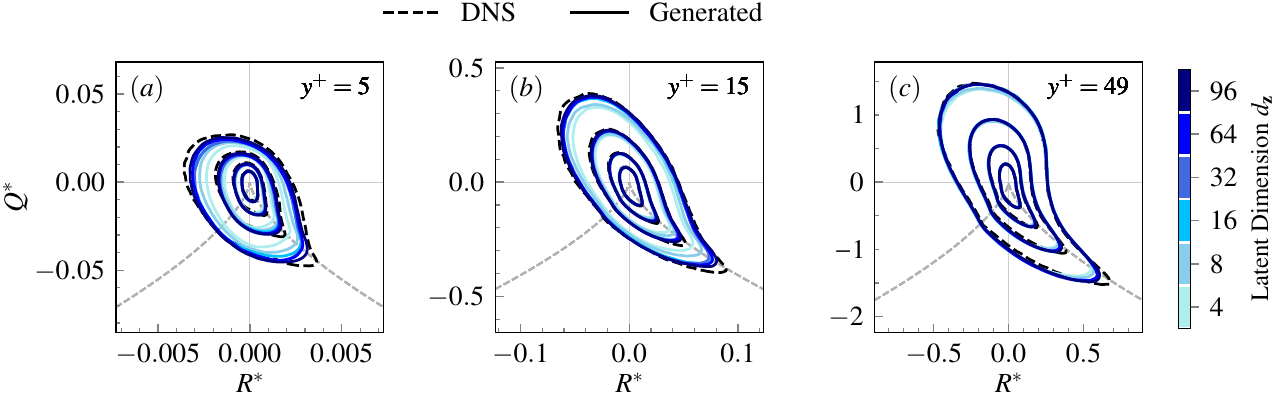}
	\caption{Strain-rotation phase alignment in the $Q$--$R$ plane of the prior samples at three wall-normal locations $(a - c)$ $y^{+} = 5$, $15$ and $49$.
		The contours are at four logarithmically spaced probability levels; the dashed grey curve is the Vieillefosse line $\tfrac{27}{4}R^{2}+Q^{3}=0$.
		We show the statistics for flows generated with varying latent dimensions $d_{\mathsf{z}}$ and compare against the DNS data.}
	\label{fig:uncond_qr}
\end{figure}

We therefore examine two phase-sensitive, higher-order measures
that are rarely reported in the generative-modelling literature. The first is the
scale-dependent probability density function (PDF) of the longitudinal velocity
increment $\delta u(r) = u(x + r\hat{\boldsymbol{e}}_x) - u(x)$, where
$\hat{\boldsymbol{e}}_x$ is the streamwise unit vector; its variation with
separation $r$ diagnoses intermittency, the progressive departure from
Gaussianity as $r$ decreases towards the dissipative scales. The second is the
joint PDF of the second and third invariants of the velocity-gradient tensor
$A_{ij} = \partial u_i / \partial x_j$, namely
$Q = -\tfrac{1}{2}\,\mathrm{tr}(A^2)$ and $R = -\tfrac{1}{3}\,\mathrm{tr}(A^3)$,
which measure the local balance of rotation against strain and the relative
orientation of the strain-rate and rotation-rate tensors. The characteristic
teardrop shape of the $Q$--$R$ plane is a direct signature of this small-scale
phase alignment and is likewise invisible to the second-order statistics.
Together these probe the intermittency and the small-scale structural
organisation of the generated fields, presented in
figures~\ref{fig:uncond_velocity_increment} and~\ref{fig:uncond_qr}.

The generated fields reproduce the intermittent, heavy-tailed statistics of the
DNS across streamwise separations, rather than collapsing towards Gaussian
behaviour. We assess this through the scale-dependent probability density
function of the velocity increment $\delta u(r)$. We show the PDF of the normalized velocity increment $\delta u^*$ in 
figure~\ref{fig:uncond_velocity_increment} at $y^+ = 5$ and $y^+ = 15$ for four streamwise
separations $r^+ \in \{10, 40, 120, 350\}$, where $\delta u^*$ is standardized to zero mean and unit variance.  
The DNS exhibits the expected intermittency signature: at the
smallest separations ($r^+ = 10, 40$) the distribution is strongly
non-Gaussian, with tails extending well above $\mathcal{N}(0,1)$ out to $\pm 6$--$8$
standard deviations, and it relaxes towards Gaussian as $r^+$ increases,
essentially collapsing onto $\mathcal{N}(0,1)$ by $r^+ = 350$. Although slight discrepancies are present around the tails at $r^+=10$ for the generated fields, the overall trend is captured. The generated
tails open up together with the DNS at small $r^+$ rather than remaining pinned
near the Gaussian core, so the model captures the scale-dependent intermittency
of the flow. The degradation at smaller latent dimensions is only modest,
mirroring the skewness and flatness results, in which all
latent dimensions likewise nearly collapse onto the DNS. This behaviour
reflects the role of normalisation: skewness, flatness, and the standardized
increment PDF all divide out the fluctuation amplitude, which is precisely the
quantity that the low-$d_\mathsf{z}$ models fail to capture (as seen in the
underestimated Reynolds stresses 
figure~\ref{fig:mean_velo_stress}$b$). The increment PDF therefore probes the
shape of the distribution rather than its amplitude and is largely insensitive
to the defect that separates $d_\mathsf{z} = 4$ from $d_\mathsf{z} = 96$.

All six models recover the teardrop at every wall-normal
location, and the latent dimension controls how faithfully its asymmetry and the
extent of the Vieillefosse tail are reproduced. We report the invariants in the
normalised form $Q^{*}=Q/\left\langle S_{ij}S_{ij}\right\rangle$ and $R^{*}=R\left\langle S_{ij}S_{ij}\right\rangle^{3/2}$,
where $S_{ij}=\tfrac{1}{2}(A_{ij}+A_{ji})$ is the strain-rate tensor and
$\langle S_{ij}S_{ij}\rangle$ is averaged over the homogeneous $x$--$z$ plane at
the wall-normal location considered and over all snapshots; since this quantity
decays away from the wall, the axis scales in
figure~\ref{fig:uncond_qr} differ between panels by more than an order of
magnitude. The generated samples place their probability mass in
the same upper-left lobe as the DNS ($Q^{*}>0$, $R^{*}<0$; vortices being
stretched) and extend a tail into the lower-right quadrant along the lower branch
of the Vieillefosse line, with no sign of the contracted,
rounded contour that an over-smoothed field would produce. Because the teardrop
is asymmetric, we quantify it by the ratio of probability mass in the
upper-left to the upper-right quadrant (vortex stretching) and in the lower-right
to the lower-left quadrant (the Vieillefosse tail), both of which tend to unity
for a phase-randomized field. The effect of the latent dimension is visible at
all three locations and is most pronounced at $y^{+}=5$ and $15$. Models with
$d_{\mathsf{z}}<16$ are systematically more symmetric than the DNS, distributing probability
mass more evenly between the quadrants and producing a visibly shorter tail. At $y^{+}=5$ the vortex-stretching ratio increases monotonically from
$1.45$ at $d_{\mathsf{z}}=4$ to $1.75$ at $d_{\mathsf{z}}=64$ against the DNS
value of $1.82$, and the Vieillefosse ratio follows the same trend
($1.24$ up to $1.42$ against $1.55$). Models with $d_{\mathsf{z}}\geq16$
collapse onto the DNS contours over most of their extent, but the tail along the
Vieillefosse line remains shorter than that of the DNS at every location and for
every model: at $y^{+}=49$ the Vieillefosse ratio spans $1.89$--$2.08$ across the
sweep against the DNS value of $2.19$, without ordering in $d_{\mathsf{z}}$. The
most intense strain-dominated events are therefore not fully recovered even at
the largest latent dimensions considered. 
Overall, the generated fields capture the small-scale structural organisation of the flow, with a
clear improvement across the $d_{\mathsf{z}}<16$ threshold identified from the single-point statistics.

\subsection{Data Assimilation}
\label{subsec:DA}

We assess the performance of the conditional latent 
diffusion model in the context of data assimilation, focusing on whether 
posterior sampling reproduces the turbulent statistics of the flow when 
conditioned on pointwise observations drawn from the same distribution as 
the prior.
In the present work, data assimilation for a chaotic system such as turbulent plane Couette flow must satisfy two fundamental requirements: (i) agreement with observations at the observation locations, and (ii) consistency with the physical laws and statistical structure implied by the underlying turbulent dynamics. In addition, because turbulence is inherently chaotic, a successful assimilation procedure should generate an ensemble of plausible flow realisations that satisfy (i) and (ii), rather than collapsing to a single deterministic trajectory.
In §\ref{subsec:ROM_TurbulenceGen}, we demonstrated that the prior samples reproduce the key turbulent statistics of the flow with near DNS-level accuracy for the resolved scales, with minor degradation observed only in the energy spectra. The principal challenge in the conditional setting is therefore not the recovery of turbulence per se, but the preservation of its statistical and physical structure under observational constraints. Imposing these constraints, especially for a larger amount of observational data \citep{amoros2026guiding}, and maintaining the physical fidelity of the generated fields represent conflicting objectives for the generative model. Satisfying these two objectives simultaneously is by no means trivial, even when the data is drawn from the learned diffusion prior. 
In what follows, we show that the proposed approach of indirectly imposing statistical quantities can satisfy both observational consistency and physical fidelity, provided that the observational data are sufficiently sparse and appropriately distributed.

\begin{figure}[t!]
	\centering
	\includegraphics[width=\textwidth]{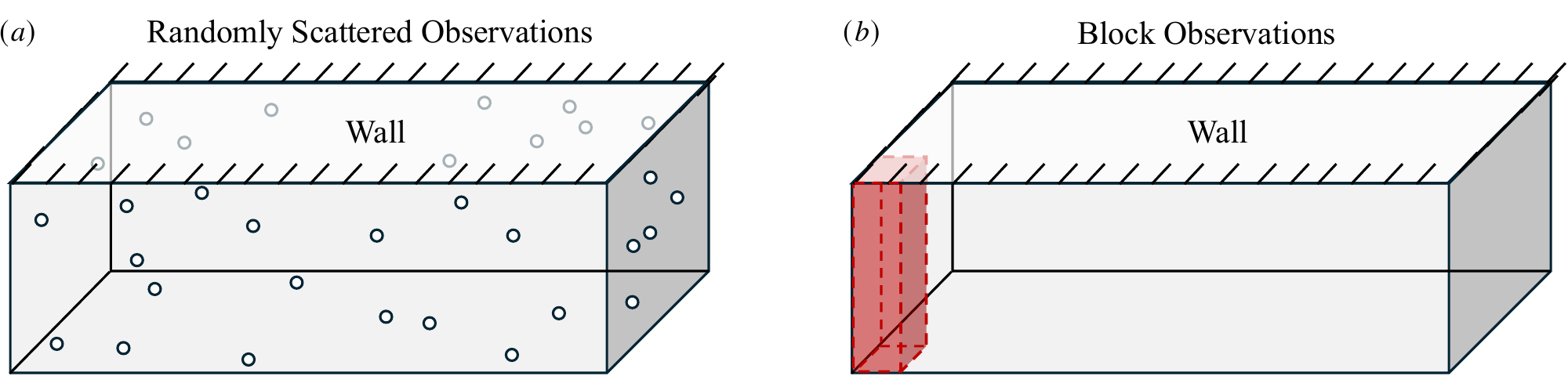}
	\caption{Schematic of the assimilated data in the $(a)$ randomly scattered and $(b)$ block observation tasks. The illustrated domains are scaled by factors of $0.2$, $1$, and $0.2$ in the $x$-, $y$-, and $z$-directions, respectively, relative to the generated domain. For clarity, every 50$^\text{th}$ data points in the baseline scattered task ($0.1\%$ of the total data points) are shown. For the block task, the baseline block of size $L_x/12 \times L_y \times L_z/8$ is depicted, while the $11 \times 64 \times 15$ data points within the block are omitted.}
	\label{fig:da_constraint_domains}
\end{figure}

\begin{figure}[t!]
	\centering
	\includegraphics[width=\textwidth]{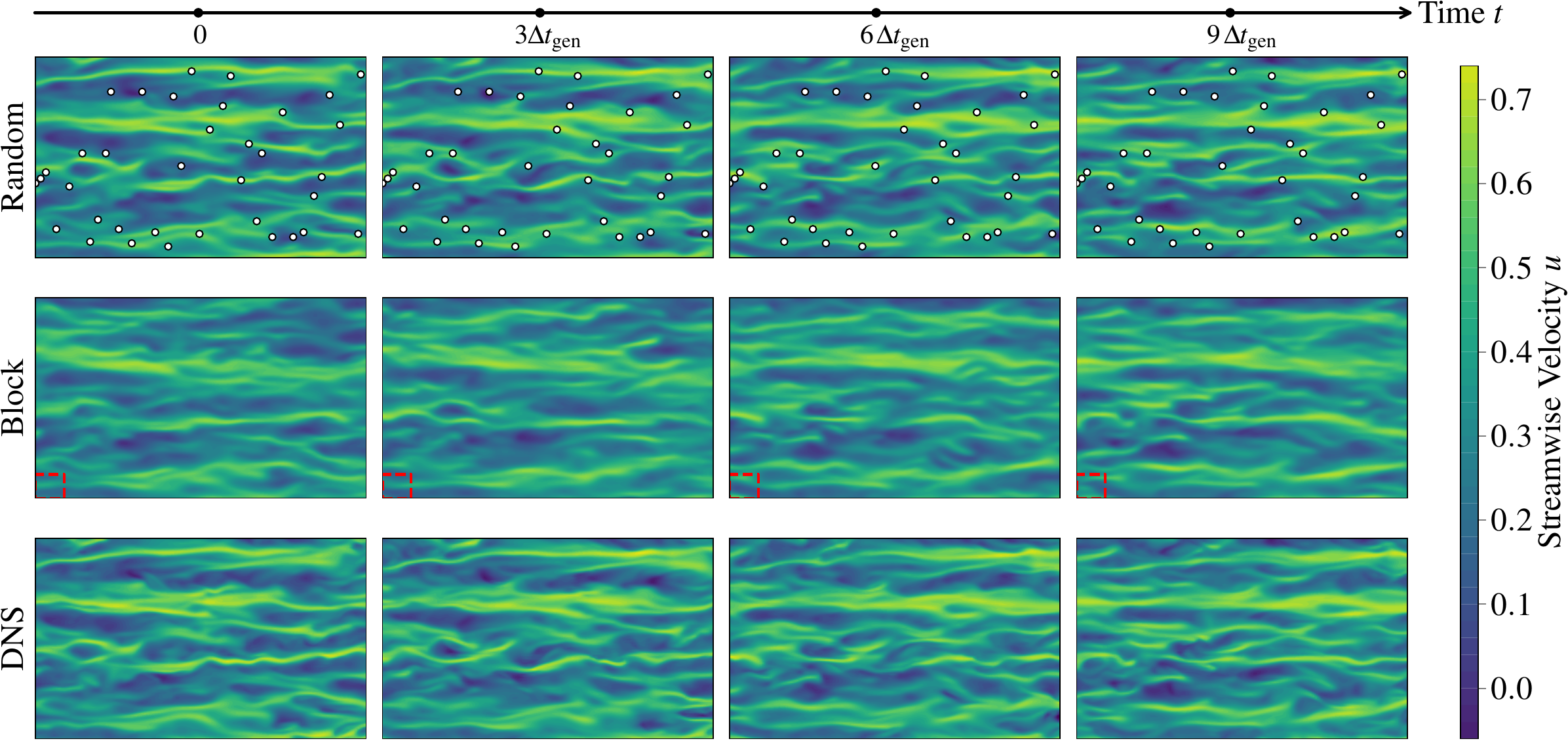}
	\caption{Assimilation of instantaneous observations to turbulence generation. Contours of streamwise velocity at $y^{+}=15$ are shown. Top row: generated samples with randomly scattered observations; middle row: generated samples with block observations; bottom row: DNS data. Sensor locations are indicated by circles (random) and a dashed box (block). In the scattered observations task, every fifth sensor location is displayed for clarity.}
	\label{fig:da_block_and_sparse}
\end{figure}

We consider two data assimilation configurations designed to reflect prototypical real-world observation scenarios. The first consists of data points randomly scattered across the full spatial domain (hereafter referred to as the ``scattered observations task''; figure~\ref{fig:da_constraint_domains}$a$). The second consists of sensors densely distributed within a rectangular subdomain (the ``block task''; figure~\ref{fig:da_constraint_domains}$b$). The block spans the full wall-normal extent of the domain and extends from $x=z=0$ in the streamwise and spanwise directions, respectively. In both configurations, the total number of observations is denoted by $N_{\text{obs}}$.
\begin{table}
	\begin{center}
		\small
		\centering
		\begin{tabular}{lrcccc}
			Case Name & Total Data Percentage & Block Size & Coarsening & Data Grid \\[3pt]
			R-50  & $50\%$  &  --      & --  &  -- \\
			R-10  & $10\%$  &  --      & --  &  -- \\
			\textbf{R-1} (baseline)  & $\mathbf{1\%}$   & \textbf{--}  & \textbf{--} & \textbf{--} \\
			R-0.1  & $0.1\%$  & --    & --  & --  \\
			R-0.01  & $0.01\%$  & --     & --  & --  \\
			R-0.001  & $0.001\%$  & --     & --  & --  \\[3pt]

			B-6  & $6\%$  & $L_x/4 \times L_y \times L_z/4$   &  --  &  $31 \times 64 \times 31$  \\
			\textbf{B-1} (baseline)   & $\mathbf{1\%}$  &  $\mathbf{L_x/12 \times L_y \times L_z/8}$  & \textbf{--}  & $\mathbf{11 \times 64 \times 15}$  \\
			B-1s3$\times$2  & $1\%$  &  $L_x/4 \times L_y \times L_z/4$              & $3 \times 1 \times 2$   & $11 \times 64 \times 15$   \\
			B-1s6$\times$4  & $1\%$  &   $L_x/2 \times L_y \times L_z/2$                & $6 \times 1 \times 4$  & $11 \times 64 \times 16$  \\
			B-0.1  & $0.1\%$  &  $L_x/12 \times L_y \times L_z/8$             & $2 \times 2 \times 2$  & $5 \times 32 \times 8$  \\
			B-0.02  & $0.02\%$  &  $L_x/12 \times L_y \times L_z/8$              &  $5 \times 4 \times 4$  &  $4 \times 16 \times 4$
		\end{tabular}   
		\caption{Considered data assimilation cases. Random observations scattered over the full domain (R), and spatially confined block observations (B) are examined. The baseline configurations, each using 1\% of the total grid points for the randomly scattered observation and block task, are highlighted in bold.}
		\label{tab:da_cases}
	\end{center}
\end{table}
The scattered observations task emulates sparsely distributed observations, such as those obtained from UAV-based sensing in wind farm applications, whereas the block task represents spatially concentrated observations within a localized region, analogous to LiDAR observations. In both cases, observational data are sampled from direct numerical simulation (DNS) at intervals of $\Updelta t_{\text{gen}}$, without any noise addition. We report additional results with observation noise in Appendix \ref{app:Noisy_Observations}. Because conditioning is imposed pointwise in physical space, the operator $\mathcal{H}$ in the observation operator $\mathcal{F} = \mathcal{H} \circ \mathcal{D}$ acts as a masking operator that retains values at observed locations and nullifies the remainder of the domain.
The specific assimilation cases considered are summarised in table~\ref{tab:da_cases}, where baseline configurations are indicated in bold. For the scattered observations task, we consider different values of the data percentage, that is, the number of sensors as a percentage of the total number of grid points. For the block task, we explore different block sizes and data percentages as well as different levels of coarsening. Coarsening refers to selecting every $n^\text{th}$ point in each coordinate direction to create a sparser, regularly spaced subset of the original block (e.g., a coarsening of $3 \times 1 \times 2$ selects every third point in the streamwise and every other point in the spanwise direction). The baseline \emph{scattered task} corresponds to observations at $1\%$ of the total spatial grid points, sampled uniformly at random. The baseline \emph{block task} also uses $1\%$ of the total grid points, but confined to a subdomain of dimensions $L_x/12$ and $L_z/8$, with all grid points within this block treated as observations. 
This observation density is representative of the observation-to-state ratios encountered in operational forecasting applications~\citep{MACK2020113291}. For all cases, we employ the latent representation introduced in §\ref{subsec:ROM_TurbulenceGen}. To focus on the impact of observation configuration and statistical conditioning, we adopt the 32-dimensional latent space, which exhibits improved posterior sampling performance compared with the 16-dimensional representation. Unless otherwise specified, the conditioning strength parameter is set to $\rho=0.5$. To impose statistical quantities indirectly, we generate 300 samples using the sampling approach described in \S\ref{subsec:Methods_cDM}, with the starting time of each observation segment separated by $30\Updelta t$.
Streamwise velocity contours of generated samples for the baseline configuration of the two tasks are shown in figure \ref{fig:da_block_and_sparse}. The velocity contours for both tasks show good agreement with the DNS sequence, with small-scale discrepancies visible upon closer inspection, especially for the block task. 

To quantify the consistency of the generated samples with the imposed data we use the average normalised root mean squared error (nRMSE). We define the average nRMSE as 
\begin{equation}
	\text{nRMSE}(\boldsymbol{u})=\sqrt{\frac{\sum_{i=1}^{N_p}\lVert\tilde{\boldsymbol{u}}_i-\boldsymbol{u}_i\rVert_2^2}{\sum_{i=1}^{N_p}\lVert\boldsymbol{u}_i\rVert_2^2}},
	\label{eq:rsme}
\end{equation}
where $N_p$ denotes the number of samples over which the average is taken and $\boldsymbol{u}$, $\tilde{\boldsymbol{u}}$ denote the DNS and generated velocity fields respectively. 
While nRMSE is reported here for completeness and comparison with prior studies, its interpretation in the context of chaotic turbulent flows requires careful consideration. We therefore defer a detailed discussion of trajectory- and distribution-level metrics to \S\ref{sec:discussion_rmse}.
For evaluating the physical consistency of the generated samples, we employ the statistical quantities presented in \S\ref{subsec:ROM_TurbulenceGen}.

\begin{figure}[t!]
	\centering
	\includegraphics[width=\textwidth]{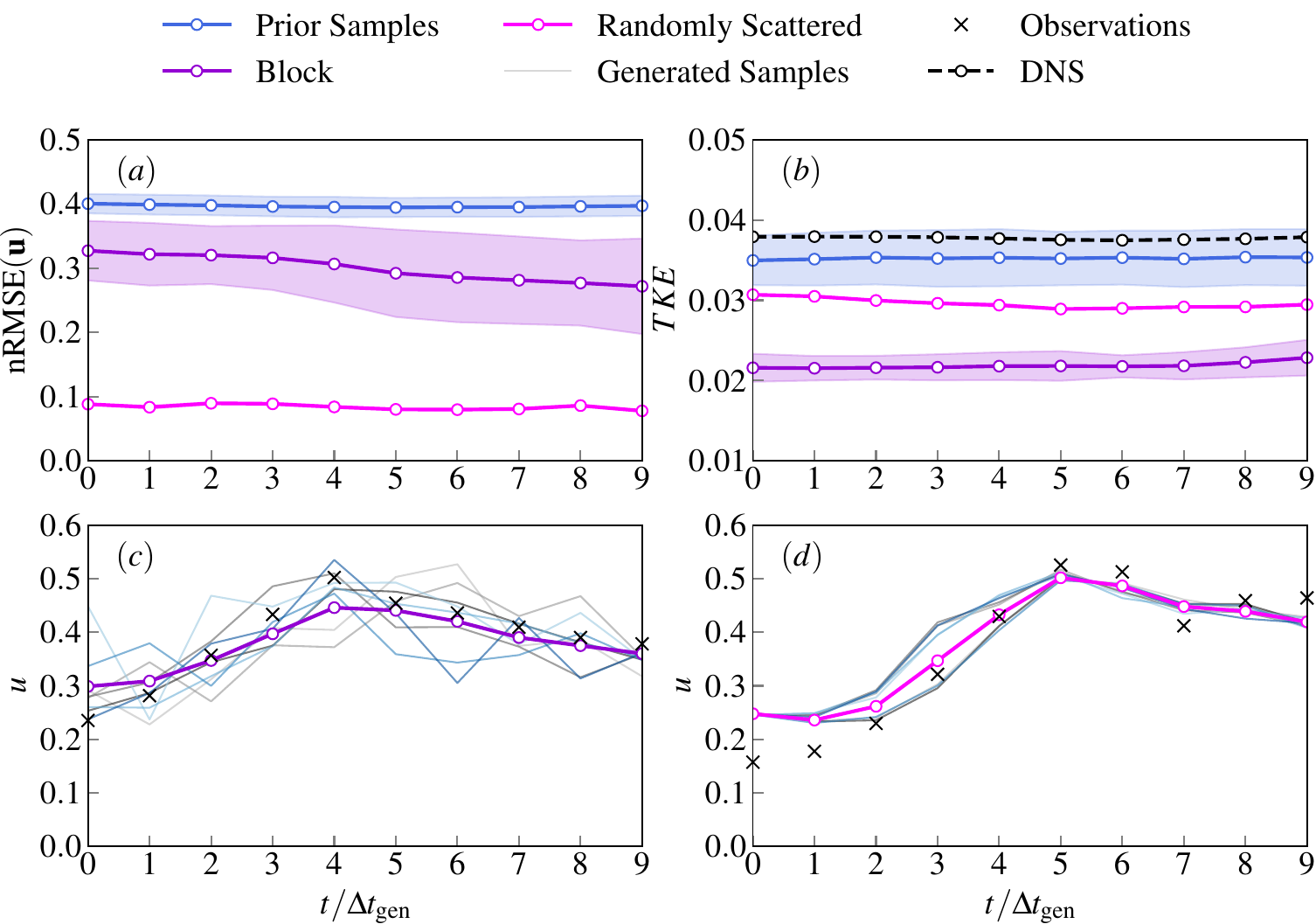}
	\caption{Temporal evolution of trajectory-level deviations from the DNS reference for the baseline configurations of the two assimilation tasks. Conditionally generated flows with scattered and block observations, as well as prior samples, are shown. $(a)$ Spatially averaged normalised root mean squared error (RMSE) in velocity between the generated fields and DNS as a function of time. $(b)$ Time evolution of instantaneous turbulent kinetic energy. The shaded region denotes one standard deviation computed from 50 samples conditioned on the same data. Time evolution of the streamwise velocity at an observed location for the block $(c)$ and randomly scattered $(d)$ observation tasks, respectively. Individual sample trajectories are shown with shades of blue and grey. For clarity, 8 samples are shown. The ensemble mean is indicated by circles ($\circ$). 
	}
	\label{fig:da-rmse-spatially-averaged}
\end{figure}
\begin{figure}[h!]
	\centering
	\includegraphics[width=\textwidth]{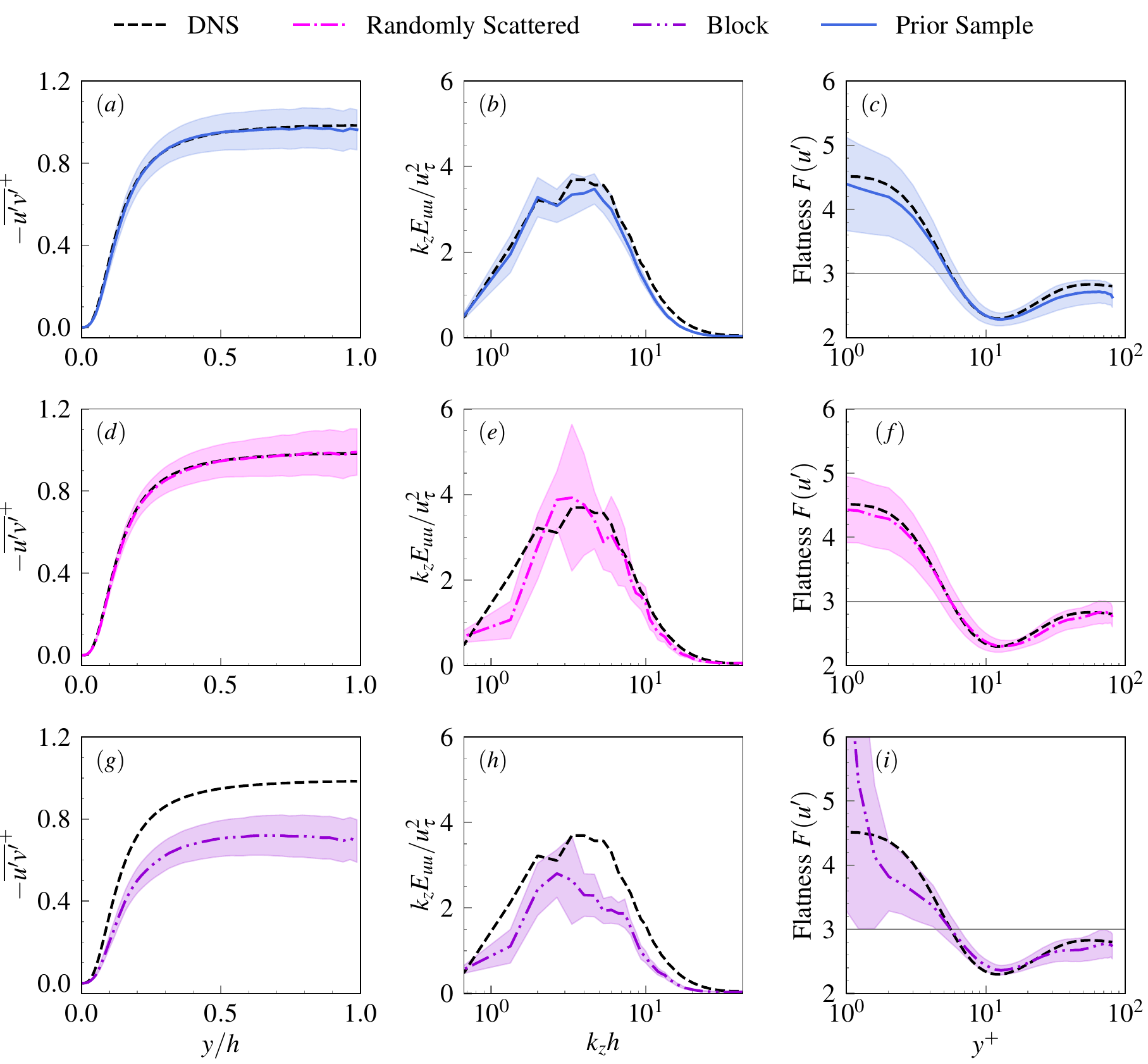}
	\caption{Performance of the conditionally generated flow for the scattered and block data assimilation tasks. The resulting statistics are shown for $(a,d,g)$ Reynolds shear stress, $(b,e,h)$ premultiplied energy spectra for the streamwise velocity as a function of the spanwise wavenumber at $y^+=15$, and $(c,f,i)$ flatness of the streamwise velocity fluctuation. We compare against prior samples with latent dimension $d_{\mathsf{z}}=32$ and the DNS reference. The shaded region denotes one standard deviation computed from 300 samples conditioned on observation segments separated by $30\Updelta t$.}
	\label{fig:da_sparse_block_statistics_with_uncertainty_bands}
\end{figure}

The difference in the spatial distribution of the observations between the two tasks lead to significant differences in the generated samples due to the highly anisotropic correlation structure of the flow.
In wall-bounded turbulence such as Couette flow, the streamwise correlation length scale is substantially larger than those in the wall-normal and spanwise directions, as illustrated in figure~\ref{fig:da_block_and_sparse}. Spatially distributed observations (circles in figure~\ref{fig:da_block_and_sparse}, \textit{top row}) therefore impose constraints that propagate efficiently in the streamwise direction, leading to a relatively homogeneous influence across the domain. In contrast, block observations (red block in figure~\ref{fig:da_block_and_sparse}, \textit{middle row}) concentrated within a localized region provide strongly localized conditioning, with constraints that weaken as the correlation with distant regions decreases. This distinction in conditioning geometry is expected to produce qualitatively different spatial patterns in the reconstruction deviation.

The baseline configurations for the two tasks show varying levels of success in terms of data consistency, sample diversity and physical fidelity. The comparison of data consistency between the two tasks is provided in figure~\ref{fig:da-rmse-spatially-averaged}$(a)$, which shows the variation of spatially averaged nRMSE over time. The results in this figure are obtained by imposing the same data 50 times to assess sample diversity as well as data consistency. 
Both data assimilation tasks show smaller deviation from the DNS than prior samples, as seen in figure~\ref{fig:da-rmse-spatially-averaged}$(a)$.
However, the scattered observations task shows smaller deviations, indicating better data consistency. 
The time evolution of instantaneous turbulent kinetic energy (figure~\ref{fig:da-rmse-spatially-averaged}$b$) and evolution of streamwise velocity of individual samples (figure~\ref{fig:da-rmse-spatially-averaged}$c,d$) further illustrate this behaviour. Results are presented for a randomly selected assimilation point in each task.
Although the block task exhibits greater apparent sample diversity than the scattered-observation case, it achieves poorer data consistency. While this may suggest improved representation of flow variability, further analysis indicates that block-conditioned samples lack physical fidelity. The observed nRMSE variation therefore arises from unphysical realizations rather than meaningful distributional coverage.

In terms of physical fidelity, the randomly scattered observations task is significantly more successful than the block task, as shown in figure~\ref{fig:da_sparse_block_statistics_with_uncertainty_bands}. The results shown here are generated using the strategy outlined in \S\ref{subsec:DA} to ensure that the generated ensemble approximates the posterior distribution conditioned on observations. The ensemble mean statistics together with the standard deviations computed from 300 posterior samples are presented for the unconditional prior and both conditional data assimilation baseline cases; additional ensemble mean statistics of first, second, and third order are provided in appendix~\ref{app:Additional_results}. The Reynolds shear stress profile of the scattered observations task shows similar performance to the prior samples and much better agreement with the DNS (figure~\ref{fig:da_sparse_block_statistics_with_uncertainty_bands}$a,d,g$) compared to the block task. The premultiplied energy spectra (figure~\ref{fig:da_sparse_block_statistics_with_uncertainty_bands}$b,e,$) confirm that posterior samples can approximately match the length scales of the coherent structures in the buffer layer, though the wavenumber with the highest energy is underestimated for both tasks and the block task additionally exhibits significant damping of turbulent fluctuations across a broad band of wavenumbers. The difference in physical fidelity between the two tasks is also apparent in the flatness profiles, which show significant discrepancies near the wall (figure~\ref{fig:da_sparse_block_statistics_with_uncertainty_bands}$c,f,i$) for the block task.

In contrast to figure~\ref{fig:da-rmse-spatially-averaged}, where repeatedly conditioning on the same data reduced the ensemble spread significantly for the randomly scattered task, here the 300 posterior samples are drawn from synchronous but independent time series segments each separated by $30\Updelta t$, such that the ensemble retains a spread comparable to the prior. Accordingly, all three cases exhibit approximately the same width of the standard deviation bands across most statistics, suggesting that conditioning on observations does not substantially reduce the ensemble spread relative to the prior. As the magnitude of the statistics increases, so does the corresponding standard deviation. An exception to this trend is observed in the fourth-order flatness profile along the wall-normal direction: as the flatness decreases below the Gaussian reference value of three in the viscous sublayer and buffer layer, the standard deviation diminishes accordingly, before increasing again to comparably elevated levels in the outer layer. Particularly noticeable is the large standard deviation in the flatness profile of the block task near the wall, which likely reflects the general difficulty in accurately predicting higher-order statistics in the immediate vicinity of the wall when observations are spatially clustered.

The baseline configurations reveal distinct failure modes that highlight the intrinsic difficulty of diffusion-based data assimilation. Although diffusion posterior sampling provides a principled Bayesian-like framework for incorporating observations, our results show that practical performance depends sensitively on observation density, spatial configuration, and conditioning strength. A detailed analysis of this tension is presented, and the relation to classical ensemble-based data assimilation methods, in \S\ref{sec:discussion_enkf}.

\begin{figure}[t!]
	\centering
	\includegraphics[width=\textwidth]{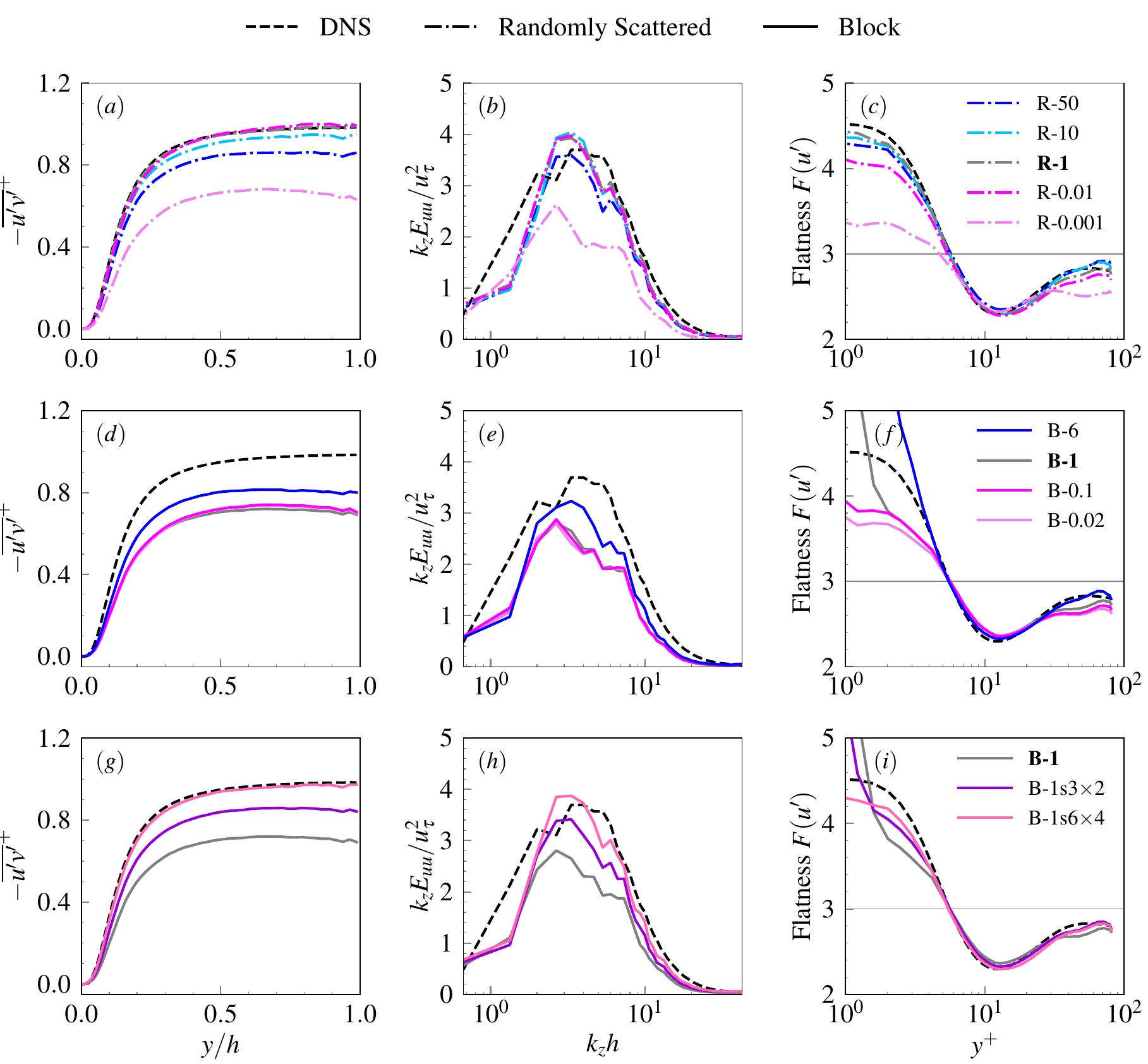}
	\caption{Effect of task parameters on conditionally generated flows for randomly scattered observations and block data assimilation. Panels ($a$--$c$) illustrate the influence of the total data percentage in the scattered observations task. 
		Panels ($d$--$f$) show the influence of total data percentage in the block task. Panels ($g$--$i$) show the influence of data grid stretching in the block task. The generated flows are evaluated using the Reynolds shear stress (second-order statistics) as a function of the wall-normal coordinate, the pre-multiplied energy spectrum of the streamwise velocity component as a function of spanwise wavenumber at $y^+=15$, and the skewness of the streamwise velocity fluctuations (fourth-order statistics) as a function of the logarithmic wall distance. Results are compared with the DNS reference data. Baseline cases are shown in grey, cases with decreasing observation density in shades of magenta/purple, and cases with increasing observation density in shades of blue. }
	\label{fig:da-studies-statistics}
\end{figure}

To elucidate the mechanisms underlying the observed degradation, we perform a series of targeted parametric studies. Specifically, we vary the block size in the streamwise and spanwise directions, the total number of observations $N_{\text{obs}}$, and the spatial configuration by coarsening the grid-arranged data points within the block region. The corresponding cases are summarised in table~\ref{tab:da_cases}.
\begin{figure}[t!]
	\centering
	\includegraphics[width=\textwidth]{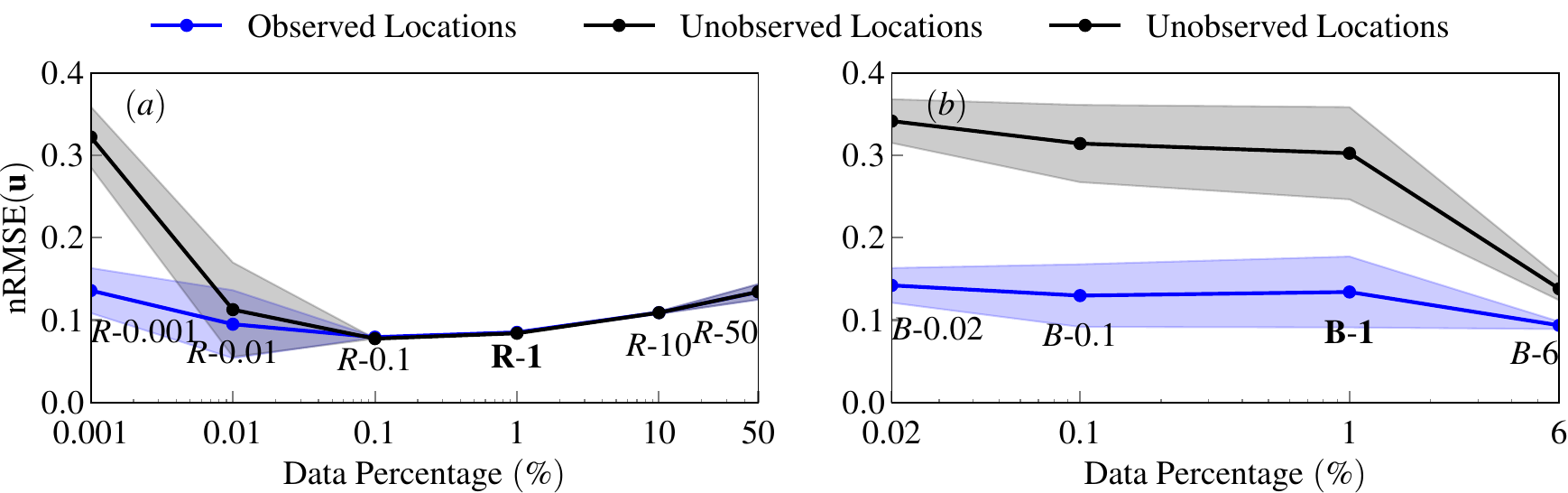}
	\caption{Effect number of observations on posterior sampling. We show the variation of spatiotemporally averaged normalised root mean squared error (nRMSE) in velocity with observation percentage for $(a)$ the scattered observations task and $(b)$ block observations. Shaded region indicates one standard deviation over 50 generated samples with the same data. The name of each case is indicated on the plot. The baseline cases are highlighted in bold.}
	\label{fig:da-rmse-parametric_study}
\end{figure}
Our studies show that increasing the number of observations tends to improve data consistency for both tasks. However, a higher sensor count alone does not necessarily lead to better agreement with the data. We examine the effect of sensor count on agreement with the data in figure~\ref{fig:da-rmse-parametric_study}, which shows the variation of nRMSE with number of observations for both tasks at observed and unobserved locations.
For randomly scattered observations, the nRMSE outside the measured locations tends to decrease as the data percentage increases. As more observations are added and the model receives more information about the state of the system, the agreement with the DNS sequence improves even in unobserved locations. 
For the block task, the nRMSE at measured and unmeasured locations stays more or less constant until a observation percentage of 6\%. Note that the B-0.02, B-0.1 and B-1 cases have the same block size and the data percentage is increased by adding more observations within this block (see table~\ref{tab:da_cases}). Since the newly added observations come from the same constrained region, they are highly correlated with the existing observations and provide limited information gain. For the block case with 6\% total data percentage (B-6) the data grid is extended and the model receives information from a wider region of the domain, leading to a sharp drop in the nRMSE. The slight increase in the nRMSE from R-1 onward cases can again be attributed to the observations becoming more correlated with increasing sensor density.
The deviation from the DNS is much lower at measured locations than the unmeasured locations for the R-0.001 case, indicating that this case can satisfy the observation constraints, even though it fails to reproduce the turbulence statistics as we demonstrate later in figure \ref{fig:da-studies-statistics}$(a$--$c)$. 
Beyond a data percentage of 0.01\%, the error at unmeasured and measured locations converge for the scattered task. This is in contrast to the block task, where error at measured locations is consistently lower at measured locations. This difference between the two tasks can be explained by the spatial correlation structure of the flow and its interactions with the conditioning geometry. In the scattered observations task, the relatively even spatial distribution of observations, together with the long streamwise correlation length scale, imposes a more homogeneous constraint across the periodic domain. Therefore, the deviation becomes more uniformly distributed across the domain. In the block task, however, the data is localized to a small portion of the domain. As the correlation with the localized block data decreases, the conditioning becomes progressively weaker, leading to larger deviations in unobserved locations. 

Conditioning with randomly scattered observations yields near DNS-level accuracy in turbulent statistics, provided that the observation density lies within an appropriate range. We examine the behaviour of  turbulent statistics in figure~\ref{fig:da-studies-statistics}($a$--$c)$ for scattered observation cases in which the total number of observations is systematically increased from $0.001\%$ to $50\%$ (R-0.001 to R-50) of all grid points in the domain.
We present three representative turbulence statistics: the Reynolds shear stress, the premultiplied spanwise energy spectrum of the streamwise velocity component, and the flatness of the streamwise velocity fluctuations. Near DNS-level accuracy is achieved when the observation density lies approximately between $0.01\%$ and $1\%$, although degradations in the pre-multiplied energy spectra for low spanwise wavenumbers and in the flatness near the wall remain. The R-0.01\% shows good agreement with the DNS statistics, while exhibiting satisfactory data consistency and sample diversity (see figure~\ref{fig:da-rmse-parametric_study}). Therefore, this case satisfies the requirements for successful data assimilation.  

To examine the limits of this method, two extreme cases with 0.001\% and 50 \% data points were additionally carried out. From these, we detect two failure modes where the turbulent statistics degrade. Intuitively, in the extreme sparse case (R-0.001), the likelihood contribution in equation~\ref{eq:dps_first} should be negligible, as it scales with only a very small number of discrepancy terms proportional to $N_{\text{obs}}$. One would therefore expect the generated statistics to converge towards those of the unconditional model. However, the turbulent statistics in the R-0.001 case deviate significantly from the prior (unconditional) samples. This apparent contradiction arises from the normalisation used in equation~\ref{eq:dps_alt}. 
Because the gradient of the $L^2$ norm rescales the update by the inverse of the innovation norm $\|\mathbf{y} - H(\mathbf{x})\|$ (i.e., the magnitude of the model-observation discrepancy), the very small number of sparse observations disproportionately determine the correction direction.
In such settings, isolated observations may be overweighted and observational noise amplified, leading to unphysical samples and degraded statistics. The failure in the R-0.001 case therefore reflects not insufficient conditioning, but rather instability induced by normalised sparse updates.
In~\citet{chung2023diffusion}, the conditioning strength was calibrated for settings with relatively fixed observation density. In contrast, the present study spans four orders of magnitude in $N_{\text{obs}}$, significantly altering the balance between likelihood and prior contributions. 
The instability in the R-0.001 case thus exposes a regime in which the original normalisation becomes unreliable. This observation aligns with broader critiques of the diffusion posterior sampling formulation~\citep{rozet2023score} and indicates that its normalisation warrants careful reconsideration when the number of observations varies widely.
On the other hand, the case with 50\% observations (R-50) exhibits a distinct failure mode associated with overly dense conditioning. This same mechanism underlies the degradation observed in the baseline block case (B-1) and is examined further in the block data assimilation study.

In the block task, increasing the number of observations by refining the sensor grid within a fixed block yields little change in the turbulent statistics. In contrast, enlarging the block, thereby increasing both the number and spatial extent of constraints, produces modest improvement in Reynolds shear stress and energy spectra (figure~\ref{fig:da-studies-statistics}$d$--$f$). However, the flatness profile deteriorates as the observation percentage increases. This behaviour can be interpreted within the Bayesian-like framework of diffusion posterior sampling. Conditioning modifies the learned diffusion prior through a likelihood term that steers the generative process towards regions of the high-dimensional state space consistent with the observations. As the number of observations increases, the admissible region of the distribution becomes progressively restricted. When the conditioning becomes sufficiently dense, particularly in a correlated manner, the posterior becomes increasingly concentrated in regions of low prior probability or outside the well-represented manifold of the learned prior. In this regime, the generative process struggles to reconcile strict local data consistency with preservation of the global statistical structure, resulting in a degradation of the turbulent statistics.

Enlarging the spatial coverage of the block while keeping the total number of observations fixed substantially improves recovery of the turbulent statistics. Figures~\ref{fig:da-studies-statistics}($g$--$i$) isolate this effect at constant observation density ($1\%$), comparing three configurations with identical $N_{\text{obs}}$ but different grid stretching (B-1, B-1s3$\times$2, and B-1s6$\times$4). Increasing the stretching enlarges the block and distributes the same number of data points over a wider region. As the block coverage increases, the Reynolds shear stress and premultiplied energy spectrum approach DNS-level accuracy, whereas a small, localized block systematically underestimates these statistics and exhibits significant deviation in flatness near the wall.
This behaviour reflects the role of spatial correlation in the conditioning process. Localized blocks impose highly correlated constraints, particularly in the streamwise direction, leading to strong but spatially concentrated posterior conditioning and degraded global statistics. Distributing the same observations over a larger region reduces inter-point correlation and spreads the constraint more evenly across the domain. In this regime, the block task increasingly resembles the scattered-observation scenario, enabling partial recovery of near DNS-level turbulent statistics, although clear degradations in the pre-multiplied energy spectra and the flatness persist.

The behaviour observed here also bears strong resemblance to classical challenges in ensemble-based data assimilation. In particular, the effective information content of observations depends not only on their number but also on their spatial correlation. We discuss the connection between diffusion-based conditioning and traditional ensemble Kalman filtering in \S\ref{sec:discussion_enkf}.

\section{Discussions}
\label{sec:discussions}
The adoption of generative models introduces a methodological shift in the simulation and data assimilation of turbulent flows. Rather than evolving turbulent trajectories through explicit integration of the governing equations, these models learn an approximation of the underlying probability distribution of flow states. As a consequence, several principles traditionally associated with turbulence simulation and classical data assimilation require reinterpretation within this probabilistic framework. Accordingly, we examine three conceptual aspects of this shift: (i) the role of reduced computational domains in probabilistic turbulence modelling, (ii) the interpretation of trajectory-level error metrics for evaluating generative models, and (iii) the balance between conditioning and prior structure in conditional diffusion models.

\subsection{Reduced Generation Domain versus DNS Domain Requirements}
\label{sec:discussion_domain}
The reduced physical domain employed in this work reflects a fundamental conceptual difference between probabilistic generative modelling and traditional direct numerical simulation (DNS).
Turbulence simulations conventionally rely on large computational domains to represent the full range of turbulent motions. In contrast, the reduced domain employed here is not intended as a numerical approximation of DNS, but as a deliberate modelling choice aligned with the goal of learning turbulence statistics rather than time-resolved dynamics.

Conventional DNS of statistically stationary wall-bounded turbulence requires a computational domain sufficiently large to resolve the largest dynamically relevant flow structures and avoid artificial confinement of large scale motions. The subdomain considered in this work does not satisfy this requirement in the streamwise direction, as indicated by the absence of a roll-off in the premultiplied energy spectra at the lowest resolved wavenumbers (see figure~\ref{fig:premult_energy_spectra_spanwise}) and by the two-point correlations shown in figure~\ref{fig:two-point-correlation-components}, which remain nonzero at the largest streamwise separations.
Consequently, this subdomain would be too short for a conventional DNS to accommodate statistically independent large-scale motions.

However, the generative model is not restricted by this domain-size 
requirement. Its objective is to learn the joint probability distribution 
of flow states rather than their explicit temporal evolution dynamics, and the 
influence of the surrounding flow is implicitly embedded in each 
subdomain through the training data, which are extracted from a properly 
resolved full-domain DNS. The ensemble of subdomains therefore encodes 
the turbulent motions across the range of scales captured within the 
window.
It is useful to distinguish what the generative model can and cannot 
capture. Scales exceeding the generation window $L_x \times L_y \times 
L_z$ (representing $1/8 \times 1/2 \times 1/2$ of the full DNS domain, 
respectively) cannot be represented in individual samples. Within the window, however, the statistics 
remain faithful to the full system for the resolved scales. This is a fundamental distinction 
from performing DNS in a reduced computational domain, where removing 
the largest scales would alter the dynamics of the smaller scales 
themselves through nonlinear coupling. In the generative setting, the 
small-window samples instead inherit the statistics of the full system, 
even though no individual sample displays the largest coherent 
structures in their entirety. The generative model should therefore be 
interpreted as a statistical surrogate for the turbulence at the 
subdomain scale, designed to reproduce local statistics faithfully 
rather than to replace a full-domain DNS. Accordingly, the generated samples should be understood as representing the learned subdomain-scale distribution: the small- and intermediate-scale turbulence within the generation window is well reproduced, whereas the largest streamwise coherent structures and the long-range streamwise organization of the flow lie outside the demonstrated capability of the present model. Consistent with this view, 
all small-scale structures are resolved down to the DNS grid spacing, 
and the statistics computed from generated samples reproduce the DNS 
reference with near DNS-level accuracy for the scales resolved by the subdomain -- an \textit{a posteriori} 
indication that, for the configuration considered here, the chosen 
subdomain captures the scales that dominate the relevant turbulent 
statistics.

\subsection{Trajectory-Level Metrics versus Distributional Fidelity}
\label{sec:discussion_rmse}
The use of root mean squared error (RMSE) as an evaluation metric in generative turbulence modelling warrants careful interpretation. While RMSE has been adopted in several related studies~\citep[e.g.,][]{LiBiferale2023,du2024conditional,li2024learning}, it fundamentally measures instantaneous, trajectory-level (pointwise) deviation relative to a reference realisation. However, in turbulent flows, trajectory matching is generally not an appropriate criterion for model quality. Turbulence is chaotic, and two DNS or LES starting from nearly identical initial conditions may diverge exponentially, with a growth rate characterised by the largest Lyapunov exponent~\citep{Ott2002}, while still producing statistically indistinguishable turbulence. Consequently, large pointwise differences between two realizations do not necessarily imply poor physical fidelity.

For this reason, DNS and LES are evaluated in the turbulence community primarily based on the statistical quantities they reproduce (e.g.\ mean profiles, Reynolds stresses, energy spectra), rather than instantaneous field agreement. Two trajectories that diverge substantially in an $L^2$ sense may nevertheless represent equally plausible realizations of the same underlying turbulent flow if their statistics coincide. Since diffusion models aim to approximate the full data distribution rather than match individual samples~\citep{ho2020denoising,song2021scorebased}, their assessment should likewise emphasise distributional fidelity rather than trajectory agreement alone. Adequate sample diversity and sufficient coverage of the underlying distribution are therefore essential for faithful representation of turbulent dynamics.

Within this perspective, RMSE should be interpreted as quantifying \emph{deviation} from a particular realisation, rather than \emph{error} in the sense of physical inaccuracy. That said, for strongly conditioned tasks—such as filling gappy data when a large portion of the flow field is prescribed, as in~\citet{LiBiferale2023} (``in-painting'' in computer vision)—trajectory-level metrics can be justified. In such cases, the conditioning substantially reduces uncertainty, leading to a sharply concentrated posterior distribution in which pointwise agreement becomes a meaningful indicator of performance.

\subsection{Relation to classical ensemble-based data assimilation}
\label{sec:discussion_enkf}
Conditional diffusion models implement data assimilation through the Bayesian decomposition in equation~\ref{eq:cdm-bayes}, in which the posterior score is expressed as the sum of a prior term and a likelihood correction. This structure is formally analogous to ensemble-based methods such as the EnKF, where a forecast model provides a prior ensemble that is subsequently updated by incorporating observations through a likelihood-based correction. In both frameworks, the central challenge is to reconcile data consistency with preservation of the dynamical or statistical structure encoded in the prior.

From this perspective, conditional generation corresponds to sampling from a restricted subset of a high-dimensional probability distribution while attempting to retain the intrinsic correlations learned by the unconditional model. However, our results demonstrate that this theoretical mechanism does not automatically yield robust performance. Conditioning does not simply ``insert'' observational information into the model; it biases the sampler towards a subregion of state space consistent with the data, while the diffusion dynamics promote global consistency with the learned prior distribution. This induces a structural tension between local agreement with observations and preservation of global turbulent statistics.

The failure modes identified in \S\ref{subsec:DA} reflect different manifestations of this tension. For sparse observations, the normalisation of the likelihood term can overweight isolated observations and amplify noise, effectively distorting the posterior correction relative to the prior structure. For dense or strongly correlated observations, the posterior becomes increasingly concentrated, restricting the admissible region of the learned distribution and potentially pushing the sampling dynamics towards regions that are poorly represented by the prior. In both regimes, the difficulty arises from the delicate balance required between the likelihood correction and the prior manifold.

These observations closely mirror classical data assimilation challenges. In EnKF, increasing the number of observations within a spatially correlated region does not necessarily increase effective information content, since redundant observations primarily constrain the same directions in state space. Similarly, in the diffusion-based setting, highly correlated or spatially concentrated conditioning can disproportionately restrict specific modes of the learned distribution. A key distinction, however, lies in how physical consistency is maintained. In traditional ensemble methods, the governing equations explicitly propagate ensemble members in physical time, ensuring dynamical coherence. In contrast, diffusion models encode physical structure implicitly within the learned prior. Excessively strong or localized conditioning can therefore distort this learned manifold, degrading statistical fidelity even when local data consistency is enforced.

A complementary physical perspective is provided by the turbulence synchronisation and scale modulation literature~\citep{yoshida2005regeneration,nikolaidis2022synchronization,vela2021synchronisation,he2024four,he2026enhancing}, which suggests that when large-scale information is sufficiently constrained, small-scale dynamics may become effectively slaved to the large scales through nonlinear turbulence interactions. In the limit of very dense conditioning, this behaviour may emerge in our framework as well: the turbulent system loses its chaotic variability and the probabilistic ensemble collapses towards a unique trajectory. The present framework is however fundamentally probabilistic rather than deterministic. For turbulence synchronisation, the spatial observation density, temporal observation density and time span of the data assimilation play a crucial role \citep{he2026enhancing}. The conditioning configurations considered here are far from this synchronisation limit. While the denser randomly scattered cases exceed the critical spatial density ($\mathcal{S} = N_{\text{obs}}/(N_x \times N_y \times N_z)^{1/3} > 0.33$) reported by \citet{he2026enhancing} for plane channel flow at $Re_\tau = 180$, the observation interval of $\Delta t_{\text{gen}}^+ \approx 5$ is considerably larger than their value of $\Delta t^+ \approx 1$, and our generation horizon of $t^+ \approx 46$ falls well short of the full synchronisation time of $t^+ > 200$ reported in the same study. 
The learned prior encodes cross-scale statistical dependencies that allow observational constraints to influence the posterior distribution over unobserved scales, but without slaving small scales uniquely to the large scales. The identification of an observation density and temporal resolution threshold beyond which synchronisation-like behaviour emerges in diffusion-based data assimilation represents an interesting direction for future work.

\section{Conclusions}
\label{sec:conclusions}
In this work, we present a latent diffusion framework for reduced-order modelling and data assimilation in wall-bounded turbulent flows. Traditional data assimilation methods such as the ensemble Kalman filter (EnKF) coupled to a numerical solver require repeated evaluations of high-fidelity numerical solvers, which limits their applicability to high-Reynolds-number, wall-bounded turbulent flows encountered in real-world settings such as wind farms. Likelihood-based generative models such as diffusion models offer a complementary approach by learning complex high-dimensional probability distributions directly from data. The present work establishes a foundation for generative data assimilation in such settings.

Our framework combines a $\beta$-variational autoencoder ($\beta$-VAE) with a diffusion transformer (DiT) to generate four-dimensional spatiotemporal flow samples. We demonstrate the approach on plane Couette flow at $Re_h=1300$. The prior samples closely reproduce DNS statistics up to fourth-order moments, as well as two-point correlations and energy spectra, using only 16 latent degrees of freedom. Minor degradation is observed in the pre-multiplied energy spectra and flatness profiles. The accuracy degrades towards LES-level predictions when fewer than 16 latent degrees of freedom are employed. Furthermore, the model performs two distinct data assimilation tasks without retraining.

Conditioning on pointwise time-series observations is sufficient to reproduce the turbulent statistics encoded in the prior without explicit statistical observation operators, provided that the observations are consistent with that prior. Conditional diffusion-based data assimilation therefore requires a delicate balance between observation density and conditioning strength~\citep{chung2023diffusion}. When properly tuned, the method recovers near DNS-level statistical fidelity; when this balance is not achieved, two failure modes emerge. Sparse observations can overweight isolated observations and amplify noise, whereas overly dense or strongly correlated constraints may lead to an overly concentrated posterior and degrade turbulent statistics.

While the present work establishes a conceptually and methodologically promising foundation, several developments are required for deployment in practical applications. Most importantly, the model must generalise across varying flow conditions. 
The present work demonstrates the framework only within the training distribution; whether the proposed conditioning strategy can steer the posterior towards statistics that depart from the prior, thereby supporting interpolation or controlled extrapolation across operating conditions, remains to be established and is a primary direction for future work.
In addition, the current temporal generation window is too short for realistic operational use; this limitation may be addressed through autoregressive rollouts~\citep{shysheya2024conditional}. Finally, practical wind-farm applications require the ability to handle complex geometries such as turbines. Generalisation to varying obstacle geometries may be achieved by treating them as additional conditioning variables~\citep{hu2025generative}.

\begin{bmhead}[Acknowledgments.]
We thank Dr. Hossein Gorji of EMPA, Dr. Ruifeng Hu of Lanzhou University, Dr. Zhiye Zhao of HKUST, and Prof. Chuangxin He of SJTU for their insightful discussions, and the anonymous reviewers for their constructive comments.
\end{bmhead}

\begin{bmhead}[Funding.]
The authors acknowledge support from
the Deutsche Forschungsgemeinschaft (DFG, German Research Foundation) under Germany’s Excellence Strategy - EXC 2075 - 390740016 and support from the Stuttgart Center for Simulation Science (SimTech). 
F.S. and B.T. are supported by the Carl Zeiss Foundation (CZS Project Number P2021-04012), which is gratefully acknowledged.
The authors also acknowledge support by the state of Baden--Württemberg through bwHPC and the German Research Foundation (DFG) through grant INST 35/1597-1 FUGG.
\end{bmhead}

\begin{bmhead}[Data availability statement.]
The data and code that support the findings of this study will be made openly available at \url{https://github.com/ITLR-DDSim/latent-diffusion-turbulence.git}.
\end{bmhead}

\begin{bmhead}[Declaration of interests.]
The authors report no conflict of interest.
\end{bmhead}

\begin{appen}
\section{The $\beta$-VAE and diffusion transformer implementation details}\label{appA}
We provide a detailed description of the $\beta$-VAE and diffusion transformer model implementation details such as architecture, training and hyperparameter choices. Both the $\beta$-VAE and the DiT are trained with the Adam optimizer \citep{DBLP:journals/corr/KingmaB14} at a learning rate of $1\times10^{-4}$. The $\beta$-VAE, which only needs to be trained once for diffusion models sharing the same latent dimension, required approximately 100 GPU hours on an NVIDIA A100 Tensor Core GPU. Each diffusion model was then trained for about 2 GPU hours on
the same hardware. 
Generating a prior (unconditional) sample consisting of 10 snapshots takes around 2 seconds wall-clock on an NVIDIA RTX6000 Ada GPU, including the time required to decode the generated samples to the physical space. Generating a posterior sample, on the other hand, takes around 180 seconds on the same hardware. This substantial difference in computational cost between prior and posterior sampling is due to the backpropogation needed to compute the gradient in equation \ref{eq:dps_alt}. Backpropogation through the $\beta$-VAE is particularly expensive, since $\beta$-VAE has $O(10^8)$ parameters, as explained in the next subsection.
\subsection{$\beta$-variational autoencoder}
\label{subsec:app_VAE}
A snapshot of the flow at time $t$ is defined as the matrix \begin{equation}
	\boldsymbol{\phi}(t)=[\boldsymbol{\Phi}(x_1,y_1,z_1,t),\, \cdots\,,\boldsymbol{\Phi}(x_{d_1},y_{d_2},z_{d_3},t)]\in\mathbb{R}^{d_c\times d}.
\end{equation}
where $d \equiv d_1 \times d_2 \times d_3$, and $d_c = 4$ is the number of channels, corresponding to each of the field variables $u$, $v$, $w$, and $p$. 
We  store the flow snapshots in physical space as four-dimensional arrays $\boldsymbol{\phi}(t)\in\mathbb{R}^{4\times d_1 \times d_2 \times d_3}$. Similarly, the latent representation of a single flow snapshot in the bottleneck layer of the $\beta$-VAE in the implementation is stored as a four-dimensional array $\mathsf{z}\in \mathbb{R}^{\tilde{d}_c  \times \tilde{d}_{1} \times \tilde{d}_{2} \times\tilde{d}_{3}}$. The number of channels and the reduced physical dimension $\tilde{d}=\tilde{d}_{1} \tilde{d}_{2} \tilde{d}_{3}$ are both design choices, and define the number of degrees of freedom of the latent variable $d_{\latent} = \tilde{d}_c\tilde{d}$. The number of layers in the encoder and decoder  networks varies depending on the number of degrees of freedom of the latent variable and so does the number of total parameters of the $\beta$-VAE (table \ref{tab:vae_details}).

A schematic of the $\beta$-VAE architecture is shown in figure \ref{fig:detailed_beta_VAE_schematic}.
We use convolutional layers with rectified linear unit (ReLU) as activation function the $\beta$-VAE. Standard convolutions are used in all convolutional layers. In the encoder, all convolutional layers in the networks have a kernel size of three, padding of one and a stride of two. Each convolutional layer is followed by a residual block also containing convolutional layers to improve training performance and stability. In the final layer, convolutions with a kernel size of one are used to obtain the mean and variance of the variational distribution. The decoder uses trilinear interpolation to upsample the data, followed by convolutional layers with a kernel size of three, padding of one and stride of one. 
\begin{table}
	\centering
	\small
	\begin{tabular}{c  c  c  c}
		$d_{\mathsf{z}}$  & Number of Layers        & $\tilde{d}_c  \times \tilde{d}_{1} \times \tilde{d}_{2} \times\tilde{d}_{3}$  & Number of Parameters \\[3pt]
		4         & 6                              &$1\times2\times1\times2$             &\(5.28\times10^8\)\\
		8         & 6                              &$2\times2\times1\times2$             &\(5.28\times10^8\)\\
		16        & 6                              &$4\times2\times1\times2$             &\(5.28\times10^8\)\\
		32        & 5                              &$1\times4\times2\times4$             &\(1.32\times10^8\)\\
		64        & 5                              &$2\times4\times2\times4$             &\(1.32\times10^8\)\\
		96        & 5                              &$3\times4\times2\times4$             &\(1.32\times10^8\)\\
	\end{tabular}
	\caption{Details of the $\beta$-VAEs used in this work.}
	\label{tab:vae_details}
\end{table}
\begin{figure}[t!]
	\centering
	\includegraphics[width=\linewidth]{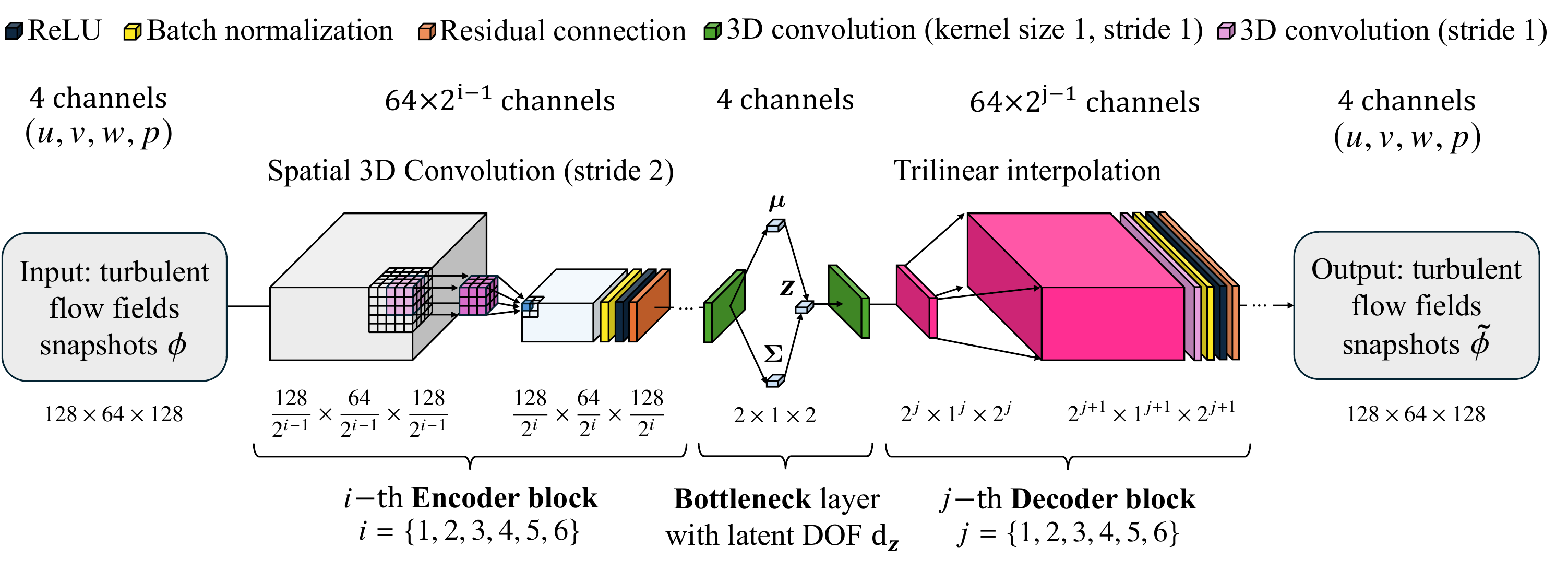}
	\caption{Schematic of the $\beta$-VAE architecture for $d_{\latent}=16$ in the first training stage. 
		In the bottleneck layer $\boldsymbol{\mu}$ and $\boldsymbol{\Sigma}$ are the mean and variance of the variational distribution. The three-dimensional convolutions in the encoder and decoder block have a kernel size of three.
	}\label{fig:detailed_beta_VAE_schematic}
\end{figure}

For training the $\beta$-VAE with the loss function described in equation~\ref{eq:vae_loss}, we set the appropriate $\beta$ value as $5 \times 10^{-4}$. As previously mentioned, the values of this scalar hyperparameter is chosen to balance the trade-off between reconstruction accuracy and disentanglement within the latent representations~\citep{higgins2017beta}. Note that the optimal $\beta$ value for our three-dimensional turbulent plane Couette channel flow is two orders of magnitude lower than the one identified for two-dimensional periodic flow at $Re=40$~\citep{solera2024beta}. Both values are noticeably lower than the ones reported for natural image tasks which are of the order $O(1)$~\citep[e.g.,][]{higgins2017beta}.

\subsection{Diffusion transformer}
\label{subsec:app_DiT}
\begin{figure}[t!]
	\centering
	\includegraphics[width=0.75\linewidth]{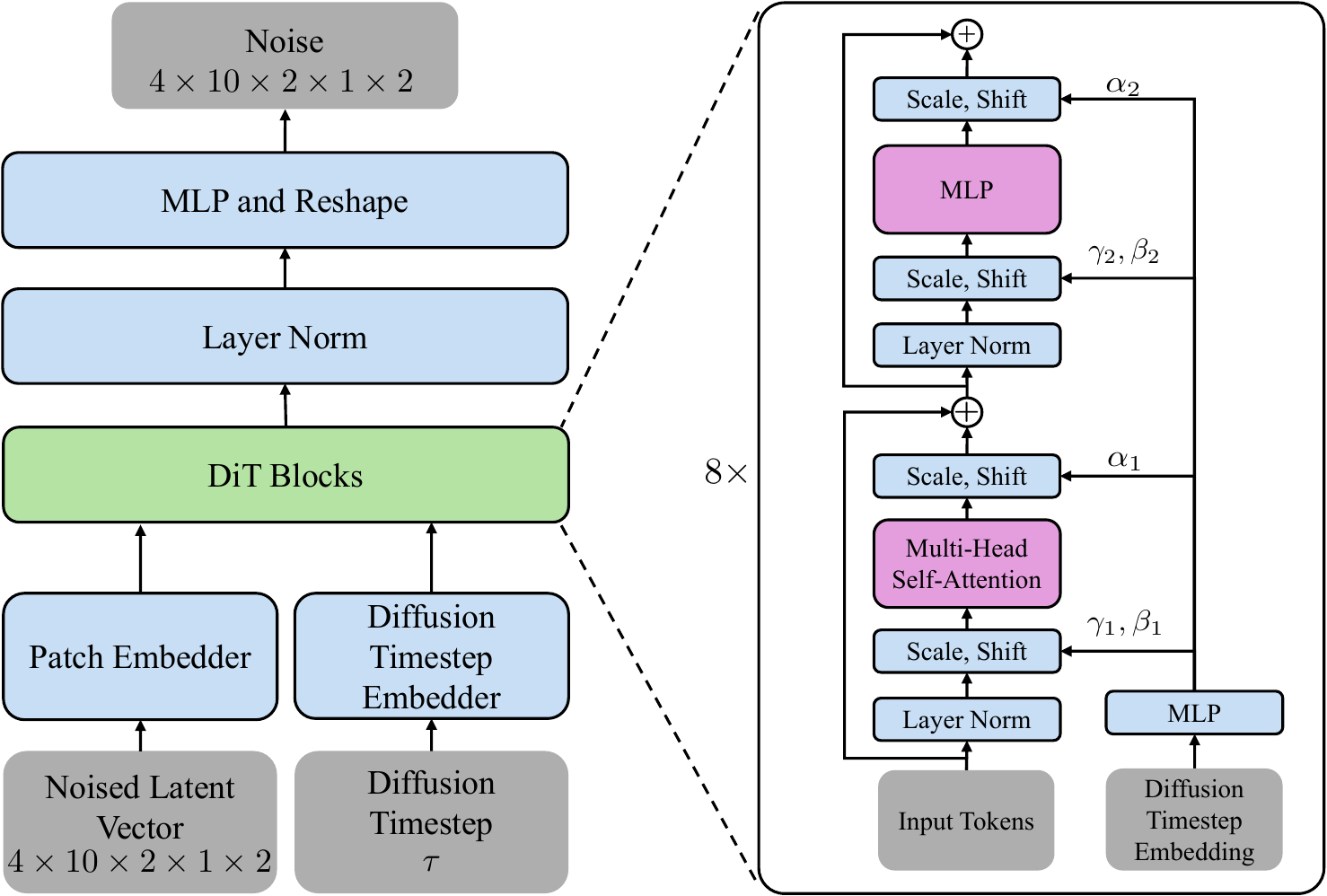}
	\caption{Schematic of the DiT architecture. The indicated shapes for the noised latent vector and the noise are for \(d_{\latent}=16\).}
	\label{fig:dit-schematic}
\end{figure}

We use the diffusion transformer (DiT) \cite{Peebles_2023_ICCV} as the neural backbone of our diffusion model. A schematic of the DiT architecture is shown in figure \ref{fig:dit-schematic}. DiT largely follows the design of the vision transformer (ViT) \citep{dosovitskiy2021an} originally developed for image classification, where the input image (or the noised image in the case of DiT) is first divided into a sequence of patches, which are then mapped to an embedding space via a patch embedding network. The dimensionality of this embedding space is termed ``hidden size" in the literature~\citet{arnab2021vivit, Peebles_2023_ICCV}, and is distinct from the dimensionality $d_\mathsf{z}$ of the latent space. The diffusion timestep \(\tau\) is also mapped to the same space via a feed-forward network. The resulting tokens are then fed to the DiT blocks. Each DiT block consists of a multi-head self-attention layer and a feed-forward  layer. The layer normalisation in the standard transformer encoder is replaced with an adaptive layer norm (adaLN), with the scaling and shifting parameters \(\gamma_{1,2}\) and \(\beta_{1,2}\) being learned from the diffusion timestep embeddings via a feed-forward network. 
A final feed-forward layer recovers the noise added to the image after the DiT blocks have processed the tokens. 

The ViT architecture was extended to the video domain by \cite{arnab2021vivit}, as was DiT itself \citep{lu2024vdt, ma2025latte}. In both cases, the extension to videos requires the patch embedding process described above to be performed in 3D (frame,height,width) instead of 2D. In this work, we adapt the DiT architecture to four-dimensional generation by treating sequences of three-dimensional latents as a 4D vector and dividing them into patches . These tokens are then fed to the DiT blocks, where the temporal and (latent) spatial dimensions are processed simultaneously, similar to ``Model 1'' in \cite{arnab2021vivit}. We also employ sinusoidal embeddings from \cite{vaswani2017attention} to encode position and time information into the tokens. Other hyperparameters are listed in Table \ref{tab:dit-hyperparameters}. Note that we use a patch size of 1 in every dimension, which is the lowest possible value for this hyperparameter. \cite{Peebles_2023_ICCV} report that decreasing the patch size leads to better performance. However, this benefit comes at an increased computational cost, as a smaller patch size leads to longer sequences. This effect is compounded by the quadratic scaling of the cost of self-attention with sequence length. In our setting, a patch size of 1 is still tractable because the latent space is extremely low-dimensional. Specifically, with a patch size of 1, the length of the sequence processed by the transformer is given by $N_\text{seq}=\tilde{d}_{1}\tilde{d}_{2}\tilde{d}_{3}d_t$ where $\tilde{d}_{1},\tilde{d}_{2},\tilde{d}_{3}$ denote the reduced physical dimensions and $d_t=10$ is the number of time snapshots generated by the model. Hence, for $d_\mathsf{z}=4,8,16$, which have $\tilde{d}_1=2$, $\tilde{d}_2=1$ and $\tilde{d}_3=2$, we have $N_\text{seq}=40$. For $d_\mathsf{z}=32,64,96$, which have $\tilde{d}_1=4$, $\tilde{d}_2=2$ and $\tilde{d}_3=4$, the sequence length is $N_\text{seq}=320$.

\begin{table}
	\small
	\centering
	\begin{tabular}{c  c}
		Hyperparameter  & Value\\[3pt]
		Number of DiT Blocks          & 8 \\
		Number of Heads         & 8 \\
		Hidden Size        & 240 \\
		Patch Size         & \(1\times1\times1\times1\)\\
		Total Number of Parameters & \(8.56\times10^6\)\\
	\end{tabular}
	\caption{Hyperparameters of the DiT. Hidden size refers to the dimension of the embedding space. Patch size refers to the dimension of each of the patches in the 4D latent space.}
	\label{tab:dit-hyperparameters}
\end{table}

\section{Derivation and Interpretation of the Likelihood Approximation}
\label{subsec:derivations}
The conditional score $\nabla_{\mathsf{z}_\tau}\log{\mathsf{p}(\mathsf{z}_\tau|\Psi)}$ can be rewritten using Bayes' rule as 
\begin{equation}
	\nabla_{\mathsf{z}_\tau}\log{\mathsf{p}(\mathsf{z}_\tau|\Psi)}=\nabla_{\mathsf{z}_\tau}\log{\mathsf{p}(\mathsf{z}_\tau)} +\nabla_{\mathsf{z}_\tau} {\log \mathsf{p}(\Psi|\mathsf{z}_\tau)}.
	\label{eq:cdm-bayes}
\end{equation}
As stated in \S\ref{subsec:Methods_cDM}, the first term in this equation is the unconditional (prior) score function learned by the trained unconditional model. The second term, corresponding to the likelihood in Bayes' rule, is intractable and thus has to be approximated. To this end, the conditional probability $\mathsf{p}(\Psi|\mathsf{z}_\tau)$ can be factorized as 
\begin{equation}
	\mathsf{p}(\Psi|\mathsf{z}_\tau)=\int \mathsf{p}(\Psi \vert \mathsf{z}_0,\mathsf{z}_\tau)\mathsf{p}(\mathsf{z}_0 \vert \mathsf{z}_\tau)\,d\mathsf{z}_0=\int \mathsf{p}(\Psi \vert \mathsf{z}_0)\mathsf{p}(\mathsf{z}_0 \vert \mathsf{z}_\tau)\,d\mathsf{z}_0,
\end{equation}
which yields
\begin{equation}
	\label{eq:dps_approx}
	\mathsf{p}(\Psi|\mathsf{z}_\tau)=\mathbb{E}_{\mathsf{z}_0\sim p\left[\mathsf{z}_0 \vert \mathsf{z}_\tau\right]}[\mathsf{p}(\Psi \vert \mathsf{z}_0)]\simeq p(\Psi|\hat{\mathsf{z}}_0),
\end{equation}
where 
$\hat{\mathsf{z}}_0
=
\mathbb{E}_{\mathsf{z}_0\sim \mathsf{p}(\mathsf{z}_0 \vert \mathsf{z}_\tau)}
[\mathsf{z}_0]$
is the posterior-mean estimate of the clean latent variable
$\mathsf{z}_0$. In DDPM, this estimate is given by
\begin{equation}    
	\hat{\mathsf{z}}_0
	= 
	\frac{1}{\sqrt{\bar{\alpha}_\tau}}
	\left(
	\mathsf{z}_\tau
	- \sqrt{1 - \bar{\alpha}_\tau}
	\boldsymbol{\hat{\epsilon}}(\mathsf{z}_\tau, \tau;\theta)
	\right),
	\label{eq:ddpm-posterior-mean}
\end{equation}
where $\boldsymbol{\hat{\epsilon}}(\mathsf{z}_\tau,\tau;\theta)$ is the DDPM denoising network. Equation~\ref{eq:ddpm-posterior-mean} follows from Tweedie's posterior-mean formula applied to the DDPM forward noising process, as stated in Proposition~1 of~\citet{chung2023diffusion}.
The error in the approximation in equation \ref{eq:dps_approx} is bounded by the Jensen gap. For the Gaussian observation model $\Psi=\mathcal{F}(\mathsf{z})+\boldsymbol{\epsilon}, \text{where } \,\boldsymbol{\epsilon}\sim\mathcal{N}\!(\boldsymbol{0}, \sigma^2\mathsfbi{I})$, we have
\begin{equation}
	\label{eq:normal_dist}
	\mathsf{p}(\mathsf{\Psi \vert \mathsf{\hat{z}}_0})=\mathcal{N}\!(\Psi;\mathcal{F}(\mathsf{z}_0), \sigma^2\mathsf{I}).
\end{equation}
Substituting equation \ref{eq:normal_dist} in equation \ref{eq:dps_approx} and taking its gradient with respect to $\mathsf{z}_\tau$, we get 
\begin{equation}
	\label{eq:dps_first}
	\nabla_{\mathsf{z}_\tau} {\log \mathsf{p}(\Psi|\mathsf{z}_\tau)}\simeq-\frac{1}{\sigma^2}\nabla_{\mathsf{z}_\tau}\lVert\Psi-\mathcal{F}(\mathsf{\hat{z}}_0(\mathsf{z}_\tau))\rVert^2.
\end{equation}
Here, $\tilde{\rho} \equiv 1/\sigma^2$ controls the influence of conditioning on the generation process, which is essentially the weighting between the physical prior and the data. The gradient in equation \ref{eq:dps_first} is computed by automatic differentiation. 
\cite{chung2023diffusion} report that directly using equation~\ref{eq:dps_first}  leads to unstable generation. They suggest a normalisation of the  conditioning strength that effectively amounts a normalisation factor of
\begin{equation} \label{eq:rho-normalize}
	\rho^\star = -\frac{2}{\sigma^2} \lVert\Psi-\mathcal{F}(\hat{\mathsf{z}}_0)\rVert.
\end{equation}
Such a normalisation leads to
\begin{equation}
	\nabla_{\mathsf{z}_\tau}\log \mathsf{p}(\Psi|\mathsf{z}_\tau)\simeq -\rho\nabla_{\mathsf{z}_\tau}\lVert\Psi-\mathcal{F}(\hat{\mathsf{z}}_0)\rVert,
	\label{eq:dps_alt_app}
\end{equation}
where a hyperparameter $\rho$ is introduced, ranging from 0.3 to 1.0, which is independent of the observation noise variance~$\sigma^2$. In Bayesian data assimilation, $\sigma^2$ plays a crucial role by controlling the weighting between prior and likelihood. The instability reported, however, originates in the point-mass approximation of equation~\ref{eq:dps_approx}, which replaces the full distribution $\mathsf{p}(\mathsf{z}_0|\mathsf{z}_\tau)$ by the single estimate $\hat{\mathsf{z}}_0$ and so discards the uncertainty in $\mathsf{z}_0$ that is large at high noise levels~\citep{chung2025diffusioninverse}. With this uncertainty omitted, the bare $\sigma^2$-weighting no longer reflects the correct prior--likelihood balance, prompting them to adopt the normalisation used here, in which $\rho$ is tuned as an ad hoc hyperparameter rather than set by the physical observation noise variance $\sigma^2$. Later methods \citep[e.g., $\Pi$GDM;][]{song2023pseudoinverse} account for this
uncertainty by replacing the point-mass approximation with an isotropic
Gaussian approximation to $\mathsf{p}(\mathsf{z}_0|\mathsf{z}_\tau)$ whose
covariance depends on the diffusion time $\tau$.

Note that equation~\ref{eq:dps_alt_app} replaces the gradient of the squared $L^2$ norm in equation~\ref{eq:dps_first} with the gradient of the norm itself. While both forms reduce the residual, the squared norm produces a correction proportional to the discrepancy $(\Psi - \mathcal{F}(\hat{\mathsf{z}}_0))$, or \emph{innovation} in data assimilation terminology, whereas the norm yields a directionally similar but rescaled update controlled by~$\rho$. The motivation for this rescaling is apparent over the reverse trajectory. The squared-norm gradient
can be rewritten as~$\nabla_{\mathsf{z}_\tau}\lVert\Psi-\mathcal{F}(\mathsf{\hat{z}}_0(\mathsf{z}_\tau))\rVert^2 = 2\lVert\Psi - \mathcal{F}(\hat{\mathsf{z}}_0)\rVert \; \nabla_{\mathsf{z}_\tau}\lVert\Psi-\mathcal{F}(\hat{\mathsf{z}}_0)\rVert$ after applying the chain rule. Therefore the gradient of the innovation norm is multiplied by itself, so its magnitude $\lVert\Psi - \mathcal{F}(\hat{\mathsf{z}}_0)\rVert$ scales with the residual: early in the reverse process $\hat{\mathsf{z}}_0$ is still far from any data-consistent state and the raw update is large enough to destabilise the generation, whereas near the end the residual is small and the update would vanish before data consistency is enforced. Dividing by the innovation norm $\lVert\Psi-\mathcal{F}(\hat{\mathsf{z}}_0)\rVert$ removes this dependence and holds the guidance at a roughly constant strength set by~$\rho$; the normalisation is therefore a deliberate stabilisation rather than an approximation artefact. The same normalisation, however, responds only to the size of the residual, irrespective of its cause. For very sparse observations $\lVert\Psi-\mathcal{F}(\hat{\mathsf{z}}_0)\rVert$ is built from only a handful of terms and is small even when $\hat{\mathsf{z}}_0$ is far from the true state, so the normalisation inflates those few observations and the noise they carry, mirroring the small-residual limit above. The stabilising rescaling thus becomes a source of over-weighting in the sparse regime, consistent with the assimilation behaviour presented 
in \S\ref{subsec:DA}.

\section{Two-Point Correlation of the Prior Samples}
\label{app:two-point}
\begin{figure}[t!]
	\centering
	\includegraphics[width=\textwidth]{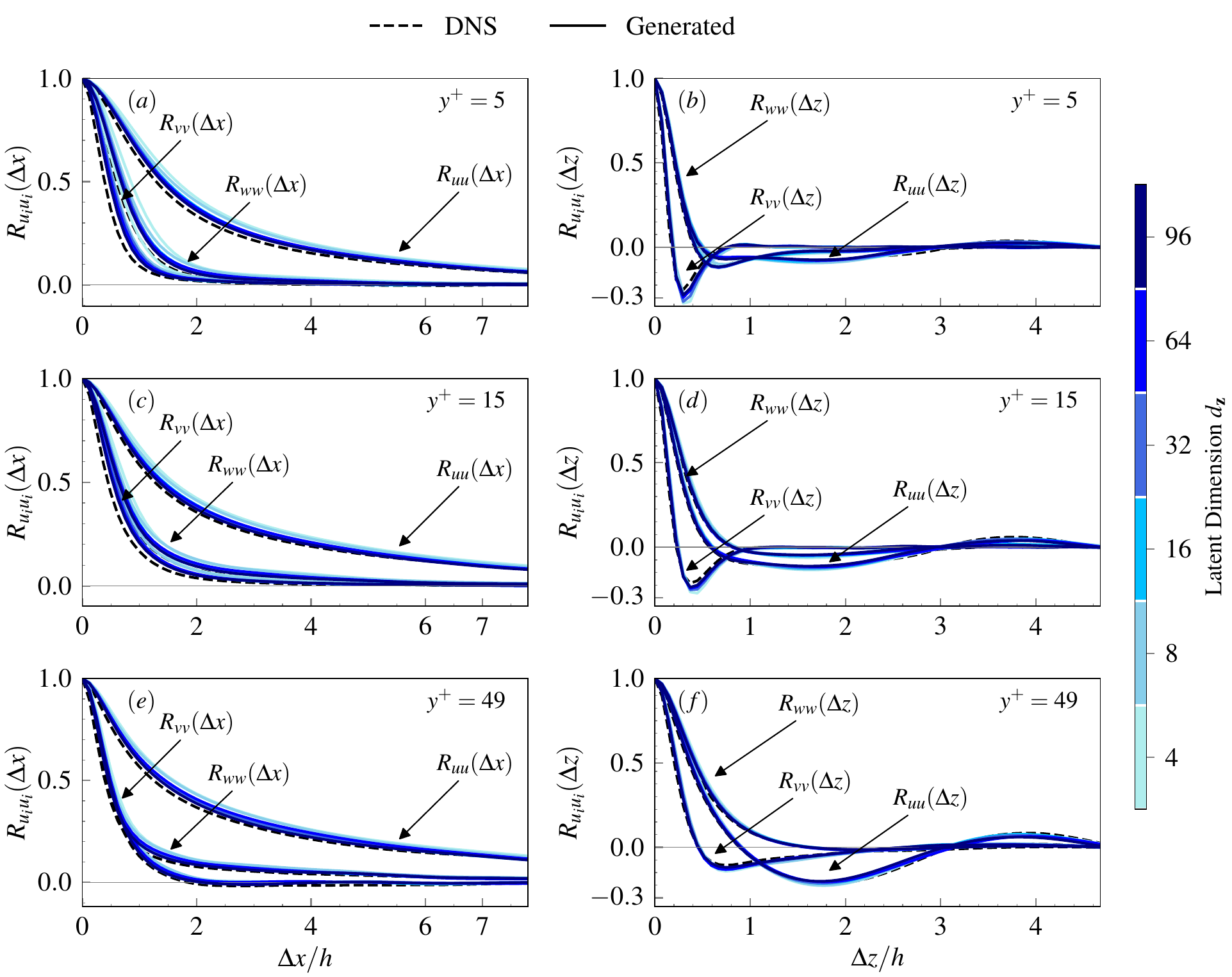}
	\caption{Two-point correlations of velocity fluctuations of the prior samples. We evaluate the correlations at three wall-normal locations: $(a,b)$ $y^+=5$, $(c,d)$ $y^+=15$, and $(e,f)$ $y^+=49$. The panels $(a,c,e)$ show streamwise two-point correlations $R_{u_i u_i}(\Updelta x)$, while $(b,d,f)$ show spanwise correlations~$R_{u_i u_i}(\Updelta z)$, for the streamwise, wall-normal, and spanwise velocity fluctuations. We report the statistics for flows generated with different latent dimensions $d_{\mathsf{z}}$ and compare with the DNS data.}
	\label{fig:two-point-correlation-components}
\end{figure}

The two-point velocity correlations provide the physical-space counterpart to the premultiplied energy spectra of figure~\ref{fig:premult_energy_spectra_spanwise} and extend the analysis to the wall-normal and spanwise velocity components and to a third wall-normal location, $y^+=5$. They are defined as
\begin{equation}
	R_{u_iu_i}(\Updelta x) = \frac{\overline{u_i'(x,z)u_i'(x+\Updelta x,z)}}{\overline{u_i'^2}}, \quad R_{u_iu_i}(\Updelta z) = \frac{\overline{u_i'(x,z)u_i'(x,z + \Updelta z)}}{\overline{u_i'^2}},
\end{equation}
where $i=\{1,2,3\}$ corresponds to the streamwise, wall-normal, and spanwise directions.
The streamwise correlations (figure~\ref{fig:two-point-correlation-components}$a,c,e$) decay slowly and do not cross zero within the domain, remaining at approximately $0.1$ in the buffer layer and $0.2$ in the core region. This is the physical-space expression of the broad low-wavenumber plateau of $k_xE_{uu}$ and confirms that the streamwise extent of the large-scale structures exceeds the streamwise period of the domain. The spanwise correlations (figure~\ref{fig:two-point-correlation-components}$b,d,f$) instead reach a first minimum within the domain, from which the dominant spanwise wavelength follows as $\lambda_z\approx2\Updelta z_{\mathrm{min}}$. At $y^+=5$ the first minimum of $R_{uu}(\Updelta z)$ lies at $\Updelta z\approx0.7$, corresponding to $\lambda_z^+\approx120$ and confirming that the near-wall streak spacing identified in the spectra persists to the viscous sublayer. At $y^+=49$ the minimum shifts to $\Updelta z\approx1.8$, giving $\lambda_z\approx3.6h$ for the superstructures of the core region, consistent with the spectral peak at $k_zh\approx2$. We note that the correlations are normalised by the variance and therefore constrain only the shape of the spectrum, not its level; the close agreement seen here is thus consistent with, and complementary to, the underestimation of the spectral peak magnitudes by the models with $d_{\mathsf{z}}<16$.

\section{Additional Results from Data Assimilation}
\label{app:Additional_results}
\begin{figure}[t!]
	\centering
	\includegraphics[width=\textwidth]{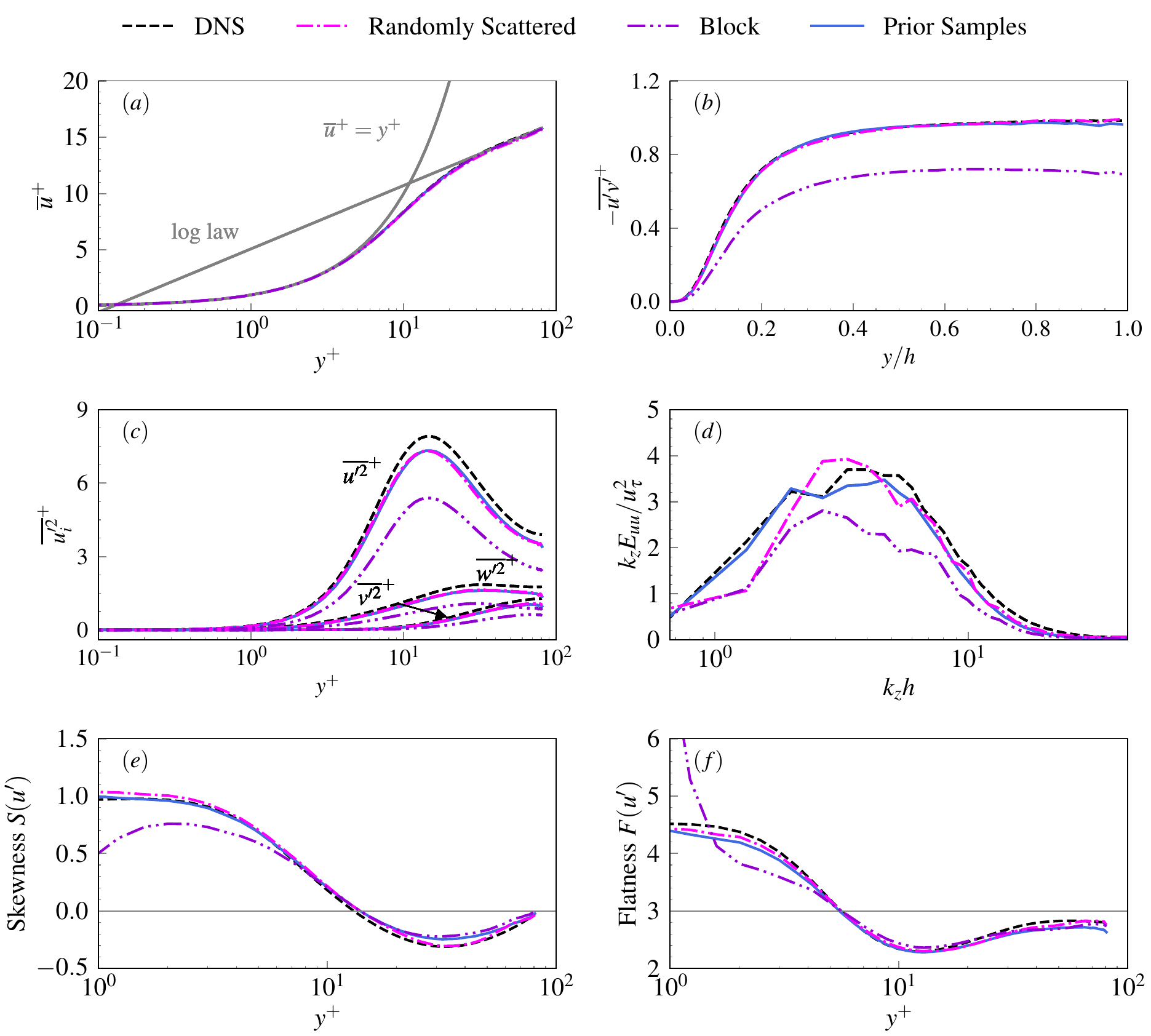}
	\caption{
		Performance of the conditionally generated flow for the scattered and block data assimilation tasks. The resulting statistics are shown for $(a)$ mean streamwise velocity, $(b)$ Reynolds shear stress, $(c)$ Reynolds normal stresses, $(d)$ pre--multiplied energy spectra for the streamwise velocity as a function of the spanwise wavenumber at $y^+=15$, $(e,f)$ skewness and flatness of the streamwise velocity fluctuation. We compare against prior samples with latent dimension $d_{\mathsf{z}}=32$ and the DNS reference.}
	\label{fig:da_sparse_block_statistics}
\end{figure}
\begin{figure}[t!]
	\centering
	\includegraphics[width=\textwidth]{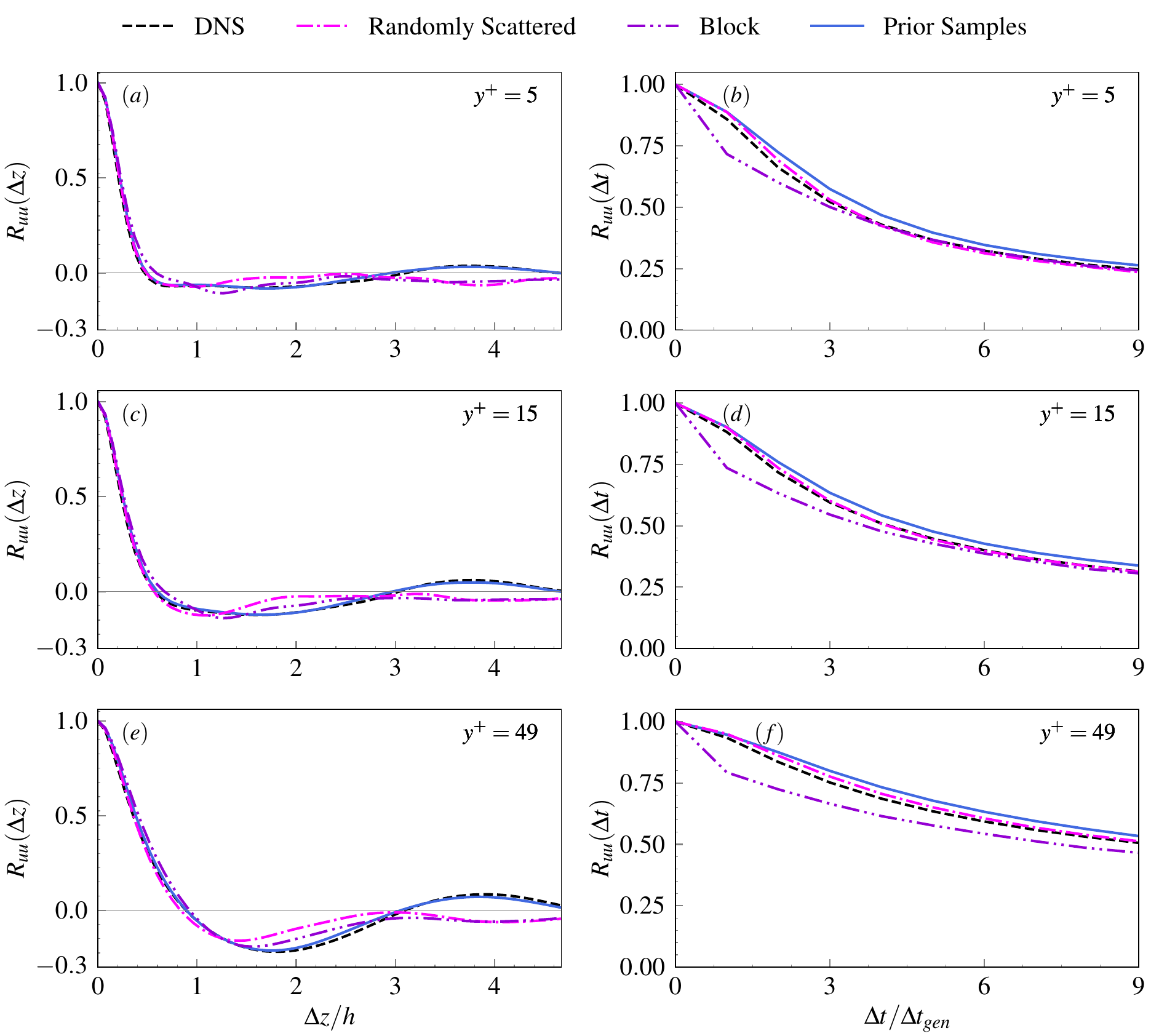}
	\caption{Statistics of the joint distribution of the conditionally generated flow for the scattered and block data assimilation tasks. $(a,c,e)$ Two-point correlation of the streamwise velocity fluctuations in the spanwise direction. $(b,d,f)$ Temporal autocorrelation of the streamwise velocity fluctuation. The results are shown for three different wall-normal locations $(a,b)$ $y^+=5$, $(c,d)$ $y^+=15$, and $(e,f)$ $y^+=49$. We compare against prior samples with latent dimension $d_{\mathsf{z}}=32$ and the DNS reference.}
	\label{fig:da_sparse_block_temp_corr}
\end{figure}
\begin{figure}[t!]
	\centering
	\includegraphics[width=\textwidth]{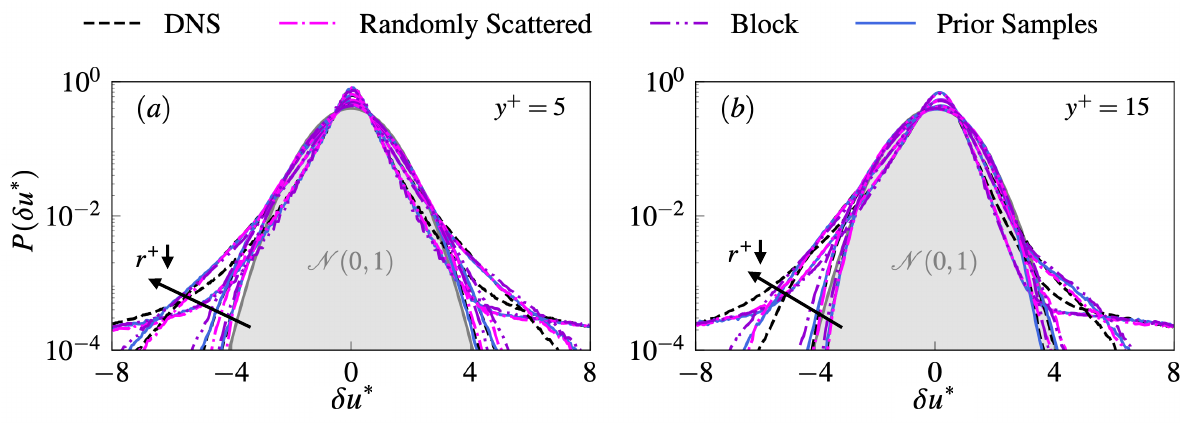}
	\caption{Probability density function of the standardized longitudinal velocity increment $\delta u^*$ for the conditionally generated flow for the scattered and block data assimilation tasks, at wall-normal stations (a) $y^+ = 5$ and (b) $y^+ = 15$. Each curve is the $\delta u^*$ distribution for one streamwise separation $r^+ = 350, 120, 40$, and $10$, standardized to zero mean and unit variance; the arrow indicates decreasing $r^+$. We compare against prior samples with latent dimension $d_{\mathsf{z}}=32$ and the DNS reference.}
	\label{fig:cond_velocity_increment}
\end{figure}
Additional results from the data assimilation experiments are presented here for both the randomly scattered observation baseline (R-1) and the block-arranged observation baseline (B-1), alongside comparisons with the prior samples. Figure \ref{fig:da_sparse_block_statistics} shows turbulent statistics averaged over 300 samples generated following the procedure described in \S\ref{subsec:DA}, which ensures that the resulting ensemble approximates the posterior distribution conditioned on the available observations.
Both configurations reproduce the mean streamwise velocity profile well (figure~\ref{fig:da_sparse_block_statistics}$a$), yet notable differences emerge for the Reynolds shear stress and Reynolds normal stresses, where the scattered observation task achieves substantially closer agreement with the DNS than the block arrangement (figure~\ref{fig:da_sparse_block_statistics}$b, c$).  The premultiplied energy spectra in figure~\ref{fig:da_sparse_block_statistics}($d$) indicate that the posterior samples broadly capture the characteristic length scales of coherent structures in the buffer layer, although the peak wavenumber is slightly underestimated in both cases. The block task additionally exhibits considerable suppression of turbulent fluctuations across a wide range of wavenumbers. The most pronounced discrepancies between the two configurations are found in the higher-order statistics. The skewness and flatness profiles of the block task deviate substantially from the DNS near the wall (figure \ref{fig:da_sparse_block_statistics}$e,f$).

Additionally, we show statistics of the joint distribution in figure~\ref{fig:da_sparse_block_temp_corr} for the same cases, specifically the two-point correlations and temporal autocorrelations of the velocity fluctuations. For the two-point correlations at $y^+ = 15$, both tasks recover the DNS and prior sample results up to separations of approximately $\Delta z/h = 1$, with small deviations appearing further from the wall. The temporal autocorrelation quantifies how velocity fluctuations at a fixed point are statistically correlated across time; its decay rate characterizes the integral timescale of the turbulence and is a key indicator of whether the model reproduces the correct temporal joint structure of the flow. Both diagnostics show good agreement with the DNS reference for the prior samples and the randomly scattered case. The block case shows a minor but clearly visible reduction in the temporal autocorrelation. Overall, the two metrics indicate that the spatial and temporal joint structure of the generated sequences is largely reproduced within the scales resolved by the subdomain, with the main deviation being the reduced temporal autocorrelation of the block case.

A further diagnostic of the scale-dependent structure is provided by the PDF of the normalised velocity increments $\delta u^*$ in figure~\ref{fig:cond_velocity_increment}, where the
scattered- and block-conditioned posteriors are compared against the prior samples
and the DNS reference. Both conditionings reproduce the DNS increment distributions
closely at either wall-normal station, recovering the heavy tails at small separations
$r^+=10,\,40$ as well as their gradual relaxation towards a Gaussian distribution as $r^+$
increases ($r^+=120, \,350$). Conditioning on pointwise observations therefore preserves the
scale-dependent intermittency of the flow. The scattered task follows the DNS most
closely and is essentially indistinguishable from the prior samples, while the block
task is only marginally worse, exhibiting small deviations in the tails but retaining
the same intermittency signature. This near-collapse of all curves follows the same
normalization argument as for the prior samples: the standardized increment
distribution probes the \emph{shape} of the increment statistics rather than their
amplitude, so that the deficiencies of the block task identified above --- the
underestimated Reynolds stresses and the broadband suppression of the energy spectra are divided out by the
standardization. The diagnostic consequently does not discriminate between the two
observation arrangements, which is to be expected since the shortcoming of the block
task lies in the magnitude of the fluctuations rather than in the shape of the
increment distribution. Importantly, neither conditioning introduces spurious
intermittency nor flattens the heavy tails relative to the DNS.

\section{Robustness to Observation Noise}
\begin{figure}[t!]
	\centering
	\includegraphics[width=\textwidth]{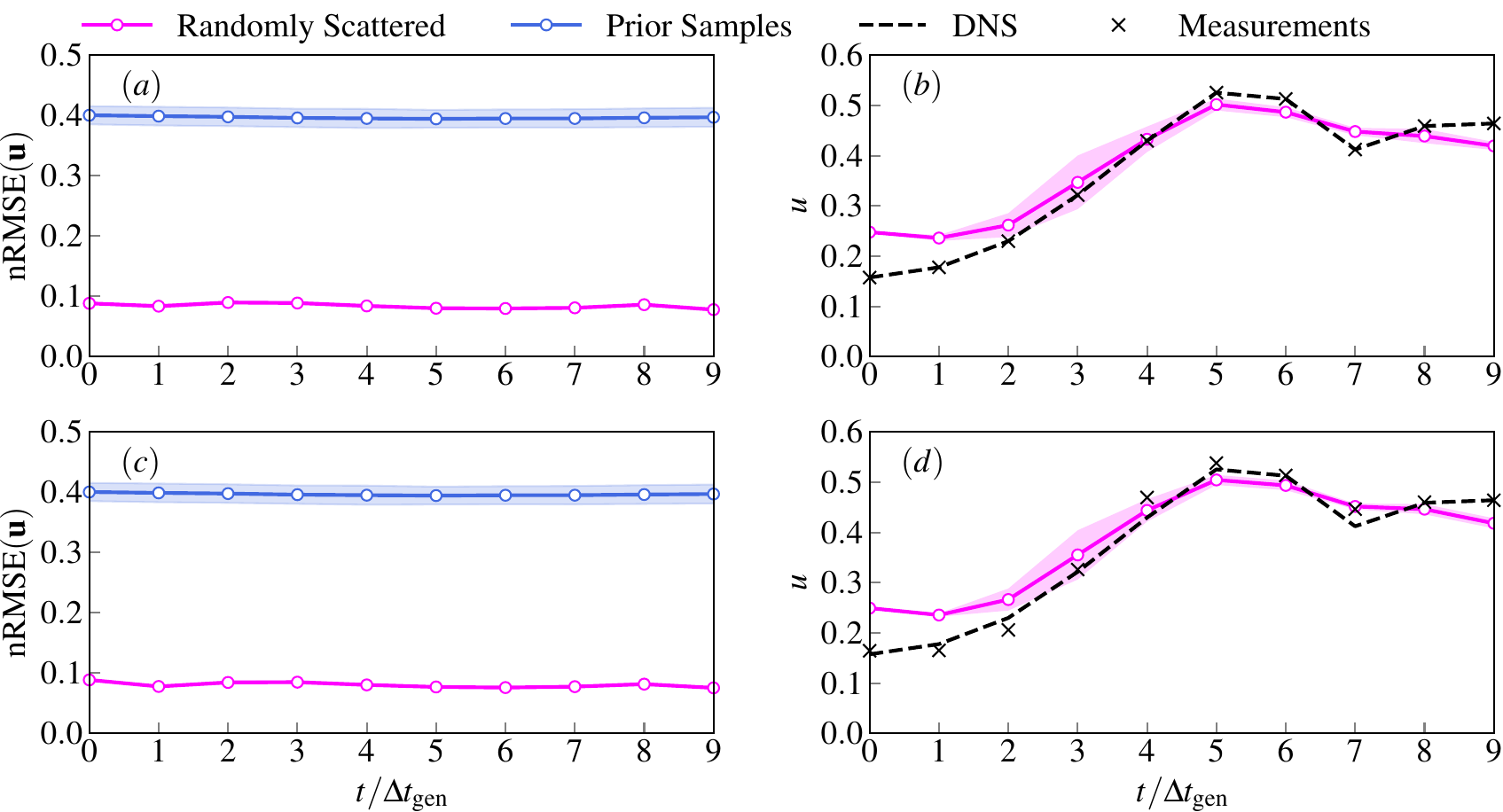}
	\caption{Effect of observation noise on posterior samples for the baseline randomly scattered observations case (R-1). We show the spatially averaged normalised root mean squared error (nRMSE) and the evolution of streamwise velocity at an observed location for $(a,b)$ noiseless and $(b,c)$ noised observations. The solid lines show ensemble means, while shaded regions indicate one standard deviation over 50 generated samples with the same observation.}
	\label{fig:noisy_obs_random}
\end{figure}
Here, we demonstrate the effect of observation noise on the posterior samples generated by our framework. We model noisy synchronous time series observations as 
\[
\Psi_{\,\text{noised}} \equiv \Psi
+ \boldsymbol{\varepsilon} \in \mathbb{R}^{d_c \times N_\text{obs} \times d_t}
\]
at $N_\text{obs}$ sensor locations. The observation noise $\boldsymbol{\varepsilon}$ is drawn from a Gaussian distribution $\boldsymbol{\varepsilon} \sim \mathcal{N}\!\left(\boldsymbol{0},\, 
\sigma^2 \mathbf{I}\right)$, with the standard deviation of the 
observation noise is given by~$\sigma=0.2\Psi'_{\text{rms}}(\boldsymbol{x})$, where $\Psi'_{\text{rms}}(\boldsymbol{x})$ denotes the local root mean squared flow field. The noise $\boldsymbol{\varepsilon}$ drawn independently for each of the four flow variables $u,v,w,p$.

\begin{figure}[t!]
	\centering
	\includegraphics[width=1\textwidth]{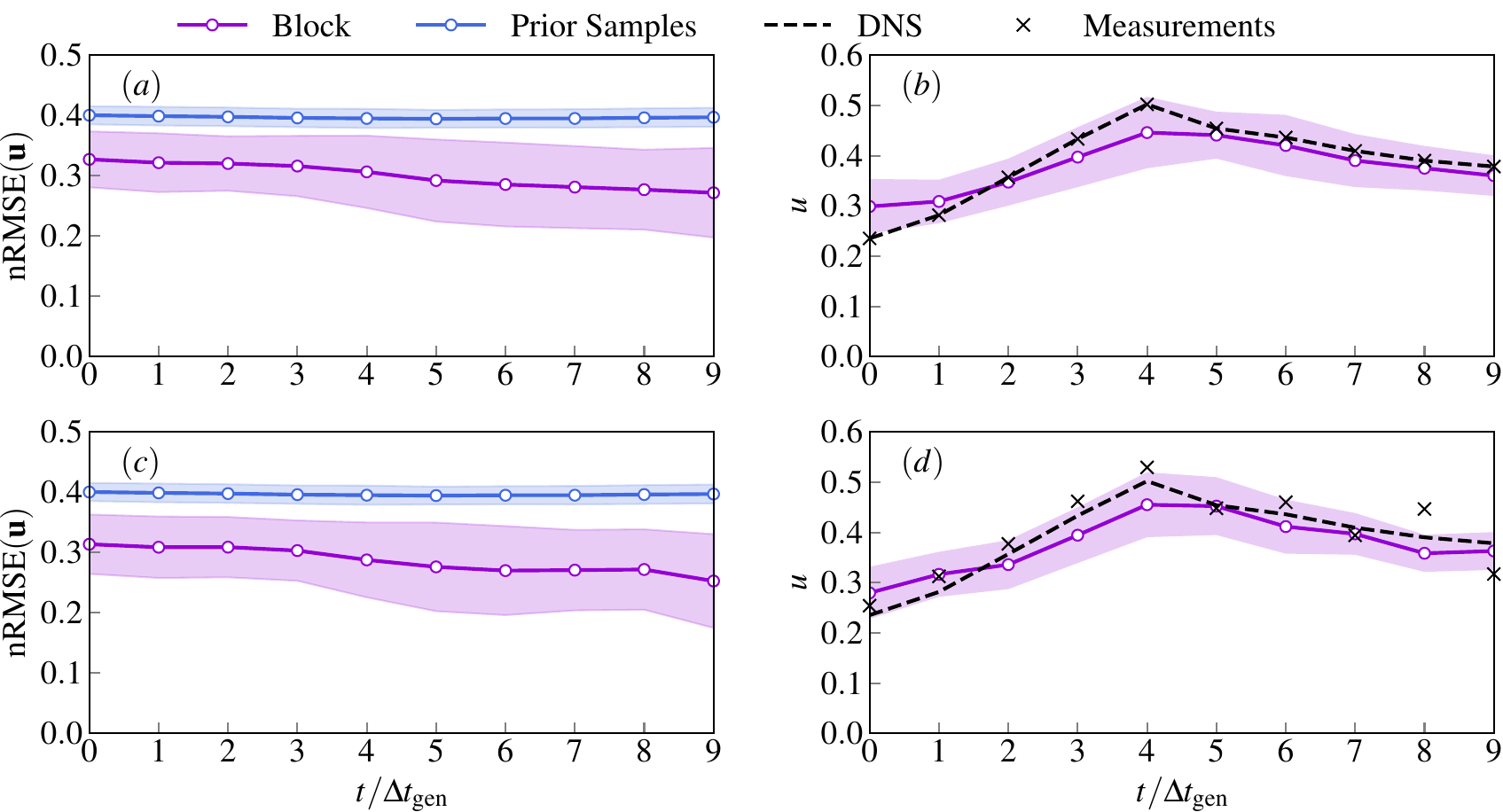}
	\caption{Effect of observation noise on posterior samples for the baseline block observations case (B-1). We show the spatially averaged normalised root mean squared error (nRMSE) and the evolution of streamwise velocity at an observed location for $(a,b)$ noiseless and $(b,c)$ noised observations. The solid lines show ensemble means, while shaded regions indicate one standard deviation over 50 generated samples with the same data.
		\label{fig:noisy_obs_block}}
\end{figure}

Our framework shows robustness to moderate observational noise. To assess this, we report the global normalised root mean squared error (nRMSE) alongside the temporal evolution of the streamwise velocity at an observed location for both noiseless and noisy observations in figure~\ref{fig:noisy_obs_random}, considering the baseline randomly scattered observation case (R-1). Neither the nRMSE nor the velocity trajectories exhibit any appreciable degradation upon the addition of noise, indicating that the framework remains effective under realistic measurement conditions. The corresponding results for the baseline block observation case (B-1) are presented in figure~\ref{fig:noisy_obs_block}, where the same conclusion holds: no notable reduction in performance is observed between the noiseless and noisy configurations.

\label{app:Noisy_Observations}

\end{appen}

\end{document}